\documentclass[pdflatex,sn-mathphys-num,oneside]{sn-jnl}

\makeatletter
\@twosidefalse
\@mparswitchfalse
\makeatother
\usepackage[utf8]{inputenc}
\usepackage[T1]{fontenc}
\usepackage{amsmath,amssymb,amsthm,mathtools}
\usepackage{changepage} 
\usepackage{bm}
\usepackage{bbm}
\usepackage{upgreek}
\usepackage{xcolor}
\usepackage{enumitem}
\usepackage{booktabs}
\usepackage{multirow}
\usepackage{tabularx}
\usepackage{siunitx}
\usepackage[capitalize]{cleveref}

\makeatletter
\def\abstractfont{\reset@font\fontsize{8.5bp}{10bp}\selectfont\leftskip=0pt\rightskip=0pt\parfillskip=0pt plus 1fil}
\makeatother
\usepackage{etoolbox}
\makeatletter
\patchcmd{\@maketitle}{\vskip21pt}{\vskip8pt}{}{}
\patchcmd{\@maketitle}{\vskip20pt}{\vskip10pt}{}{}
\patchcmd{\@maketitle}{\vskip20pt}{\vskip10pt}{}{}
\patchcmd{\@maketitle}{\vskip24pt}{\vskip10pt}{}{}
\patchcmd{\@maketitle}{\vskip24pt}{\vskip10pt}{}{}
\patchcmd{\@maketitle}{\vskip36pt}{\vskip8pt}{}{}
\patchcmd{\@maketitle}{\removelastskip\vskip10pt\ifpresentaddress}{\removelastskip\vskip0pt\ifpresentaddress}{}{}
\def\Titlefont{\reset@font\fontsize{16bp}{19bp}\selectfont\titraggedcenter}
\def\addressfont{\reset@font\fontsize{10bp}{11bp}\selectfont\titraggedcenter}
\def\keywordfont{\reset@font\fontsize{7.7bp}{8.5bp}\selectfont\leftskip=0pt\rightskip=0pt plus0.5fill}
\patchcmd{\keywords}{\addvspace{10pt}}{\addvspace{3pt}}{}{}
\makeatother

\newcommand{\rev}[1]{#1}
\newif\ifrevision \revisiontrue
\definecolor{RevBlue}{RGB}{20,70,140}
\newcommand{\revx}[1]{#1}

\DeclareRobustCommand{\BA}{{\boldsymbol{\rev{\mathbf A}}}}

\DeclareRobustCommand{\BE}{{\boldsymbol{\rev{\mathbf E}}}}
\DeclareRobustCommand{\BF}{{\boldsymbol{\rev{\mathbf F}}}}

\DeclareRobustCommand{\BI}{{\boldsymbol{\rev{\mathbf I}}}}
\DeclareRobustCommand{\BK}{{\boldsymbol{\rev{\mathbf K}}}}

\DeclareRobustCommand{\BM}{{\boldsymbol{\rev{\mathbf M}}}}
\DeclareRobustCommand{\BN}{{\boldsymbol{\rev{\mathbf N}}}}

\DeclareRobustCommand{\BR}{{\boldsymbol{\rev{\mathbf R}}}}
\DeclareRobustCommand{\BS}{{\boldsymbol{\rev{\mathbf S}}}}
\DeclareRobustCommand{\BT}{{\boldsymbol{\rev{\mathbf T}}}}

\DeclareRobustCommand{\BX}{{\boldsymbol{\rev{\mathbf X}}}}
\DeclareRobustCommand{\BY}{{\boldsymbol{\rev{\mathbf Y}}}}

\DeclareRobustCommand{\Ba}{{\boldsymbol{\rev{\mathbf a}}}}
\DeclareRobustCommand{\Bb}{{\boldsymbol{\rev{\mathbf b}}}}
\DeclareRobustCommand{\Bd}{{\boldsymbol{\rev{\mathbf d}}}}
\DeclareRobustCommand{\Be}{{\boldsymbol{\rev{\mathbf e}}}}

\DeclareRobustCommand{\Bg}{{\boldsymbol{\rev{\mathbf g}}}}
\DeclareRobustCommand{\Bn}{{\boldsymbol{\rev{\mathbf n}}}}
\DeclareRobustCommand{\Bp}{{\boldsymbol{\rev{\mathbf p}}}}
\DeclareRobustCommand{\Bq}{{\boldsymbol{\rev{\mathbf q}}}}
\DeclareRobustCommand{\Bt}{{\boldsymbol{\rev{\mathbf t}}}}

\DeclareRobustCommand{\Bu}{{\boldsymbol{\rev{\mathbf u}}}}
\DeclareRobustCommand{\Bv}{{\boldsymbol{\rev{\mathbf v}}}}
\DeclareRobustCommand{\Bw}{{\boldsymbol{\rev{\mathbf w}}}}
\DeclareRobustCommand{\Bx}{{\boldsymbol{\rev{\mathbf x}}}}

\DeclareRobustCommand{\Bzero}{{\boldsymbol{\rev{\mathbf 0}}}}

\DeclareRobustCommand{\Bsigma}{{\boldsymbol{\sigma}}}
\DeclareRobustCommand{\Bepsilon}{{\boldsymbol{\epsilon}}}

\DeclareRobustCommand{\Bvarphi}{\rev{\bm{\upvarphi}}}
\DeclareRobustCommand{\Bnabla}{{\boldsymbol{\nabla}}}

\DeclareRobustCommand{\calB}{{\mathcal B}}

\DeclareRobustCommand{\calF}{{\mathcal F}}

\DeclareRobustCommand{\calL}{{\mathcal L}}

\DeclareRobustCommand{\calN}{{\mathcal N}}
\DeclareRobustCommand{\calP}{{\mathcal P}}

\DeclareRobustCommand{\calT}{{\mathcal T}}
\DeclareRobustCommand{\calU}{{\mathcal U}}
\DeclareRobustCommand{\calV}{{\mathcal V}}
\DeclareRobustCommand{\calW}{{\mathcal W}}

\newcommand{\dV}{\mkern\medmuskip dV}
\newcommand{\dA}{\mkern\medmuskip dA}

\newcommand{\tr}{\operatorname{tr}}

\newcommand{\Uspace}[3]{\mathcal{U}\!\left(#1,\mathcal{C}^{#2},#3\right)}
\newcommand{\Pspace}[3]{\mathcal{P}\!\left(#1,\mathcal{C}^{#2},#3\right)}

\usepackage{graphicx}
\makeatletter
\def\caption@documentclass{standard}
\makeatother
\usepackage{subcaption}
\hypersetup{bookmarksdepth=subsubsection}
\usepackage{tikz}

\newlength{\CookCBHeight}
\newlength{\CookCBWidth}

\newlength{\CookQZeroCBHeight}
\newlength{\CookQZeroCBWidth}

\definecolor{kTwoColor}{HTML}{C96B12}
\definecolor{kThreeColor}{HTML}{0B6F8E}
\definecolor{kFourColor}{HTML}{5AA03C}

\newlength{\KirschCBHeight}
\newlength{\KirschCBWidth}

\newlength{\KirschCTECBHeight}
\newlength{\KirschCTECBWidth}

\definecolor{plotorange}{RGB}{191,87,0}
\definecolor{plotblue}{RGB}{0,95,134}
\definecolor{plotgreen}{RGB}{87,157,66}

\newcommand{\ktwo}{\textcolor{plotorange}{orange ($k=2$)}}
\newcommand{\kthree}{\textcolor{plotblue}{blue ($k=3$)}}
\newcommand{\kfour}{\textcolor{plotgreen}{green ($k=4$)}}

\definecolor{stablegreen}{RGB}{0,120,70}
\definecolor{unstablered}{RGB}{190,30,45}

\newcommand{\stablepair}[1]{\textcolor{stablegreen}{#1}}
\newcommand{\unstablepair}[1]{\textcolor{unstablered}{#1}}

\begin{document}

\title[Local stabilized IGA with discontinuous pressure]{A Local Stabilized Isogeometric Formulation with Discontinuous Pressure for Mesh-Conforming and Immersed Analysis of Nearly Incompressible Elasticity}

\author*[1]{\fnm{Ashkan} \sur{Ali Madadi}}\email{ashkan31194@utexas.edu}

\author[1,2]{\fnm{Thomas J.R.} \sur{Hughes}}

\affil[1]{\orgdiv{Oden Institute for Computational Engineering and Sciences}, \orgname{The University of Texas at Austin}, \orgaddress{\city{Austin}, \state{TX}, \country{USA}}}

\affil[2]{\orgdiv{Department of Aerospace Engineering and Engineering Mechanics}, \orgname{The University of Texas at Austin}, \orgaddress{\city{Austin}, \state{TX}, \country{USA}}}

\abstract{
Nearly incompressible materials arise in soft tissues, elastomers, gels, and other soft solids. Yet, their numerical modeling remains challenging because displacement-based formulations can exhibit volumetric locking near incompressibility. Mixed displacement–pressure formulations address this difficulty but require compatible approximation spaces satisfying the inf–sup condition. In mixed formulations, discontinuous pressure spaces are attractive because their pressure degrees of freedom can be condensed locally at the element level. In high-order isogeometric analysis, however, Galerkin stability of discontinuous pressure spaces requires increasing pressure-grid coarsening with polynomial degree. Here, we develop a variational multiscale (VMS) stabilized mixed isogeometric formulation that retains the locality of discontinuous pressure spaces while substantially relaxing this restriction. The key observation is that VMS stabilization controls nonconstant elementwise pressure modes and is unconditionally stable with respect to the stabilization parameter, provided the elementwise constant pressure component satisfies the inf–sup condition. Consequently, for maximally continuous displacement splines, a pressure grid coarsened by a fixed factor of two satisfies the constant-mode inf–sup criterion independently of polynomial degree. Local condensation then yields a displacement-only system, facilitating incorporation into existing displacement solvers. The formulation is further extended to immersed isogeometric analysis using extended B-splines, pressure-cell aggregation, subdivision-based cut-cell quadrature, and weak enforcement of essential boundary conditions. Discrete inf–sup studies and convergence analyses in the Kirsch and Cook’s membrane benchmarks demonstrate the stability behavior and show that the immersed treatment preserves the mixed stability properties. The proposed framework therefore provides a local, high-order, and robust approach for nearly incompressible problems in both mesh-conforming and immersed settings.
}

\keywords{Stokes problem, Mixed formulation, Discontinuous pressure, Isogeometric analysis, Inf-sup stability, Residual-based stabilization, VMS stabilization, Immersed methods, Nitsche method, Cut-cell stability}

\maketitle

\section{Introduction}
\label{sec:introduction}

\noindent Nearly incompressible elasticity presents a classical difficulty for
displacement-based finite element formulations, which can become excessively
stiff and exhibit volumetric locking as the incompressible limit is
approached \cite{MalkusHughes1978,Hughes1980}. Mixed formulations provide a
natural alternative by introducing the pressure as an independent variable
and enforcing the incompressibility constraint through a
displacement--pressure saddle-point problem. This formulation has the same
fundamental mathematical structure as the velocity--pressure formulation of
the incompressible Stokes equations and has been extensively studied in the
finite element literature \cite{MalkusHughes1978,Brezzi1974,kadapa2016subdivision,madadi_subdivision-stabilized_2024}. The stability of the resulting
discrete problem depends critically on the compatibility of the displacement
and pressure approximation spaces, characterized by the
Ladyzhenskaya--Babu\v{s}ka--Brezzi (LBB), or inf--sup, condition
\cite{Brezzi1974}. Discretizations that do not satisfy this condition can
admit spurious pressure modes, which may manifest as nonphysical oscillations
in the pressure and, consequently, in the computed stress field. The
construction of stable and computationally efficient mixed approximation
spaces therefore remains a central consideration in the numerical treatment
of nearly incompressible problems.

Isogeometric analysis (IGA), introduced by Hughes et al.
\cite{hughes2005isogeometric,HUGHES20054135,cottrell2009isogeometric},
provides a high-order computational framework in which the spline and NURBS
basis functions commonly used for geometric representation in computer-aided
design (CAD) are also employed to approximate the solution fields, thereby
seeking to bridge the gap between geometric design and numerical analysis.
In addition to representing many CAD geometries exactly, IGA enables the
systematic construction of high-order and highly continuous approximation
spaces, which can provide high accuracy per degree of freedom and improved
approximation properties compared with conventional \(\mathcal{C}^0\) finite elements
\cite{THOMAS201555,HUGHES20054135}. These features have made IGA particularly
attractive for problems that benefit from smooth, high-order approximations.
For mixed problems, however, the increased continuity of spline spaces makes
the construction of inf--sup-stable displacement--pressure pairs more
restrictive than in the conventional finite element setting. For example,
the classical Taylor--Hood strategy of reducing the pressure degree by one
relative to the displacement does not, when both fields are represented by
maximally continuous spline spaces on the same mesh, generally yield an
inf--sup-stable isogeometric pair
\cite{kadapa2016subdivision,madadi_subdivision-stabilized_2024}.
A successful alternative is to retain the high continuity of both fields
while defining the pressure approximation on a mesh coarsened by a factor
of two relative to the displacement mesh. Such subdivision-based mixed
discretizations have been shown to satisfy the inf--sup condition for
both Taylor--Hood and equal-order pairs and have been successfully applied
to incompressible and nearly incompressible problems
\cite{ruberg2012subdivision,ruberg2014fixed,kadapa2016subdivision,
madadi_subdivision-stabilized_2024}.
More generally, theoretical analyses have established that the stability of
mixed isogeometric discretizations depends on the interplay among the
polynomial degrees, inter-element continuities, and relative mesh resolutions
of the displacement and pressure spaces \cite{8179196,Sangaliref2017}.
Consequently, familiar degree-based pairing rules from \(\mathcal{C}^0\) finite
elements cannot be transferred directly to highly continuous spline
discretizations, and the displacement and pressure spaces must instead be
constructed with their degree, continuity, and relative mesh structure
considered together.

Discontinuous pressure approximations are particularly attractive for mixed
formulations because the associated pressure degrees of freedom have
element-local support. Consequently, the pressure block of the discrete
mixed system is block diagonal at the element level, allowing the pressure
unknowns to be eliminated locally through static condensation. The globally
assembled system then contains only displacement degrees of freedom, while
the pressure field can be recovered locally after the displacement solution
has been obtained. This locality is especially appealing from an
implementation perspective, since mixed incompressibility treatments can be
incorporated into computational frameworks based on a single-field
displacement solver without requiring a globally coupled
displacement--pressure solution procedure. Several approaches have been
proposed to achieve such locality in isogeometric formulations. The classical isogeometric \(\bar{B}\) formulation, which is equivalent to
the condensed form of the corresponding mixed
formulation, employs a projection onto a continuous lower-order spline space
and provides an effective high-order treatment of nearly incompressible
problems \cite{elguedj2008f,ELGUEDJ20082732}; however, the corresponding
\(L^2\) projection is global and introduces additional coupling and
computational cost \cite{antolin2017isogeometric}. Diagonal mass lumping has been
considered as one approach for avoiding the inversion of the global
projection operator, but this approximation destroys consistency and its
accuracy can deteriorate away from the nearly incompressible limit
\cite{antolin2017isogeometric}. Local quasi-interpolation provides another strategy for avoiding the global
projection, with B\'ezier projection offering a particularly convenient
local construction for spline and NURBS spaces
\cite{THOMAS201555,MIAO2018273}. Symmetric and nonsymmetric local
\(\bar{B}\) formulations based on such projection operators have been
investigated by Antolin et al.~\cite{antolin2017isogeometric}; however, these
formulations are not consistent with the underlying mixed variational
problem. The symmetric formulation exhibits suboptimal convergence, whereas
the nonsymmetric formulation recovers optimal convergence despite this lack
of variational consistency \cite{antolin2017isogeometric}. As an alternative, Antolin et al.~\cite{antolin2017isogeometric}
introduced a discontinuous \(\bar{B}\) formulation in which the volumetric
field is projected onto a lower-degree discontinuous space. The
discontinuity of the projection space makes the projection element-local,
preserves the sparsity of the resulting displacement system, and retains a
Galerkin variational structure. However, the complete discontinuous
pressure space must then satisfy the corresponding discrete inf--sup
condition. For maximally continuous displacement spaces and discontinuous
pressure spaces, satisfying this condition requires progressively greater
coarsening of the pressure discretization as the polynomial degree is
increased \cite{antolin2017isogeometric}. This creates an increasing mismatch between
the approximation capabilities of the displacement and pressure spaces and
can limit the accuracy attainable for the pressure field at higher
polynomial orders. Thus, existing approaches either require a global
projection, sacrifice properties of the underlying variational formulation,
or impose increasingly restrictive pressure-space requirements at higher
polynomial orders. These limitations motivate the development of a
formulation that preserves the locality and static condensation afforded by
discontinuous pressure approximations while relaxing the inf--sup
restrictions associated with a fully Galerkin treatment.

Residual-based stabilization provides an alternative approach for relaxing
the inf--sup restrictions imposed on Galerkin mixed discretizations.
Originally developed in the context of incompressible flow and mixed finite
element formulations, methods such as the Galerkin/least-squares (GLS),
SUPG/PSPG-type, and variational multiscale (VMS) formulations introduce
consistent residual-based terms that stabilize otherwise incompatible
displacement--pressure approximation spaces
\cite{HUGHES1989173,HUGHES1986223,HughesFrancaBalestra1986,franca1991convergence,FrancaRolf1991,FrancaHughesStenberg1993,HUGHES1995387}. Although closely related, these formulations
possess different stability properties: the GLS and SUPG/PSPG-type
formulations are conditionally stable and require an appropriate upper bound
on the stabilization parameter, whereas the VMS formulation considered here
is unconditionally stable with respect to a positive stabilization
parameter \cite{franca1991convergence,HUGHES1986223,FrancaRolf1991,FrancaHughesStenberg1993,HughesFrancaBalestra1986}. These approaches are particularly attractive in the present
context because they provide a mechanism for stabilizing higher-order
discontinuous pressure approximations without requiring the complete
pressure space to satisfy the increasingly restrictive Galerkin inf--sup
condition. An important distinction, however, arises for discontinuous
pressure spaces. The pressure-gradient contribution appearing in the
residual-based stabilization controls the nonconstant elementwise pressure
modes, whereas the elementwise constant mode has identically zero gradient
and is therefore unaffected by this stabilization. The constant pressure
component must consequently be controlled separately, either by augmenting
the stabilization with suitable interelement pressure-jump terms or by
requiring inf--sup stability between the displacement space and the
elementwise constant pressure space. In the present work, the latter
approach is adopted, thereby retaining a purely element-local
residual-based stabilization \cite{HUGHES1986223,FrancaRolf1991,FrancaHughesStenberg1993}. This separation between the constant and
nonconstant pressure modes forms the basis of the stabilized
discontinuous-pressure formulation developed herein.

Beyond the construction of stable and computationally efficient mixed
approximation spaces, an additional objective is to retain these properties
when the computational discretization is not required to conform to the
physical geometry. Immersed computational methods provide an attractive
framework for this purpose by embedding the physical domain within an
independently defined background mesh, thereby avoiding the construction of
boundary-fitted meshes and simplifying the treatment of complex geometries.
Immersed isogeometric analysis combines this geometric flexibility with the
high-order and high-continuity approximation properties of spline spaces
\cite{casquero2016hybrid,casquero2021divergence,schillinger2012isogeometric,
fromm2023interpolation,fromm2024interpolation,casquero2015nurbs}.
The nonconforming representation, however, introduces several numerical
challenges that are absent from mesh-conforming discretizations. Essential
boundary conditions must generally be imposed weakly, for example using
Nitsche's method
\cite{fernandez-mendez_imposing_2004,annavarapu2012robust,
annavarapu2014nitsche,embar_imposing_2010,nitsche1971variationsprinzip,
yamaguchi2021extended,Madadi_2,ALIMADADI2027119339}, while elements intersected by the physical
boundary require specialized quadrature procedures for accurate integration
over the resulting irregular cut-cell domains
\cite{kamensky2015immersogeometric,divi2020error,verhoosel2023scan}.
In addition, basis functions having only a small fraction of their support
inside the physical domain can lead to severe ill-conditioning and loss of
robustness, requiring appropriate stabilization or basis-function treatment
\cite{Madadi_2,ALIMADADI2027119339,davydov_two-stage_2013,ruberg2012subdivision,
yamaguchi2021extended,marussig2017stable,marussig2018review}.
For mixed formulations, these difficulties are compounded by the requirement
that the cut-cell integration and the treatment of basis functions near the
immersed boundary preserve the inf--sup stability of the underlying
displacement--pressure discretization. This issue is particularly important
for the discontinuous pressure spaces considered here, since the treatment
of cut pressure cells must preserve both the accuracy and the inf--sup
control of the elementwise constant pressure mode. Although few mixed immersed
isogeometric formulations employing continuous pressure approximation spaces
have been considered in the literature within both Galerkin \cite{ruberg2012subdivision,Madadi_2,yamaguchi2021extended} and VMS
frameworks \cite{RiichiSugai2025}, corresponding formulations based on discontinuous
pressure spaces have received considerably less attention. In particular,
an immersed formulation that retains the locality and static condensation
afforded by discontinuous pressure approximations while preserving their
mixed stability properties appears to be lacking. Thus, extending the
stabilized discontinuous-pressure formulation to immersed discretizations
requires simultaneously addressing weak enforcement of boundary conditions,
accurate cut-cell integration, cut-cell-induced ill-conditioning, and
preservation of the intended mixed stability properties.

The present work develops a stabilized mixed isogeometric framework for
nearly incompressible elasticity with particular emphasis on local
discontinuous pressure approximations and their extension to immersed
discretizations. The principal contribution is a residual-based stabilized
discontinuous-pressure formulation that substantially relaxes the
polynomial-order-dependent inf--sup restriction of the corresponding
Galerkin construction. In the Galerkin discontinuous formulation of
Antolin et al.~\cite{antolin2017isogeometric}, stability requires
progressively greater coarsening of the pressure grid as the spline
polynomial degree is increased. In the proposed formulation, residual-based
stabilization controls the nonconstant elementwise pressure modes, leaving
only the elementwise constant pressure component to be controlled by the
inf--sup compatibility of the displacement and pressure spaces. As a
consequence, a pressure grid coarsened by a fixed factor of two relative to
the displacement grid is sufficient for all polynomial orders considered
here, eliminating the increasing disparity between the displacement and
pressure resolutions required by the Galerkin formulation. Importantly, this
improvement is obtained without sacrificing the principal computational
advantage of discontinuous pressure approximations: the pressure unknowns
remain element-local and can be statically condensed, resulting in a global
system involving displacement degrees of freedom only. The formulation is
developed within a unified residual-based setting encompassing GLS,
SUPG/PSPG-type, and VMS forms, with the VMS formulation providing
unconditional stability with respect to a positive stabilization parameter.
A second contribution is the extension of this stabilized
discontinuous-pressure formulation to immersed isogeometric analysis, for
which a corresponding local discontinuous-pressure formulation has received
little attention. Extended B-splines
\cite{ruberg2012subdivision,Madadi_2,yamaguchi2021extended} are employed for
the displacement approximation, whereby basis functions associated with
small cut supports are eliminated by extending the neighboring interior
basis functions into the cut region. For the discontinuous pressure space,
cut pressure cells are aggregated with neighboring interior pressure cells
to avoid poorly supported pressure degrees of freedom while preserving the
local character of the pressure approximation. Integration over the
irregular portion of cut cells is performed using a subdivision-based
octree quadrature \cite{kamensky2015immersogeometric,divi2020error,verhoosel2023scan}, in which cut regions are recursively subdivided and
integrated over their material portions. Essential boundary conditions on
the nonconforming immersed boundary are imposed weakly using Nitsche's
method \cite{nitsche1971variationsprinzip}, avoiding the need for boundary-fitted displacement degrees of
freedom. Together, these ingredients provide a robust immersed
implementation while retaining the locality of the discontinuous pressure
approximation.
Of particular interest is whether these immersed treatments preserve the
mixed stability properties of the underlying approximation spaces. The
discrete inf--sup studies demonstrate that the extended B-spline and
pressure-cell aggregation procedures preserve the intended stability of the
displacement--pressure pairs, including the inf--sup control of the
elementwise constant pressure mode. The resulting formulations are assessed
through discrete inf--sup studies, convergence analyses, and pressure and
stress fields for the Kirsch and Cook's membrane benchmarks in both
mesh-conforming and immersed settings.

\revx{The theoretical development is deliberately focused on the ingredients that are specific to the present formulation. The energy identities of \Cref{subsec:residual_based_stabilization}, the elementwise pressure decomposition of \Cref{subsec:discontinuous_stabilized_condition}, and the resulting stability conditions are stated and derived explicitly, while standard stabilized-method results are cited rather than re-proved. In particular, the classical discontinuous-pressure argument of Franca, Hughes, and Stenberg~\cite{FrancaHughesStenberg1993} extends to the present spline setting once the corresponding spline inverse estimate and constant-mode inf--sup condition are supplied; the latter is established here through the macroelement analysis.}

The remainder of this paper is organized as follows.
\Cref{sec:theory} presents the theoretical formulation and numerical
framework developed in this work. The Stokes-type mixed problem for nearly
incompressible elasticity is introduced in
\Cref{subsec:stokes_type_mixed_problem}, followed by the Galerkin
discretization and pressure-elimination procedure in
\Cref{subsec:galerkin_pressure_elimination}. The B-spline and NURBS
displacement--pressure approximation spaces are introduced in
\Cref{subsec:iga_approximation_spaces}, and their Galerkin inf--sup
requirements for both continuous and discontinuous pressure spaces are examined in \Cref{subsec:inf_sup_galerkin_IGA}. Residual-based GLS,
SUPG/PSPG-type, and VMS formulations are developed and analyzed in
\Cref{subsec:residual_based_stabilization}, while the choice and scaling of
the stabilization parameter are discussed in \Cref{subsec:tau_parameter}.
Local pressure elimination for the stabilized formulation is presented in
\Cref{subsec:pressure_elimination_stabilized_formulation}, followed by the
stability analysis of discontinuous pressure spaces in
\Cref{subsec:discontinuous_stabilized_condition}, where the distinct
requirements for the elementwise constant and nonconstant pressure modes are
established. Finally, the extension to nonconforming geometries is presented
in \Cref{subsec:immersed_iga_theory}, including cut-cell integration, weak
enforcement of essential boundary conditions, extended B-spline displacement
approximation, and discontinuous-pressure aggregation.
\Cref{subsec:kirsch_problem} considers the Kirsch problem as a smooth
benchmark with an analytical solution and examines convergence, discrete
inf--sup stability, and pressure and stress fields for continuous pressure,
discontinuous constant-pressure, and higher-order discontinuous-pressure
spaces in both mesh-conforming and immersed settings.
\Cref{subsec:cooks_membrane} then considers Cook's membrane as a more
challenging nearly incompressible benchmark with singular behavior and
examines the same three classes of pressure approximation, with particular
emphasis on spurious pressure and stress oscillations, discrete inf--sup
stability, and the effectiveness and parameter dependence of the
residual-based stabilization. Finally, \Cref{sec:conclusions} summarizes
the principal findings and conclusions of the work.


\section{Theory}
\label{sec:theory}

\noindent This section gives the theoretical formulation used in this work. The paper follows the linear Stokes-type mixed structure that appears both in incompressible flow and in nearly incompressible linear elasticity. The physical interpretation of the variables is different in the two settings, but the algebraic and stability structure is the same. In incompressible flow, the pressure is a Lagrange multiplier enforcing the exact incompressibility constraint. In nearly incompressible elasticity, the pressure is an auxiliary volumetric variable introduced to avoid volumetric locking and to represent the nearly constrained volumetric response.

The emphasis is on four questions. First, how does the mixed displacement-pressure formulation lead to a saddle-point problem? Second, when can pressure degrees of freedom be eliminated locally? Third, which isogeometric displacement-pressure pairs are inf-sup stable? Fourth, how can residual-based stabilization and immersed cut-cell treatments be combined with discontinuous pressure spaces while retaining a local formulation?

\subsection{Stokes-type mixed problem}
\label{subsec:stokes_type_mixed_problem}

\noindent Let $\Omega\subset\mathbb{R}^d$, $d=2$ or $3$, denote the physical
domain with boundary
    $\Gamma=\Gamma_D\cup\Gamma_N,
    \
    \Gamma_D\cap\Gamma_N=\emptyset$.
The outward unit normal is denoted by $\Bn$. Essential boundary conditions are
prescribed on $\Gamma_D$, while tractions are prescribed on $\Gamma_N$. The
unknowns are a vector field $\Bu$ and a scalar pressure-like field $p$.

The corresponding trial, test, and pressure spaces are
\begin{align}
    \calU
    &=
    \left\{
        \Bu\in[H^1(\Omega)]^d:
        \Bu=\bar{\Bu}\ \text{on }\Gamma_D
    \right\},
    \\
    \calV
    &=
    \left\{
        \Bv\in[H^1(\Omega)]^d:
        \Bv=\Bzero\ \text{on }\Gamma_D
    \right\},
    \\
    \calP
    &=
    L^2(\Omega).
    \label{eq:continuous_mixed_spaces}
\end{align}

The Stokes equations provide the canonical mixed structure for incompressible
flow, and an analogous saddle-point structure arises in nearly incompressible
elasticity. In the solid-mechanics interpretation adopted in this work,
$\Bu$ denotes the displacement field and $p$ denotes an auxiliary
pressure-like variable associated with the volumetric response. We define
\begin{equation}
    p
    :=
    -\lambda\,
    \tr\left(\Bepsilon(\Bu)\right)
    =
    -\lambda\,
    \Bnabla\cdot\Bu,
    \label{eq:p_volumetric_relation}
\end{equation}
where $\lambda$ is the first Lam\'e parameter and
\begin{equation}
    \Bepsilon(\Bu)
    =
    \frac{1}{2}
    \left(
    \Bnabla\Bu+\Bnabla\Bu^T
    \right)
    \label{eq:epsilon_definition}
\end{equation}
is the symmetric gradient. The Cauchy stress is written as
\begin{equation}
    \Bsigma(\Bu,p)
    =
    2\mu\Bepsilon(\Bu)-p\BI,
    \label{eq:sigma_mixed_linear}
\end{equation}
where $\mu$ is the shear modulus in the solid-mechanics interpretation. In
terms of the conventional displacement formulation, elimination of $p$ from
\cref{eq:p_volumetric_relation,eq:sigma_mixed_linear} gives
\begin{equation}
    \Bsigma(\Bu)
    =
    2\mu\Bepsilon(\Bu)
    +
    \lambda
    \left(
    \Bnabla\cdot\Bu
    \right)\BI,
    \label{eq:standard_linear_elastic_stress}
\end{equation}
which is the standard constitutive relation for isotropic linear elasticity.
In three dimensions, the bulk modulus is related to the Lam\'e parameters by
\begin{equation}
    \kappa
    =
    \lambda+\frac{2}{3}\mu.
    \label{eq:bulk_lame_relation}
\end{equation}
Thus, the auxiliary pressure variable used here is associated with the
$\lambda$-weighted volumetric strain rather than directly with the hydrostatic
stress defined using the bulk modulus.

The mixed formulation may be obtained from the saddle-point functional
\begin{equation}
    \Pi(\Bu,p)
    =
    \int_{\Omega}
    \left[
    \frac{1}{2}
    \left(
    2\mu\Bepsilon(\Bu)
    \right)
    :
    \Bepsilon(\Bu)
    -
    p\,\Bnabla\cdot\Bu
    -
    \frac{p^2}{2\lambda}
    \right]
    \dV
    -
    \Pi^{\mathrm{ext}}(\Bu),
    \label{eq:mixed_potential}
\end{equation}
with
\begin{equation}
    \Pi^{\mathrm{ext}}(\Bu)
    =
    \int_{\Omega}
    \Bu\cdot\Bb_0
    \dV
    +
    \int_{\Gamma_N}
    \Bu\cdot\bar{\Bt}
    \dA.
    \label{eq:external_potential}
\end{equation}
Here $\Bb_0$ is the prescribed body force and $\bar{\Bt}$ is the prescribed
traction on $\Gamma_N$. The mixed solution is characterized by the saddle
point
\begin{equation}
    \left\{\Bu^*,p^*\right\}
    =
    \operatorname{Arg}
    \left\{
    \inf_{\Bu\in\mathcal{U}}
    \sup_{p\in\mathcal{P}}
    \Pi(\Bu,p)
    \right\},
    \label{eq:saddle_point_problem}
\end{equation}
where $\calU$ and $\calP$ are the admissible displacement and pressure spaces.

Taking the first variation of \cref{eq:mixed_potential} in the direction of
admissible test functions $(\Bv,q)\in\calV\times\calP$ gives the mixed weak
formulation: find $(\Bu,p)\in\calU\times\calP$ such that
\begin{align}
    D\Pi(\Bu,p)\cdot(\Bv,q)
    ={}&
    \underbrace{
    \left[
    \int_{\Omega}
    \Bepsilon(\Bv):
    2\mu\Bepsilon(\Bu)
    \dV
    -
    \int_{\Omega}
    p\,\Bnabla\cdot\Bv
    \dV
    -
    \int_{\Omega}
    \left(
    \Bnabla\cdot\Bu
    +
    \frac{p}{\lambda}
    \right)
    q
    \dV
    \right]
    }_{\mathcal{B}(\Bu,p;\Bv,q)}
    \nonumber\\
    &
    -
    \underbrace{
    \left[
    \int_{\Omega}
    \Bv\cdot\Bb_0
    \dV
    +
    \int_{\Gamma_N}
    \Bv\cdot\bar{\Bt}
    \dA
    \right]
    }_{\mathcal{F}(\Bv)}
    =
    0,
    \qquad
    \forall(\Bv,q)\in\calV\times\calP.
    \label{eq:mixed_weak_form_combined}
\end{align}

Using integration by parts, the corresponding strong form is
\begin{align}
    -\Bnabla\cdot\Bsigma(\Bu,p)
    &=
    \Bb_0
    &&\text{in }\Omega,
    \label{eq:strong_balance_sigma}
    \\
    \Bnabla\cdot\Bu+\frac{p}{\lambda}
    &=
    0
    &&\text{in }\Omega,
    \label{eq:strong_constraint_from_potential}
    \\
    \Bu
    &=
    \bar{\Bu}
    &&\text{on }\Gamma_D,
    \label{eq:strong_dirichlet_from_potential}
    \\
    \Bsigma(\Bu,p)\Bn
    &=
    \bar{\Bt}
    &&\text{on }\Gamma_N.
    \label{eq:strong_neumann_from_potential}
\end{align}
Substituting \cref{eq:sigma_mixed_linear} into
\cref{eq:strong_balance_sigma} gives
\begin{equation}
    -\Bnabla\cdot
    \left(
    2\mu\Bepsilon(\Bu)
    \right)
    +
    \Bnabla p
    =
    \Bb_0
    \qquad
    \text{in }\Omega.
    \label{eq:strong_momentum_expanded}
\end{equation}
This form is particularly useful for the residual-based stabilization
introduced later, since the momentum operator has the same differential
structure as the Stokes operator.

The role of the finite-compressibility term
$-p^2/(2\lambda)$ can be seen directly from the mixed bilinear form. Using
the sign convention in \cref{eq:mixed_weak_form_combined}, take
$(\Bv,q)=(\Bu,-p)$. Then
\begin{align}
    \mathcal{B}(\Bu,p;\Bu,-p)
    &=
    \int_{\Omega}
    2\mu
    \Bepsilon(\Bu):
    \Bepsilon(\Bu)
    \dV
    +
    \int_{\Omega}
    \frac{1}{\lambda}p^2
    \dV .
    \label{eq:mixed_energy_test}
\end{align}
Thus, for finite $\lambda$, the mixed form provides direct control of both the
shear-scaled strain term and the $L^2$ norm of the pressure:
\begin{equation}
    \mathcal{B}(\Bu,p;\Bu,-p)
    =
    2\mu
    \|\Bepsilon(\Bu)\|_{0,\Omega}^{2}
    +
    \frac{1}{\lambda}
    \|p\|_{0,\Omega}^{2}.
    \label{eq:pressure_control_finite_lambda}
\end{equation}
However, as $\lambda\rightarrow\infty$,
\begin{equation}
    \lim_{\lambda\rightarrow\infty}
    \mathcal{B}(\Bu,p;\Bu,-p)
    =
    2\mu
    \|\Bepsilon(\Bu)\|_{0,\Omega}^{2}.
    \label{eq:pressure_control_lost}
\end{equation}
Hence, in the incompressible limit, the pressure is no longer directly
controlled by the finite-compressibility term in the preceding estimate.
Instead, $p$ acts as a Lagrange multiplier enforcing
\begin{equation}
    \Bnabla\cdot\Bu=0.
    \label{eq:incompressible_constraint}
\end{equation}
Its stability must therefore be supplied by the displacement--pressure
coupling
\begin{equation}
    b(\Bv,p)
    =
    -
    \int_{\Omega}
    p\,\Bnabla\cdot\Bv
    \dV,
    \label{eq:continuous_pressure_coupling}
\end{equation}
through the continuous inf--sup condition
\begin{equation}
    \inf_{p\in\calP}
    \sup_{\Bv\in\calV}
    \frac{
    b(\Bv,p)
    }{
    \|\Bv\|_{\calV}
    \|p\|_{0,\Omega}
    }
    \geq
    \beta
    >
    0.
    \label{eq:continuous_infsup_condition}
\end{equation}
Thus, finite compressibility and the inf--sup coupling provide two distinct
mechanisms for pressure control. The direct contribution
$\lambda^{-1}\|p\|_{0,\Omega}^2$ vanishes as
$\lambda\rightarrow\infty$, whereas the pressure control provided through the
displacement--pressure coupling remains essential.

If the corresponding inf--sup condition is not satisfied at the discrete
level, pressure stability must be recovered by a consistent stabilization
mechanism.

The same saddle-point structure appears in incompressible flow and nearly
incompressible elasticity, although the physical interpretation of the
variables differs. In the incompressible-flow setting, $\Bu$ represents the
velocity, $p$ is the hydrodynamic pressure acting as a Lagrange multiplier,
and $\mu$ is the dynamic viscosity. In the solid-mechanics setting considered
here, $\Bu$ is the displacement, $\mu$ is the shear modulus, and $p$ is the
auxiliary pressure-like variable
\begin{equation}
    p=-\lambda\Bnabla\cdot\Bu.
\end{equation}
As $\lambda/\mu\rightarrow\infty$, the elastic problem approaches the
incompressible constraint
$\Bnabla\cdot\Bu=0$, and the mathematical structure approaches that of the
incompressible Stokes problem.

\subsection{Galerkin discretization and pressure elimination}
\label{subsec:galerkin_pressure_elimination}

\noindent We now discretize the mixed weak formulation in
\cref{eq:mixed_weak_form_combined} using finite-dimensional trial and test functions. The
displacement, pressure, and their corresponding weighting functions are approximated as
\begin{align}
    \Bu_h
    &=
    \sum_A N_A^u \Bd_A, \qquad
    \Bv_h
    =
    \sum_A N_A^u \Bw_A,
    \label{eq:discrete_expansions_u_v}
    \\
    p_h
    &=
    \sum_B N_B^p p_B, \qquad
    q_h
    =
    \sum_B N_B^p q_B .
    \label{eq:discrete_expansions_p_q}
\end{align}
Here $N_A^u$ and $N_B^p$ denote the displacement and pressure basis functions, respectively;
$\Bd_A$ and $p_B$ are the unknown displacement and pressure coefficients, while $\Bw_A$ and
$q_B$ are arbitrary weighting coefficients. Substituting
\cref{eq:discrete_expansions_u_v,eq:discrete_expansions_p_q} into the weak form and using the
arbitrariness of $\Bw_A$ and $q_B$ gives the Galerkin algebraic system
\begin{equation}
    \begin{bmatrix}
        \BK_{uu} & \BK_{up} \\
        \BK_{up}^{T} & \BK_{pp}
    \end{bmatrix}
    \begin{Bmatrix}
        \Delta \Bu \\
        \Delta p
    \end{Bmatrix}
    =
    \begin{Bmatrix}
        \BF_u \\
        \BF_p
    \end{Bmatrix}.
    \label{eq:galerkin_block_system}
\end{equation}
 Here $\BK_{uu}$ is the displacement stiffness matrix and $\BK_{up}$ is the
displacement--pressure coupling matrix associated with the term
$-\int_{\Omega}p_h\,\Bnabla\cdot\Bv_h\,\dV$. With the sign convention adopted in
\cref{eq:mixed_weak_form_combined}, the pressure block is
\begin{equation}
    \BK_{pp}
    =
    -\frac{1}{\lambda}\BM_p,
    \qquad
    \left[\BM_p\right]_{AB}
    =
    \int_{\Omega}
    N_A^p N_B^p\,\dV,
    \label{eq:galerkin_pressure_mass_block}
\end{equation}
where $\BM_p$ is the pressure mass matrix.

Moreover, $\Delta\Bu$ and $\Delta p$ in \cref{eq:galerkin_block_system} denote the algebraic solution increments; for the present linear problem, they may equivalently be interpreted as the discrete unknown vectors. For the homogeneous incompressibility constraint, $\BF_p = \Bzero$,
and the lower block row in \cref{eq:galerkin_block_system} gives
\begin{equation}
    \BK_{up}^{T}\Delta \Bu
    +
    \BK_{pp}\Delta p
    =
    \BF_p
    =
    \Bzero .
    \label{eq:galerkin_lower_block_row}
\end{equation}
Assuming $\BK_{pp}$ is invertible, the pressure increment can therefore be written as
\begin{equation}
    \Delta p
    =
    -\BK_{pp}^{-1}\BK_{up}^{T}\Delta \Bu,
    \label{eq:pressure_increment_elimination_simplified}
\end{equation}
or, retaining a nonzero pressure residual,
\begin{equation}
    \Delta p
    =
    \BK_{pp}^{-1}
    \left(
    \BF_p
    -
    \BK_{up}^{T}\Delta \Bu
    \right).
    \label{eq:pressure_increment_elimination_general}
\end{equation}
Substituting \cref{eq:pressure_increment_elimination_simplified} into the upper block row in \cref{eq:galerkin_block_system} gives
the pressure-eliminated, one-field system
\begin{equation}
    \left(
    \BK_{uu}
    -
    \BK_{up}\BK_{pp}^{-1}\BK_{up}^{T}
    \right)
    \Delta \Bu
    =
    \BF_u .
    \label{eq:galerkin_reduced_u_system}
\end{equation}

This reduction highlights the difference between the fully incompressible Stokes problem and
nearly incompressible elasticity. In the incompressible limit,
\begin{equation}
    \lambda \rightarrow \infty
    \qquad \Longrightarrow \qquad
    \BK_{pp} = \Bzero ,
    \label{eq:Kpp_vanishes_incompressible_limit}
\end{equation}
and the pressure block cannot be inverted. Hence, the reduction to a one-field formulation is
not possible. For nearly incompressible solids, $\lambda$ is large but finite, so the pressure
block is regularized by the volumetric energy and the elimination in
\cref{eq:pressure_increment_elimination_simplified,eq:galerkin_reduced_u_system} is available.

Direct inversion of $\BK_{pp}$ may still be expensive if the pressure space is continuous,
because then $\BK_{pp}$ is globally coupled. This is the motivation for using discontinuous
pressure spaces. If $p_h$ is discontinuous across the pressure-cell boundaries, then
$\BK_{pp}$ is block diagonal with respect to the pressure cells
$\Omega_E^p\in\calT_h^p$:
\begin{equation}
    \BK_{pp}
    =
    \operatorname{diag}
    \left(
    \BK_{pp}^{(1)},
    \BK_{pp}^{(2)},
    \ldots,
    \BK_{pp}^{(n_p)}
    \right),
    \label{eq:block_diagonal_Kpp}
\end{equation}
where $n_p$ denotes the number of pressure cells. The pressure elimination is then performed
pressure cell by pressure cell:
\begin{equation}
    \Delta p^{(E)}
    =
    -
    \left(\BK_{pp}^{(E)}\right)^{-1}
    \left(\BK_{up}^{(E)}\right)^T
    \Delta \Bu^{(E)},
    \qquad
    \Omega_E^p\in\calT_h^p .
    \label{eq:local_pressure_elimination}
\end{equation}
Thus, a discontinuous pressure space converts the pressure Schur complement into a
pressure-cell-wise operation. This locality is computationally attractive, but it does not by
itself guarantee stability; the resulting mixed space must still satisfy the inf--sup condition,
or the formulation must be consistently stabilized.
\subsection{Isogeometric approximation spaces}
\label{subsec:iga_approximation_spaces}

\noindent We briefly recall the B-spline approximation spaces used in this work. Let
\(\Xi=\{\xi_1,\xi_2,\ldots,\xi_{n_b+k+1}\}\) be a nondecreasing knot vector, where \(k\) is the
polynomial degree and \(n_b\) is the number of univariate basis functions. The B-spline basis
functions are defined by the Cox--de Boor recursion. For degree zero,
\begin{equation}
    N_{i,0}(\xi)
    =
    \begin{cases}
    1, & \xi_i \leq \xi < \xi_{i+1}, \\
    0, & \text{otherwise},
    \end{cases}
    \label{eq:bspline_degree_zero}
\end{equation}
and, for \(k \geq 1\),
\begin{equation}
    N_{i,k}(\xi)
    =
    \frac{\xi-\xi_i}{\xi_{i+k}-\xi_i}
    N_{i,k-1}(\xi)
    +
    \frac{\xi_{i+k+1}-\xi}{\xi_{i+k+1}-\xi_{i+1}}
    N_{i+1,k-1}(\xi),
    \label{eq:cox_de_boor}
\end{equation}
where fractions with zero denominator are taken to be zero. The functions \(N_{i,k}\) are
nonnegative, form a partition of unity, and have compact support on
\([\xi_i,\xi_{i+k+1}]\). If an interior knot has multiplicity \(m\), then the basis has
continuity \(\mathcal{C}^{k-m}\) at that knot. Thus, for an open knot vector with simple interior knots,
the basis is maximally smooth, i.e., \(\mathcal{C}^{k-1}\) across element interfaces. Reduced-continuity
Spaces are obtained by increasing the multiplicity of selected interior knots. For example,
a degree-\(k\) spline space with double interior knots has continuity \(\mathcal{C}^{k-2}\), while
repeating an interior knot \(k+1\) times produces a discontinuity, denoted by \(\mathcal{C}^{-1}\).
This knot-insertion mechanism allows us to construct displacement and pressure spaces with
different degrees, continuities, and mesh sizes while retaining the same tensor-product
B-spline framework.

In multiple dimensions, the approximation is constructed by tensor products of the univariate
basis functions. In two dimensions, for example,
\begin{equation}
    N_{A}^{k_x,k_y}(\xi,\eta)
    =
    N_{i,k_x}(\xi)N_{j,k_y}(\eta),
    \qquad
    A \equiv (i,j),
    \label{eq:tensor_product_basis_2d}
\end{equation}
and similarly in three dimensions,
\begin{equation}
    N_{A}^{k_x,k_y,k_z}(\xi,\eta,\zeta)
    =
    N_{i,k_x}(\xi)N_{j,k_y}(\eta)N_{\ell,k_z}(\zeta),
    \qquad
    A \equiv (i,j,\ell).
    \label{eq:tensor_product_basis_3d}
\end{equation}
The mapping from the parametric domain to the physical domain is written as
\begin{equation}
    \Bx(\boldsymbol{\xi})
    =
    \sum_A N_A(\boldsymbol{\xi}) \BX_A,
    \label{eq:geometry_mapping_bspline}
\end{equation}
where \(\BX_A\) are the control points. When NURBS are used, the B-spline basis functions are
replaced by rational basis functions
\begin{equation}
    R_A(\boldsymbol{\xi})
    =
    \frac{w_A N_A(\boldsymbol{\xi})}
    {\sum_B w_B N_B(\boldsymbol{\xi})},
    \label{eq:nurbs_basis}
\end{equation}
where \(w_A\) are the corresponding weights. In the following, \(N_A\) denotes either the
B-spline or NURBS basis function, depending on the geometric representation.

The displacement and pressure fields are approximated using possibly different spline spaces,
denoted by
\begin{equation}
    \calU_h
    =
    \mathcal{U}(k_u,\mathcal{C}^{r_u},h_u),
    \qquad
    \calP_h
    =
    \mathcal{P}(k_p,\mathcal{C}^{r_p},h_p),
    \label{eq:mixed_iga_spaces_general}
\end{equation}
where \(k_u\) and \(k_p\) are the displacement and pressure polynomial degrees, \(\mathcal{C}^{r_u}\)
and \(\mathcal{C}^{r_p}\) denote the corresponding inter-element continuities, and \(h_u\) and \(h_p\)
are the displacement and pressure mesh sizes. The discrete fields are written as
\begin{equation}
    \Bu_h(\Bx)
    =
    \sum_A N_A^{u}(\Bx)\Bd_A,
    \qquad
    p_h(\Bx)
    =
    \sum_B N_B^{p}(\Bx)p_B .
    \label{eq:iga_up_approximation}
\end{equation}
Here \(N_A^u\) and \(N_B^p\) are basis functions associated with
\(\calU_h\) and \(\calP_h\), respectively.
The corresponding homogeneous discrete test space is
\begin{equation}
    \calV_h
    =
    \calU_{h,0}
    :=
    \left\{
        \Bv_h\in\calU_h:
        \Bv_h=\Bzero \ \text{on }\Gamma_D
    \right\}.
    \label{eq:discrete_homogeneous_displacement_space}
\end{equation}

Several pressure choices are considered in this work. A continuous pressure space has
\(r_p\geq 0\), whereas a discontinuous pressure space is denoted by
    $\mathcal{P}(k_p,\mathcal{C}^{-1},h_p)$,
meaning that the pressure basis is broken across element interfaces. The discontinuous choice
is especially important for local pressure elimination because it makes the pressure block
pressure-cell-wise block diagonal. However, as discussed below, this locality does not guarantee
inf-sup stability by itself and often works against it.
\subsection{Inf--sup stability of the Galerkin formulation for isogeometric spaces}
\label{subsec:inf_sup_galerkin_IGA}

\noindent
The mixed Galerkin formulation is a saddle-point problem. For finite
$\lambda$, the term $\|p_h\|_{0,\Omega}^{2}/{\lambda}$
provides direct pressure control through the finite-compressibility relation.
As $\lambda/\mu$ becomes large, however, this contribution becomes
progressively weaker. In the incompressible limit, pressure stability must be
provided entirely by the displacement--pressure coupling.

Accordingly, a uniformly stable Galerkin discretization must satisfy the
discrete inf--sup condition
\begin{equation}
    \inf_{q_h\in\calP_h}
    \sup_{\Bv_h\in\calU_{h,0}}
    \frac{
    b(\Bv_h,q_h)
    }{
    \|\Bv_h\|_{1,\Omega}
    \|q_h\|_{0,\Omega}
    }
    \geq
    \beta_h,
    \qquad
    \beta_h\geq\beta_0>0,
    \label{eq:inf_sup_condition}
\end{equation}
where
\begin{equation}
    b(\Bv_h,q_h)
    =
    -
    \int_{\Omega}
    q_h\,
    \Bnabla\cdot\Bv_h
    \dV,
\end{equation}
and $\beta_0$ is independent of the mesh size. In the fully incompressible
limit, the pressure space is understood to be equipped with the appropriate
pressure normalization whenever a constant-pressure null mode is present.

If \cref{eq:inf_sup_condition} is not satisfied uniformly, the pressure space
contains modes that are weakly or not at all detected by the divergence of the
displacement space. Such modes can appear as spurious pressure oscillations
and, through $\Bsigma(\Bu_h,p_h)
    =
    2\mu\Bepsilon(\Bu_h)-p_h\BI$,
as corresponding stress oscillations. For moderate values of $\lambda/\mu$,
the finite-compressibility term can partially mask this instability, but this
control vanishes progressively as the incompressible limit is approached.

The inf--sup condition therefore expresses the requirement that the
displacement space be sufficiently rich relative to the pressure space. If the
pressure approximation contains too many modes relative to those accessible
through $\Bnabla\cdot\calU_{h,0}$, the pair may become unstable. This issue is
particularly important in isogeometric analysis, where high inter-element
continuity can significantly reduce the number of locally independent
displacement modes compared with $\mathcal{C}^0$ finite-element spaces of the same
polynomial degree.

The role of the discrete inf--sup condition in controlling the pressure can
be seen directly from \cref{eq:inf_sup_condition}. For any
$q_h\in\calP_h$,
\begin{equation}
    \beta_0
    \|q_h\|_{0,\Omega}
    \leq
    \sup_{\Bv_h\in\calU_{h,0}}
    \frac{
        |b(\Bv_h,q_h)|
    }{
        \|\Bv_h\|_{1,\Omega}
    }.
    \label{eq:infsup_pressure_control}
\end{equation}
Assuming that the homogeneous essential boundary conditions eliminate the
rigid-body modes, Korn's inequality gives
\begin{equation}
    \|\Bv_h\|_{1,\Omega}
    \leq
    C_K
    \|\Bepsilon(\Bv_h)\|_{0,\Omega},
    \qquad
    \Bv_h\in\calU_{h,0},
    \label{eq:korn_inequality_discrete}
\end{equation}
where $C_K$ is independent of the mesh size. Using \cref{eq:korn_inequality_discrete} and introducing the mechanically
scaled displacement norm
\begin{equation}
    \|\Bv_h\|_{u}^{2}
    :=
    2\mu
    \|\Bepsilon(\Bv_h)\|_{0,\Omega}^{2},
\end{equation}
\cref{eq:infsup_pressure_control} implies, up to mesh-independent constants,
\begin{equation}
    \frac{1}{\sqrt{2\mu}}
    \|q_h\|_{0,\Omega}
    \lesssim
    \frac{1}{\beta_0}
    \sup_{\Bv_h\in\calU_{h,0}}
    \frac{
        |b(\Bv_h,q_h)|
    }{
        \sqrt{2\mu}
        \|\Bepsilon(\Bv_h)\|_{0,\Omega}
    }.
    \label{eq:scaled_infsup_pressure_control}
\end{equation}
Thus, the pressure scale
$(2\mu)^{-1/2}\|q_h\|_{0,\Omega}$ can be controlled through the
displacement--pressure coupling whenever the discrete inf--sup constant is
uniformly bounded away from zero. This control is distinct from the direct
finite-compressibility contribution
$\lambda^{-1/2}\|q_h\|_{0,\Omega}$.

Motivated by these two distinct pressure-control mechanisms, we introduce the
parameter-scaled mixed norm:
\begin{equation}
    \left\|\!\left|
    \Bv,q
    \right|\!\right\|_{G,\lambda}^{2}
    :=
    2\mu
    \|\Bepsilon(\Bv)\|_{0,\Omega}^{2}
    +
    \left(
    \frac{1}{2\mu}
    +
    \frac{1}{\lambda}
    \right)
    \|q\|_{0,\Omega}^{2}.
    \label{eq:galerkin_parameter_scaled_norm}
\end{equation}
The two pressure contributions in
\cref{eq:galerkin_parameter_scaled_norm} have different origins. The term
$\lambda^{-1}\|q\|_{0,\Omega}^{2}$ corresponds to the direct pressure control
provided by finite compressibility, whereas $(2\mu)^{-1}\|q\|_{0,\Omega}^{2}$ defines the natural pressure
scale associated with the displacement--pressure coupling. Uniform control
of this pressure component by the discrete mixed operator follows from the
inf--sup condition, with a stability constant depending on the uniform
lower bound $\beta_0$.

Consequently, unlike the direct energy estimate in \cref{eq:pressure_control_finite_lambda}, 
the mixed norm in
\cref{eq:galerkin_parameter_scaled_norm} remains nondegenerate as
$\lambda\rightarrow\infty$:
\begin{equation}
    \lim_{\lambda\rightarrow\infty}
    \left\|\!\left|
    \Bv,q
    \right|\!\right\|_{G,\lambda}^{2}
    =
    2\mu
    \|\Bepsilon(\Bv)\|_{0,\Omega}^{2}
    +
    \frac{1}{2\mu}
    \|q\|_{0,\Omega}^{2}.
    \label{eq:galerkin_norm_incompressible_limit}
\end{equation}

Assuming
$\Bu\in[H^{k_u+1}(\Omega)]^d$ and
$p\in H^{k_p+1}(\Omega)$, and using the standard approximation
properties of the spline spaces, the conventional error analysis for
uniformly inf--sup-stable mixed Galerkin formulations gives
\begin{equation}
    \left\|\!\left|
    e_u,e_p
    \right|\!\right\|_{G,\lambda}^{2}
    \lesssim
    2\mu C_u
    h_u^{2k_u}
    \|\Bu\|_{k_u+1,\Omega}^{2}
    +
    \left(
    \frac{1}{2\mu}
    +
    \frac{1}{\lambda}
    \right)
    C_p
    h_p^{2(k_p+1)}
    \|p\|_{k_p+1,\Omega}^{2},
    \label{eq:galerkin_mixed_error_estimate}
\end{equation}
where
\begin{equation}
    e_u=\Bu_h-\Bu,
    \qquad
    e_p=p_h-p.
\end{equation}

In the nearly incompressible regime,
$\lambda/\mu\gg1$, the direct compressibility contribution becomes small and
\cref{eq:galerkin_mixed_error_estimate} approaches
\begin{equation}
    \left\|\!\left|
    e_u,e_p
    \right|\!\right\|_{G,\infty}^{2}
    \lesssim
    2\mu C_u
    h_u^{2k_u}
    \|\Bu\|_{k_u+1,\Omega}^{2}
    +
    \frac{C_p}{2\mu}
    h_p^{2(k_p+1)}
    \|p\|_{k_p+1,\Omega}^{2}.
    \label{eq:galerkin_mixed_error_estimate_incompressible}
\end{equation}
Hence, for a uniformly inf--sup-stable pair, the mixed approximation remains
controlled as the incompressible limit is approached.

For comparable displacement and pressure mesh sizes,
$h_u\sim h_p\sim h$, the two approximation contributions in
\cref{eq:galerkin_mixed_error_estimate} are balanced when
$
    k_u=k_p+1$.
This corresponds to a Taylor--Hood-type degree choice. In this case,
$
    \left\|\!\left|
    e_u,e_p
    \right|\!\right\|_{G,\lambda}
    =
    O\!\left(h^{k_u}\right)$,
and the estimate gives the compatible approximation orders
\begin{equation}
    \|\Bepsilon(\Bu-\Bu_h)\|_{0,\Omega}
    =
    O\!\left(h^{k_u}\right),
    \qquad
    \|p-p_h\|_{0,\Omega}
    =
    O\!\left(h^{k_p+1}\right).
    \label{eq:galerkin_balanced_component_rates}
\end{equation}
Since $k_u=k_p+1$, the displacement $H^1$-type error and the pressure
$L^2$-error have the same asymptotic order for the balanced pair.
For general degree combinations, however, the mixed estimate contains
approximation contributions from both spaces.

The preceding error estimate relies essentially on a uniform discrete
inf--sup condition. If $\beta_h$ decreases toward zero under mesh refinement,
the pressure control associated with the $(2\mu)^{-1}\|p_h\|_{0,\Omega}^{2}$
contribution in the mixed norm deteriorates, and the constant implicit in the
error estimate may no longer remain uniformly bounded. Weakly controlled
pressure modes may then appear, particularly as $\lambda/\mu$ becomes large
and the direct compressibility contribution
$\lambda^{-1}\|p_h\|_{0,\Omega}^{2}$ becomes progressively weaker.


 We now specialize the discrete inf-sup requirement in
\cref{eq:inf_sup_condition} to the isogeometric spaces introduced in
\cref{subsec:iga_approximation_spaces}. Consider the mixed pair
\begin{equation}
    \calU_h/\calP_h
    =
    \mathcal{U}(k_u,\mathcal{C}^{r_u},h)/\mathcal{P}(k_p,\mathcal{C}^{r_p},nh),
    \label{eq:generic_iga_pair}
\end{equation}
where \(k_u\) and \(k_p\) are the displacement and pressure degrees, \(\mathcal{C}^{r_u}\) and
\(\mathcal{C}^{r_p}\) are the corresponding continuities, and \(h\) is the displacement mesh size. The
pressure mesh size is taken as \(nh\). In other words, the displacement and pressure meshes are assumed to be aligned,
nested tensor-product meshes with integer mesh ratio $n$. Thus, \(n=1\) corresponds to equal displacement and
pressure meshes, while \(n=2\) corresponds to a pressure mesh twice as coarse as the
displacement mesh. 

For this family of tensor-product isogeometric spaces, the stability condition of Bressan and
Sangalli \cite{8179196} can be written as
\begin{equation}
    n\left(k_u-r_u\right)
    >
    k_p-r_p .
    \label{eq:bressan_sangalli_condition}
\end{equation}
\revx{\Cref{eq:bressan_sangalli_condition} is a \emph{sufficient} condition. It is established by the macroelement technique under the following hypotheses, which are stated here because they are load bearing for everything that follows: a single patch; tensor-product spaces on aligned, nested meshes; quasi-uniform knot vectors; and a geometry map whose derivatives are uniformly bounded and whose Jacobian is bounded away from zero. Both benchmarks of \Cref{subsec:kirsch_problem,subsec:cooks_membrane} satisfy these hypotheses: the quarter-annulus of the Kirsch problem is a rational map with Jacobian bounded away from zero on the annulus, and the Cook trapezoid is bilinear. The immersed discretizations satisfy them more simply still, since the background approximation spaces are Cartesian and the physical geometry enters only through the quadrature. Because the condition is sufficient and not necessary, a pair that fails \cref{eq:bressan_sangalli_condition} is not thereby shown to be unstable. 
}

The condition in \cref{eq:bressan_sangalli_condition} shows explicitly how degree, continuity, and relative mesh size affect
inf-sup stability. The quantity \(k-r\) measures the number of new basis functions introduced
per element in one parametric direction. Increasing continuity decreases \(k-r\), thereby
reducing the local richness of the space. Conversely, coarsening the pressure mesh increases
\(n\), making the displacement space richer relative to the pressure space over each pressure
cell.

The preceding inf--sup condition also admits a useful heuristic
interpretation through a degree-of-freedom count. This count does not by
itself constitute a proof of inf--sup stability, but it illustrates why
degree, continuity, and relative mesh size enter the rigorous macroelement
condition. In one dimension,
a spline space of degree \(k\) and continuity \(\mathcal{C}^r\) over \(m\) elements has approximately
\begin{equation}
    N(k,r,m)
    =
    m(k-r)+r+1
    \label{eq:one_d_dof_count}
\end{equation}
degrees of freedom. If the pressure grid has \(m\) elements and the displacement grid has
\(nm\) elements, then the leading-order counts are
\begin{align}
    N_p
    &\sim
    m\left(k_p-r_p\right),
    \label{eq:pressure_dof_count_leading}
    \\
    N_u
    &\sim
    nm\left(k_u-r_u\right).
    \label{eq:displacement_dof_count_leading}
\end{align}
Thus, the displacement space can dominate the pressure space locally only if
\begin{equation}
    n\left(k_u-r_u\right)
    >
    k_p-r_p,
    \label{eq:counting_condition}
\end{equation}
which is the condition in \cref{eq:bressan_sangalli_condition}. In multiple dimensions, the
tensor-product structure leads to the same one-directional criterion for the structured spline
pairs considered here.

This criterion is especially important in IGA because high continuity reduces the number of
local displacement modes. Therefore, a pair that resembles a stable Taylor--Hood finite
element pair in polynomial degree may still be unstable if the displacement space is too
smooth relative to the pressure space. Stable IGA pairings can be obtained either by reducing
the displacement continuity, by coarsening the pressure mesh, or by using a pressure space
that is sufficiently small relative to the displacement space.

\revx{The discrete stability of the spline pairs considered here is examined
numerically using the Kirsch and Cook's membrane examples presented in
\cref{subsec:kirsch_problem,subsec:cooks_membrane}. In particular, the
discrete inf--sup constant \(\beta_h\) is evaluated under mesh refinement to
determine whether it remains uniformly bounded away from zero for a family
satisfying the sufficient criterion in
\cref{eq:bressan_sangalli_condition}, or decreases toward zero for a family
that fails the criterion. The values of \(\beta_h\) are computed using the
generalized eigenvalue procedure of Chapelle and Bathe~\cite{chapelle1993inf}.
The corresponding matrix formulation of this test is provided
in~\ref{app:numerical_inf_sup_test}.}

\subsubsection{Continuous pressure examples}
\label{subsubsec:continuous_pressure_examples}

\noindent The condition \cref{eq:bressan_sangalli_condition} distinguishes finite-element-like and
isogeometric pairings that may look superficially similar but behave differently.

\begin{figure*}[!htb]
    \centering
    \includegraphics[width=\textwidth]
    {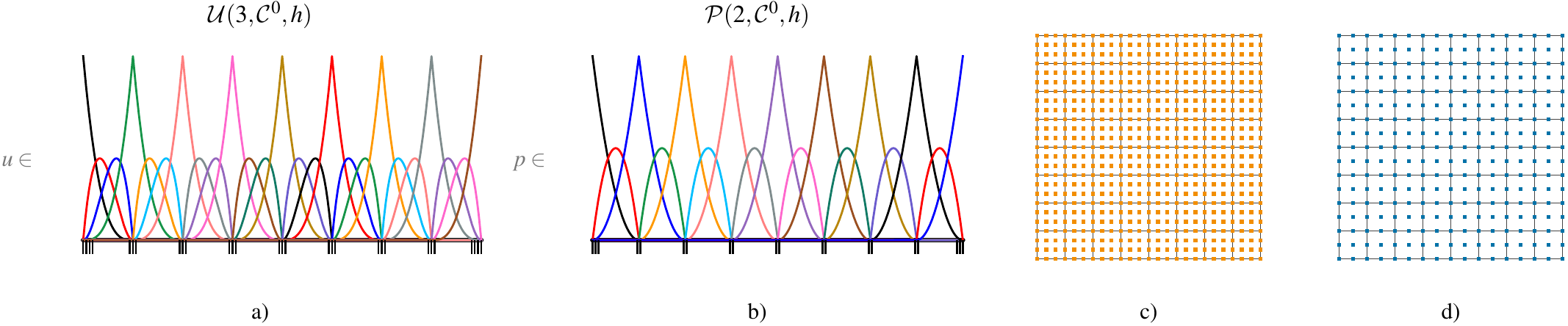}
    \caption{Finite-element Taylor--Hood-type pair for the representative cubic case,
    \(\stablepair{\mathcal{U}(3,\mathcal{C}^0,h)/\mathcal{P}(2,\mathcal{C}^0,h)}\). Panels show: (a) displacement basis functions,
    (b) pressure basis functions, (c) displacement mesh and control points, and
    (d) pressure mesh and control points.}
    \label{fig:displacement_pressure_spaces_T}
\end{figure*}

First, consider the finite-element Taylor--Hood-type pair
$
    \stablepair{\mathcal{U}(k,\mathcal{C}^0,h)/\allowbreak\mathcal{P}(k-1,\mathcal{C}^0,h)}
$.
Here \(n=1\), and therefore
$1(k-0)>(k-1)-0$,
so the pair satisfies the stability condition. The corresponding
displacement and pressure spaces are illustrated in
\cref{fig:displacement_pressure_spaces_T} for the representative cubic case.

\begin{figure*}[!htb]
    \centering
    \includegraphics[width=\textwidth]
    {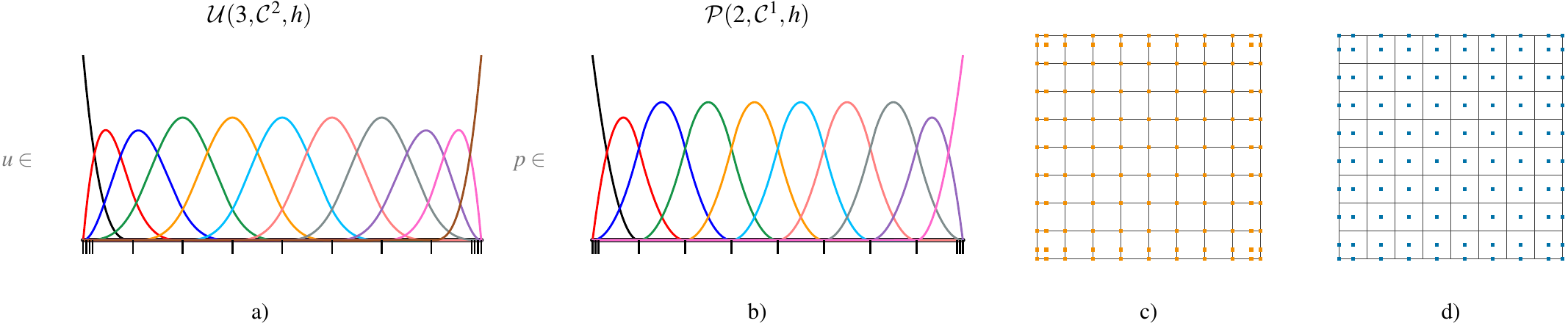}
    \caption{Maximally continuous isogeometric Taylor--Hood analogue for the representative
    cubic case, \(\unstablepair{\mathcal{U}(3,\mathcal{C}^2,h)/\mathcal{P}(2,\mathcal{C}^1,h)}\). Although the displacement degree is one order
    higher than the pressure degree, the pair fails the strict IGA inf-sup count because both
    spaces are maximally continuous.}
    \label{fig:displacement_pressure_spaces_TU}
\end{figure*}

Second, consider the maximally continuous isogeometric analogue
$
    \unstablepair{\mathcal{U}(k,\mathcal{C}^{k-1},h)/\allowbreak\mathcal{P}(k-1,\mathcal{C}^{k-2},h)}
$.
Although the displacement degree is still one order higher than the pressure degree, the
continuity is also higher. The stability count gives
$1(k-(k-1)) \not> (k-1)-(k-2)$.
The two sides are equal, and the strict inequality fails. \revx{This pair therefore fails the sufficient condition, and its lack of uniform inf--sup stability is established by the discrete inf--sup tests of \Cref{subsec:cooks_membrane}.} The corresponding
displacement and pressure spaces are illustrated in
\cref{fig:displacement_pressure_spaces_TU} for the representative cubic case.

\begin{figure*}[!htb]
    \centering
    \includegraphics[width=\textwidth]
    {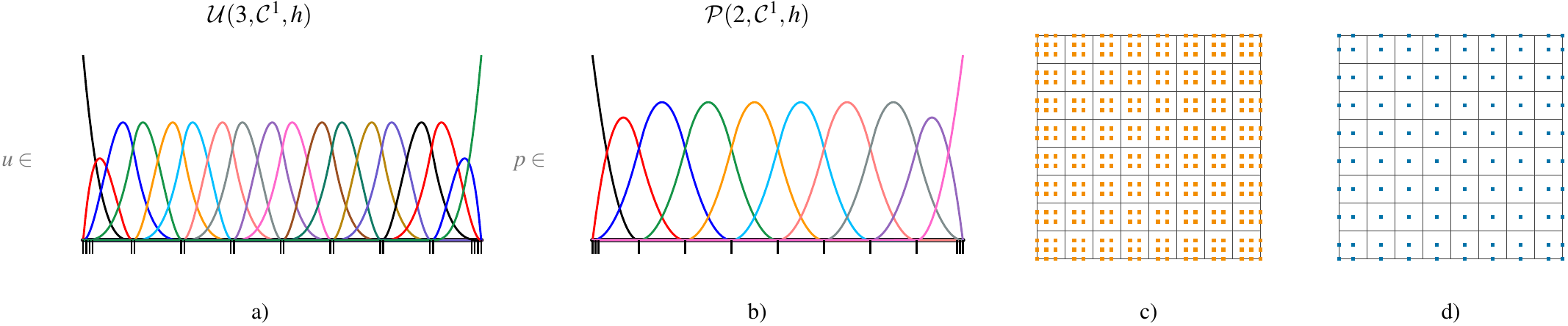}
    \caption{Reduced-continuity isogeometric pair for the representative cubic case,
    \(\stablepair{\mathcal{U}(3,\mathcal{C}^1,h)/\mathcal{P}(2,\mathcal{C}^1,h)}\). Reducing the displacement continuity increases the local
    richness of the displacement space relative to the pressure space and restores the
    inf-sup condition.}
    \label{fig:displacement_pressure_spaces_R}
\end{figure*}

A stable IGA alternative is obtained by reducing the displacement continuity by one order
while keeping the same pressure mesh:
$\stablepair{\mathcal{U}(k,\mathcal{C}^{k-2},h)/\allowbreak\mathcal{P}(k-1,\mathcal{C}^{k-2},h)}$.
In this case,
$1(k-(k-2))>(k-1)-(k-2)$,
and the pair satisfies the stability condition. The corresponding
displacement and pressure spaces are illustrated in
\cref{fig:displacement_pressure_spaces_R} for the representative cubic case.

\begin{figure*}[!htb]
    \centering
    \includegraphics[width=\textwidth]
    {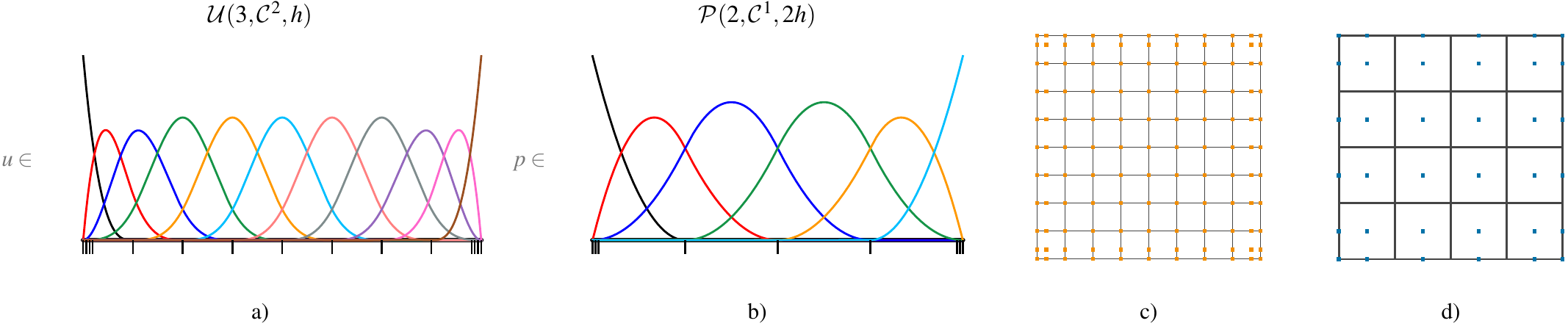}
    \caption{Subdivision pressure pair for the representative cubic case,
    \(\stablepair{\mathcal{U}(3,\mathcal{C}^2,h)/\mathcal{P}(2,\mathcal{C}^1,2h)}\). The displacement space remains maximally continuous, while
    the pressure mesh is twice as coarse, making the displacement space sufficiently rich over
    each pressure cell.}
    \label{fig:displacement_pressure_spaces_SUB}
\end{figure*}

Another stable alternative is to keep maximum continuity in the displacement space but use a
coarser pressure mesh: $\stablepair{    \mathcal{U}(k,\mathcal{C}^{k-1},h)/\allowbreak\mathcal{P}(k-1,\mathcal{C}^{k-2},2h)}$.
Then
$2(k-(k-1))>(k-1)-(k-2)$,
so the subdivision pressure pair is stable while preserving maximum continuity in the
displacement space. The corresponding
displacement and pressure spaces are illustrated in
\cref{fig:displacement_pressure_spaces_SUB} for the representative cubic case.

These examples show why standard finite-element intuition does not directly transfer to IGA.
In IGA, polynomial degree alone does not determine stability; continuity and relative mesh
size are equally important.
\subsubsection{Discontinuous pressure spaces}
\label{subsec:discontinuous_pressure_spaces}

\begin{figure*}[!htb]
    \centering
    \includegraphics[width=\textwidth]
    {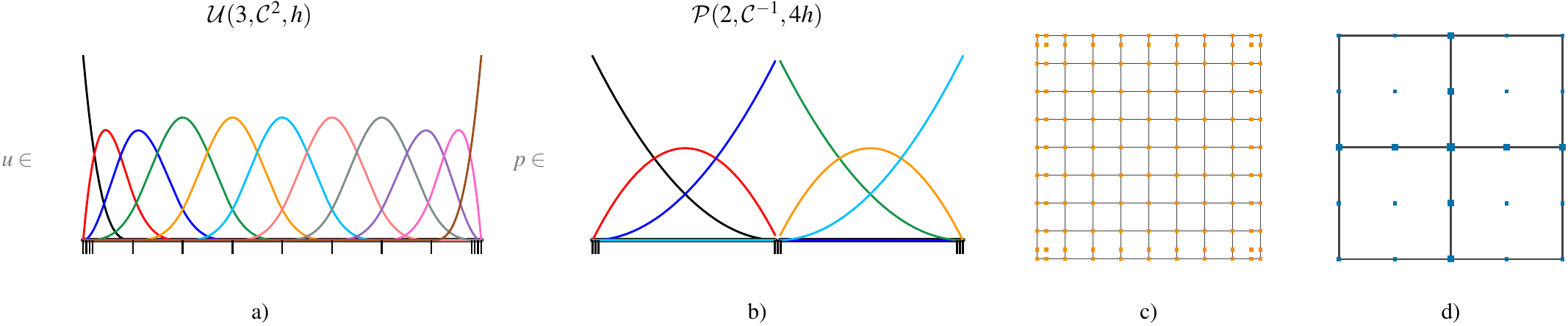}
    \caption{Illustration of the displacement and pressure approximation spaces and their corresponding meshes and control points for the stable Galerkin pair $\stablepair{\Uspace{3}{2}{h}/\Pspace{2}{-1}{4h}}$ with a discontinuous pressure space.}
    \label{fig:displacement_pressure_spaces}
\end{figure*}

\noindent For local pressure elimination, the pressure space should be discontinuous across element boundaries. In the notation of \cref{eq:generic_iga_pair}, this means
$r_p=-1$. For discontinuous pressure spaces, the macroelement stability framework
was extended to the local pressure-projection setting by
Antol\'in et al.~\cite{antolin2017isogeometric}. In the notation
adopted here, for $
    \Uspace{k_u}{r_u}{h}/\Pspace{k_p}{-1}{nh}$ pairs of spaces, the corresponding \revx{sufficient} condition becomes:
\begin{equation}
    n\left(k_u-r_u\right) > k_p+1.
    \label{eq:discontinuous_stability_condition}
\end{equation}
For full pressure accuracy in a mixed formulation, a common choice is
$
    k_p=k_u-1.
$
Then \cref{eq:discontinuous_stability_condition} becomes
\begin{equation}
    n\left(k_u-r_u\right)>k_u.
    \label{eq:discontinuous_full_accuracy_condition}
\end{equation}
For a maximally continuous displacement space, $r_u=k_u-1$, and hence $k_u-r_u=1$. The condition reduces to
\begin{equation}
    n>k_u.
    \label{eq:discontinuous_max_continuity_condition}
\end{equation}
For example, with a cubic maximally continuous displacement space, the stable discontinuous pressure pair requires
\begin{equation}
    \stablepair{\Uspace{3}{2}{h}/\Pspace{2}{-1}{4h}}.
\end{equation}
This pair, illustrated in \cref{fig:displacement_pressure_spaces}, is stable and allows local pressure elimination. Its drawback is that the pressure mesh is very coarse relative to the displacement mesh. The pressure field may then lose spatial resolution, and the assembly is less convenient than a same-mesh elementwise local formulation.

The central motivation of this work is to retain the computational advantages of discontinuous pressure spaces while avoiding the severe coarsening implied by \cref{eq:discontinuous_max_continuity_condition}. Residual-based stabilization can control many unstable pressure modes, but, as discussed below, it does not remove the need for an inf-sup condition against the elementwise constant pressure component.

\subsection{Residual-based stabilization}
\label{subsec:residual_based_stabilization}

\noindent The Galerkin formulation discussed above is stable only when the displacement and pressure
spaces satisfy the discrete inf-sup condition. This requirement is restrictive for several
attractive isogeometric pairs, especially equal-order pairs and discontinuous pressure spaces
used for local pressure elimination. To relax this restriction, we use residual-based
stabilization. The idea is to add element-wise consistent terms based on the strong residual of
the momentum equation.

The strong momentum operator is defined as
\begin{equation}
    \calL(\Bu,p)
    =
    -\Bnabla\cdot\left(2\mu\Bepsilon(\Bu)\right)
    +
    \Bnabla p .
    \label{eq:strong_operator}
\end{equation}
Thus, the strong form of the momentum equation is
\begin{equation}
    \calL(\Bu,p)
    =
    \Bb_0
    \qquad \text{in } \Omega ,
    \label{eq:strong_operator_equation}
\end{equation}
and the corresponding element-wise residual is
\begin{equation}
    \BR_m(\Bu,p)
    =
    \Bb_0-\calL(\Bu,p)
    =
    \Bb_0
    +
    \Bnabla\cdot\left(2\mu\Bepsilon(\Bu)\right)
    -
    \Bnabla p .
    \label{eq:strong_residual}
\end{equation}

Following a residual-based construction, we introduce the test operator
\begin{equation}
    \calL_{\chi}(\Bv,q)
    =
    \chi\,
    \Bnabla\cdot\left(2\mu\Bepsilon(\Bv)\right)
    -
    \Bnabla q ,
    \label{eq:test_operator_chi}
\end{equation}
where the parameter \(\chi\) selects the stabilization variant:
\begin{align}
    \chi &= +1,\qquad \text{GLS: symmetric stiffness matrix,}
    \nonumber\\
    &\qquad \text{conditional stability with respect to }\tau_e,
    \label{eq:chi_gls}
    \\
    \chi &= 0,\qquad \text{SUPG/PSPG-type: nonsymmetric stiffness matrix,}
    \nonumber\\
    &\qquad \text{conditional stability with respect to }\tau_e,
    \label{eq:chi_supg}
    \\
    \chi &= -1,\qquad \text{VMS: nonsymmetric stiffness matrix,}
    \nonumber\\
    &\qquad \text{unconditional stability with respect to }\tau_e.
    \label{eq:chi_vms}
\end{align}

The stabilized formulation is: find \((\Bu_h,p_h)\in\calU_h\times\calP_h\) such that
\begin{equation}
    \calB_{\tau}(\Bu_h,p_h;\Bv_h,q_h)
    =
    \calF_{\tau}(\Bv_h,q_h)
    \qquad
    \forall (\Bv_h,q_h)\in\calV_h\times\calP_h ,
    \label{eq:stabilized_weak_problem}
\end{equation}
where
\begin{align}
    \calB_{\tau}(\Bu_h,p_h;\Bv_h,q_h)
    &=
    \calB(\Bu_h,p_h;\Bv_h,q_h)
    +
    \sum_{e\in\calT_h}
    \tau_e
    \int_{\Omega_e}
    \calL_{\chi}(\Bv_h,q_h)
    \cdot
    \calL(\Bu_h,p_h)
    \dV ,
    \label{eq:stabilized_bilinear_form}
    \\
    \calF_{\tau}(\Bv_h,q_h)
    &=
    \calF(\Bv_h)
    +
    \sum_{e\in\calT_h}
    \tau_e
    \int_{\Omega_e}
    \calL_{\chi}(\Bv_h,q_h)
    \cdot
    \Bb_0
    \dV .
    \label{eq:stabilized_linear_form}
\end{align}
Equivalently, the stabilized weak equation can be written as
\begin{equation}
    \calB(\Bu_h,p_h;\Bv_h,q_h)
    -
    \calF(\Bv_h)
    -
    \sum_{e\in\calT_h}
    \tau_e
    \int_{\Omega_e}
    \calL_{\chi}(\Bv_h,q_h)
    \cdot
    \BR_m(\Bu_h,p_h)
    \dV
    =
    0 .
    \label{eq:stabilized_residual_form}
\end{equation}
This form makes consistency explicit: if the exact solution satisfies
\(\calL(\Bu,p)=\Bb_0\), then \(\BR_m(\Bu,p)=\mathbf{0}\), and the added residual-based term
vanishes.
The stabilizing effect can be seen by testing the homogeneous stabilized bilinear form with
\((\Bv_h,q_h)=(\Bu_h,-p_h)\). For compactness, we suppress the subscript \(h\) and first state
the element-wise inverse estimate used in the analysis:
\begin{equation}
    2\mu
    \left\|
    \Bnabla \cdot \Bepsilon(\Bv_h)
    \right\|_{0,\Omega_e}^{2}
    \leq
    \Lambda_e
    \left\|
    \Bepsilon(\Bv_h)
    \right\|_{0,\Omega_e}^{2},
    \qquad
    \forall \Bv_h \in \calV_h .
    \label{eq:element_inverse_estimate_stabilization}
\end{equation}

The cross terms that arise from the residual-based stabilization are bounded element-wise
using Young's inequality:
\begin{equation}
    \left|
    (\Ba,\Bb)_{\Omega_e}
    \right|
    \leq
    \|\Ba\|_{0,\Omega_e}\|\Bb\|_{0,\Omega_e}
    \leq
    \frac{1}{2\gamma}
    \|\Ba\|_{0,\Omega_e}^{2}
    +
    \frac{\gamma}{2}
    \|\Bb\|_{0,\Omega_e}^{2},
    \qquad
    \gamma>0 .
    \label{eq:young_inequality_stabilization}
\end{equation}
The resulting stability bounds are written element-wise in terms of the product
$t_e := \tau_e \Lambda_e $.
In the computations, we choose \(t_e=t\), where \(t\) is a prescribed constant independent
of the element.

\revx{Before the three estimates are derived, it is worth recording the algebraic fact from which they all follow. Writing \(\BS:=\Bnabla\cdot\left(2\mu\Bepsilon(\Bu)\right)\), the energy pairing gives \(\calL_\chi(\Bu,-p)=\chi\BS+\Bnabla p\) and \(\calL(\Bu,p)=-\BS+\Bnabla p\), so the stabilization integrand is
\begin{equation}
    \calL_\chi(\Bu,-p)\cdot\calL(\Bu,p)
    =
    \begin{cases}
        |\Bnabla p|^2-|\BS|^2, & \chi=+1,\\[2pt]
        |\Bnabla p|^2-\BS\cdot\Bnabla p, & \chi=0,\\[2pt]
        |\Bnabla p-\BS|^2=|\calL(\Bu,p)|^2, & \chi=-1.
    \end{cases}
    \label{eq:chi_integrand_identity}
\end{equation}
The three variants are therefore three different algebraic objects: a difference of squares, an indefinite cross term, and a perfect square. The last case is an identity, not an inequality. It gives, without any inverse estimate or auxiliary parameter,
\begin{equation}
    \calB_{\tau,-1}(\Bu,p;\Bu,-p)
    =
    2\mu\|\Bepsilon(\Bu)\|_{0,\Omega}^2
    +
    \frac{1}{\lambda}\|p\|_{0,\Omega}^2
    +
    \sum_{e\in\calT_h}\tau_e\|\calL(\Bu,p)\|_{0,\Omega_e}^2 ,
    \label{eq:vms_energy_identity}
\end{equation}
which is the structural reason the VMS form is unconditionally stable. The estimates below are consequences of \cref{eq:chi_integrand_identity};\linebreak[1] the VMS estimate in particular is what remains of \cref{eq:vms_energy_identity} after the viscous part of the residual is deliberately discarded in exchange for explicit control of the pressure gradient.}

\begin{itemize}
    \item \textbf{GLS: \(\chi=+1\).}  
    For the GLS choice, one obtains
    \begin{align}
        \calB_{\tau,+1}(\Bu,p;\Bu,-p)
        &\geq
        \sum_{e\in\calT_h}
        \left[
        2\mu
        \left(1-\tau_e\Lambda_e\right)
        \|\Bepsilon(\Bu)\|_{0,\Omega_e}^{2}
        +
        \tau_e
        \|\Bnabla p\|_{0,\Omega_e}^{2}
        +
        \frac{1}{\lambda}
        \|p\|_{0,\Omega_e}^{2}
        \right].
        \label{eq:gls_stability_estimate}
    \end{align}
    Therefore, the GLS form is stable if
    \begin{equation}
        0 < \tau_e\Lambda_e < 1
        \qquad
        \forall e\in\calT_h .
        \label{eq:gls_tau_condition_local}
    \end{equation}
    Equivalently, if \(t_e=t\), the condition is
    \begin{equation}
        0<t< 1 .
        \label{eq:gls_t_condition}
    \end{equation}
    Hence, GLS is conditionally stable with respect to the stabilization parameter.

    \item \textbf{SUPG/PSPG-type form: \(\chi=0\).}  
    For the SUPG/PSPG-type choice, Young's inequality gives
    \begin{align}
    \calB_{\tau,0}(\Bu,p;\Bu,-p)
    &\geq
    \sum_{e\in\calT_h}
    \biggl[
        2\mu\left(1-\frac{\tau_e\Lambda_e}{2\gamma}\right)
        \|\Bepsilon(\Bu)\|_{0,\Omega_e}^{2}
    \nonumber\\
    &\qquad
        +\tau_e\left(1-\frac{\gamma}{2}\right)
        \|\Bnabla p\|_{0,\Omega_e}^{2}
        +\frac{1}{\lambda}\|p\|_{0,\Omega_e}^{2}
    \biggr].
    \label{eq:supg_stability_estimate}
\end{align}
    The right-hand side is positive if
    \begin{equation}
        0<\gamma< 2,
        \qquad
        0<\tau_e\Lambda_e < 2\gamma
        \qquad
        \forall e\in\calT_h .
        \label{eq:supg_tau_condition_general_local}
    \end{equation}
    Equivalently, if \(t_e=t\), the condition is
    \begin{equation}
        0<\gamma<2,
        \qquad
        0<t< 2\gamma .
        \label{eq:supg_t_condition_general}
    \end{equation}
    The largest admissible upper bound is obtained by taking \(\gamma=2^-\), giving
    \begin{equation}
        0<t<4 .
        \label{eq:supg_t_condition}
    \end{equation}
    Thus, the SUPG/PSPG-type form is conditionally stable.

    \item \textbf{VMS: \(\chi=-1\).}  
    For the VMS choice, the corresponding estimate is
    \begin{align}
    \calB_{\tau,-1}(\Bu,p;\Bu,-p)
    &\geq
    \sum_{e\in\calT_h}
    \biggl[
        2\mu\left\{1-\tau_e\Lambda_e
        \left(\frac{1}{\gamma}-1\right)\right\}
        \|\Bepsilon(\Bu)\|_{0,\Omega_e}^{2}
    \nonumber\\
    &\qquad
        +\tau_e(1-\gamma)\|\Bnabla p\|_{0,\Omega_e}^{2}
        +\frac{1}{\lambda}\|p\|_{0,\Omega_e}^{2}
    \biggr].
    \label{eq:vms_stability_estimate}
\end{align}
    The two coefficients multiplying
    \(\|\Bepsilon(\Bu)\|_{0,\Omega_e}^{2}\) and
    \(\|\Bnabla p\|_{0,\Omega_e}^{2}\) are nonnegative if
    \begin{equation}
        \frac{\tau_e\Lambda_e}{\tau_e\Lambda_e+1}
        <
        \gamma
        <
        1
        \qquad
        \forall e\in\calT_h .
        \label{eq:vms_gamma_condition_local}
    \end{equation}
    Equivalently, if \(t_e=t\), this becomes
    \begin{equation}
        \frac{t}{t+1}
        <
        \gamma
        <
        1 .
        \label{eq:vms_gamma_condition_t}
    \end{equation}
    For every \(t>0\), such a \(\gamma\) exists. Therefore, the VMS form is unconditionally
    stable with respect to the stabilization parameter:
    \begin{equation}
        \tau_e>0
        \qquad
        \forall e\in\calT_h .
        \label{eq:vms_tau_condition}
    \end{equation}
\end{itemize}

These estimates show the different roles of the nondimensional stabilization level
\(t_e=\tau_e\Lambda_e\) for GLS, SUPG/PSPG, and VMS. The GLS and SUPG/PSPG choices require
an upper bound on \(t_e\), whereas the VMS choice remains stable for any positive
\(\tau_e\), or equivalently any positive \(t_e\). Nevertheless, very large values of
\(t_e\) can still lead to over-stabilization and loss of accuracy.

For pressure spaces that are at least \(\mathcal{C}^0\)-continuous, the pressure-gradient contribution
$
    \sum_{e\in\calT_h}
    \tau_e
    \|\Bnabla p_h\|_{0,\Omega_e}^{2}
$
provides stabilization for arbitrary displacement--pressure pairs, including equal-order
pairs that do not satisfy the Galerkin inf-sup condition. For discontinuous pressure spaces,
however, this term only controls the nonconstant pressure modes inside each element. The
elementwise constant pressure mode has zero elementwise gradient and is therefore not controlled
by the residual-based stabilization. Consequently, discontinuous pressure spaces require an
additional stability criterion for the constant pressure subspace, as will be discussed in
\cref{subsec:discontinuous_stabilized_condition}.

The stability estimates also indicate the appropriate mesh-dependent norm in which the
stabilized error is controlled. Since the coercive lower bound depends on the choice of
\(\chi\), we define
\begin{equation}
    \left\|\!\left| \Bv,q \right|\!\right\|_{\tau,\chi,\lambda}^{2}
    :=
    \sum_{e\in\calT_h}
    \left[
    2\mu A_{\chi,e}
    \|\Bepsilon(\Bv)\|_{0,\Omega_e}^{2}
    +
    \tau_e B_{\chi,e}
    \|\Bnabla q\|_{0,\Omega_e}^{2}
    +
    \frac{1}{\lambda}
    \|q\|_{0,\Omega_e}^{2}
    \right],
    \label{eq:method_dependent_stabilized_norm}
\end{equation}
where the coefficients \(A_{\chi,e}\) and \(B_{\chi,e}\) are those appearing in the stability
bounds:
\begin{align}
    \chi=+1 \quad \text{(GLS):}
    \qquad
    A_{+1,e}
    &=
    1-\tau_e\Lambda_e,
    &
    B_{+1,e}
    &=
    1,
    \label{eq:gls_norm_weights}
    \\
    \chi=0 \quad \text{(SUPG/PSPG):}
    \qquad
    A_{0,e}
    &=
    1-\frac{\tau_e\Lambda_e}{2\gamma},
    &
    B_{0,e}
    &=
    1-\frac{\gamma}{2},
    \label{eq:supg_norm_weights}
    \\
    \chi=-1 \quad \text{(VMS):}
    \qquad
    A_{-1,e}
    &=
    1-\tau_e\Lambda_e
    \left(
    \frac{1}{\gamma}-1
    \right),
    &
    B_{-1,e}
    &=
    1-\gamma .
    \label{eq:vms_norm_weights}
\end{align}
The admissible ranges of \(\tau_e\Lambda_e\) and \(\gamma\) stated above are precisely the
conditions that make these weights nonnegative. Thus, the error should be interpreted in the
method-dependent stabilized norm induced by the corresponding stability estimate.

For sufficiently smooth solutions and \(\mathcal{C}^0\)-continuous displacement and
pressure spaces, the classical residual-based Stokes analysis gives an
estimate of the form
\begin{equation}
    \left\|\!\left| e_u,e_p \right|\!\right\|_{\tau,\chi,\lambda}^{2}
    \lesssim
    2\mu C_u h_u^{2k_u}
    \|\Bu\|_{k_u+1,\Omega}^{2}
    +
    C_p
    \left(
    \frac{1}{2\mu}
    +
    \frac{1}{\lambda}
    \right)
    h_p^{2(k_p+1)}
    \|p\|_{k_p+1,\Omega}^{2},
    \label{eq:method_dependent_error_estimate}
\end{equation}
where \(C_u\) and \(C_p\) are independent of \revx{\(h_u\) and \(h_p\)}. \revx{The two mesh sizes are kept distinct because the discretizations of interest deliberately use \(h_p=nh_u\). In the direct two-mesh analogue of the classical residual-based analysis, the pressure approximation gives rise to terms scaling as \(h_p^{2(k_p+1)}\) and \(h_u^2h_p^{2k_p}\). The latter originates from the pressure-gradient stabilization through the scaling \(\tau_e=O(h_u^2/(2\mu))\). Since the nested pressure grids considered here satisfy \(h_p\ge h_u\) $\rightarrow$
$
h_u^2h_p^{2k_p}\leq h_p^{2(k_p+1)},
$
so both contributions can be bounded by the \(h_p^{2(k_p+1)}\) term appearing in \cref{eq:method_dependent_error_estimate}. The factor \(1/\lambda\) in this estimate represents the direct \(L^2\)-pressure control provided by finite compressibility, whereas the \(1/(2\mu)\) contribution represents pressure control associated with the stabilized mixed formulation. Consequently, for any fixed mesh ratio \(n=h_p/h_u\), pressure-grid coarsening affects the error constant but does not change the asymptotic convergence rate.}

In the nearly incompressible limit,
\(\lambda/\mu\rightarrow\infty\), the compressibility contribution vanishes,
and \cref{eq:method_dependent_error_estimate} reduces to
\begin{equation}
    \left\|\!\left| e_u,e_p \right|\!\right\|_{\tau,\chi,\infty}^{2}
    \lesssim
    2\mu C_u h_u^{2k_u}
    \|\Bu\|_{k_u+1,\Omega}^{2}
    +
    \frac{C_p}{2\mu}
    h_p^{2(k_p+1)}
    \|p\|_{k_p+1,\Omega}^{2}.
    \label{eq:method_dependent_error_estimate_incompressible}
\end{equation}
The estimate is stated here only to indicate the expected convergence
behavior; no proof is repeated. For conforming continuous pressure spaces,
the residual-based pressure-gradient term supplies additional pressure
control and permits stable discretizations, including equal-order pairs that
need not satisfy the Galerkin inf--sup condition.

The estimate also clarifies the convergence behavior of the displacement
and pressure gradients. Since
$\tau_e=O(h_u^2/(2\mu))$, the pressure-gradient contribution to the
mesh-dependent norm is weighted by one additional factor of $h_u^2$.
Consequently, \revx{for \(h_p=nh_u\) with fixed \(n\),} \cref{eq:method_dependent_error_estimate_incompressible}
implies, up to parameter-independent constants,
\begin{equation}
    \|\Bepsilon(e_u)\|_{0,\Omega}
    =
    O\!\left(
        h_u^{\min(k_u,k_p+1)}
    \right),
    \qquad
    \|\Bnabla e_p\|_{0,\Omega}
    =
    O\!\left(
        h_u^{\min(k_u-1,k_p)}
    \right).
    \label{eq:stabilized_gradient_convergence_rates}
\end{equation}
Thus, when
$k_u=k_p+1$, the displacement and pressure gradients attain their
respective optimal approximation rates,
\begin{equation}
    \|\Bepsilon(e_u)\|_{0,\Omega}
    =
    O(h_u^{k_u}),
    \qquad
    \|\Bnabla e_p\|_{0,\Omega}
    =
    O(h_u^{k_p}).
\end{equation}
This includes Taylor--Hood-type pairs, for which the displacement degree is
one order higher than the pressure degree.

For equal-order approximations,
$k_u=k_p=k$, the estimate instead gives
\begin{equation}
    \|\Bepsilon(e_u)\|_{0,\Omega}
    =
    O(h_u^{k}),
    \qquad
    \|\Bnabla e_p\|_{0,\Omega}
    =
    O(h_u^{k-1}).
\end{equation}
Hence, the displacement gradient retains its optimal rate, whereas the
available error estimate predicts a one-order loss in the pressure-gradient
convergence rate. In particular, for first-order equal-order
approximations, the estimate guarantees only boundedness of the pressure
gradient. These conclusions are consistent with the classical
residual-based Stokes analysis of
Hughes et al.~\cite{HughesFrancaBalestra1986}.
\subsection{Stabilization parameter}
\label{subsec:tau_parameter}

\noindent The stability estimates in the previous section show that the relevant quantity is not
\(\tau_e\) alone, but the nondimensional product
$    t_e
    :=
    \tau_e \Lambda_e ,
$
where \(\Lambda_e\) is the local inverse constant on element \(\Omega_e\). The parameter
\(\tau_e\) must be chosen large enough to control pressure-gradient modes, but not so large
that the solution becomes over-stabilized.

For every displacement test function restricted to \(\Omega_e\), we use the element-wise inverse
estimate in \cref{eq:element_inverse_estimate_stabilization}.
The constant \(\Lambda_e\) depends on the local spline space, the polynomial degree, the
element geometry, and the quadrature domain. With this definition, the stability restrictions
derived above are expressed directly in terms of \(t_e=\tau_e\Lambda_e\). In particular,
GLS requires \(0<t_e< 1\), the SUPG/PSPG-type form requires \(0<t_e< 4\) when the
largest admissible Young parameter is used, and the VMS form is stable for any \(t_e>0\).

The sharp local inverse constant is computed from a generalized eigenvalue problem. Let
\(\BY_e\) denote the vector of local displacement coefficients on \(\Omega_e\), and define
\begin{align}
    \left[\BA_e\right]_{AB}
    &=
    \int_{\Omega_e}
    \Bnabla\cdot\left(2\mu\Bepsilon(N_A^u)\right)
    \cdot
    \Bnabla\cdot\left(2\mu\Bepsilon(N_B^u)\right)
    \dV,
    \label{eq:Ae_inverse_matrix}
    \\
    \left[\BM_e\right]_{AB}
    &=
    \int_{\Omega_e}
    2\mu
    \Bepsilon(N_A^u)
    :
    \Bepsilon(N_B^u)
    \dV .
    \label{eq:Me_inverse_matrix}
\end{align}
Then
\begin{equation}
    \Lambda_e
    =
    \max_{\BY_e\neq\Bzero}
    \frac{
    \BY_e^T\BA_e\BY_e
    }{
    \BY_e^T\BM_e\BY_e
    },
    \label{eq:rayleigh_inverse_constant}
\end{equation}
or equivalently
\begin{equation}
    \BA_e\BY_e
    =
    \Lambda_e
    \BM_e\BY_e .
    \label{eq:generalized_eigenproblem_tau}
\end{equation}
\revx{The matrix \(\BM_e\) is singular on the rigid-body motions of the element, on which \(\BA_e\) also vanishes, so the Rayleigh quotient in \cref{eq:rayleigh_inverse_constant} is indeterminate there. The maximum is taken over the orthogonal complement of the null space of \(\BM_e\), exactly as for the boundary inverse constant in \Cref{subsubsec:nitsche_stability}. Since \(\BA_e\) annihilates every affine displacement field while \(\BM_e\) annihilates only the rigid motions, this restriction does not affect the maximum, but it is required for the quotient to be well defined.}
In the numerical studies, the stabilization parameter is chosen as
$
    \tau_e
    =
    {t}/{\Lambda_e},
$
so that
$
    \tau_e\Lambda_e=t
$
on every element. Thus, \(t\) is the nondimensional stabilization level varied in the computations. This choice
normalizes the stabilization by the local inverse constant and makes the parameter comparable
across different mesh sizes, polynomial degrees, and element geometries.


\subsection{Pressure elimination in the stabilized formulation}
\label{subsec:pressure_elimination_stabilized_formulation}

\noindent As in the Galerkin discretization of
\cref{subsec:galerkin_pressure_elimination}, we use the same finite-dimensional trial and
test expansions introduced in
\cref{eq:discrete_expansions_u_v,eq:discrete_expansions_p_q}. Substituting these approximations
into the stabilized weak form in \cref{eq:stabilized_weak_problem} gives the stabilized
algebraic system
\begin{equation}
    \begin{bmatrix}
        \BK_{uu}-\chi \BK_{uu}^{\tau}
        &
        \BK_{up}-\chi \BK_{up}^{\tau}
        \\
        \BK_{up}^{T}+\BK_{pu}^{\tau}
        &
        \BK_{pp}+\BK_{pp}^{\tau}
    \end{bmatrix}
    \begin{Bmatrix}
        \Delta \Bu \\
        \Delta p
    \end{Bmatrix}
    =
    \begin{Bmatrix}
        \BF_u+\chi \BF_u^{\tau} \\
        \BF_p^{\tau}
    \end{Bmatrix}.
    \label{eq:stabilized_block_system_chi}
\end{equation}
Here the matrices without superscript are the Galerkin contributions, while the superscript
\(\tau\) denotes the residual-based stabilization contributions. The parameter \(\chi\)
selects the stabilization variant: \(\chi=+1\) gives GLS, \(\chi=0\) gives the
SUPG/PSPG-type form, and \(\chi=-1\) gives the VMS form. The stabilized formulation is
generally nonsymmetric because \(\BK_{pu}^{\tau}\) is not, in general, the transpose of
\(\BK_{up}^{\tau}\).

From the lower block row of \cref{eq:stabilized_block_system_chi}, we have
\begin{equation}
    \left(
    \BK_{up}^{T}
    +
    \BK_{pu}^{\tau}
    \right)\Delta \Bu
    +
    \left(
    \BK_{pp}
    +
    \BK_{pp}^{\tau}
    \right)\Delta p
    =
    \BF_p^{\tau}.
    \label{eq:stabilized_lower_block_row_chi}
\end{equation}
If the stabilized pressure block
\(\BK_{pp}+\BK_{pp}^{\tau}\) is invertible, the pressure increment can be eliminated:
\begin{equation}
    \Delta p
    =
    \left(
    \BK_{pp}
    +
    \BK_{pp}^{\tau}
    \right)^{-1}
    \left[
    \BF_p^{\tau}
    -
    \left(
    \BK_{up}^{T}
    +
    \BK_{pu}^{\tau}
    \right)
    \Delta \Bu
    \right].
    \label{eq:stabilized_pressure_elimination_chi}
\end{equation}
Substituting \cref{eq:stabilized_pressure_elimination_chi} into the upper block row gives the
reduced one-field system
\begin{align}
    &
    \left[
    \BK_{uu}
    -
    \chi \BK_{uu}^{\tau}
    -
    \left(
    \BK_{up}
    -
    \chi \BK_{up}^{\tau}
    \right)
    \left(
    \BK_{pp}
    +
    \BK_{pp}^{\tau}
    \right)^{-1}
    \left(
    \BK_{up}^{T}
    +
    \BK_{pu}^{\tau}
    \right)
    \right]
    \Delta \Bu
    \nonumber \\
    &\qquad =
    \BF_u
    +
    \chi \BF_u^{\tau}
    -
    \left(
    \BK_{up}
    -
    \chi \BK_{up}^{\tau}
    \right)
    \left(
    \BK_{pp}
    +
    \BK_{pp}^{\tau}
    \right)^{-1}
    \BF_p^{\tau}.
    \label{eq:stabilized_reduced_system_chi}
\end{align}
Thus, the stabilized mixed formulation can again be reduced to a displacement-only system,
provided that the stabilized pressure block is invertible.

In the fully incompressible Stokes case,
\(
    \BK_{pp}=\Bzero
\).
Then the pressure block is only \(\BK_{pp}^{\tau}\). Since this stabilization block typically
controls pressure-gradient modes, it may still possess a pressure nullspace. Therefore, exact
pressure elimination is not automatically available in the fully incompressible Stokes case.

In nearly incompressible elasticity, the finite first Lam\'e parameter contributes a genuine pressure
block,
\(
    \BK_{pp}\neq \Bzero
\).
Consequently,
\(
    \BK_{pp}
    +
    \BK_{pp}^{\tau}
\)
is invertible under the stability conditions discussed above, and the pressure can be
eliminated exactly.

For a discontinuous pressure space, both \(\BK_{pp}\) and
\(\BK_{pp}^{\tau}\) are local to the pressure cells
\(\Omega_E^p\in\calT_h^p\). Hence,
\(
    \BK_{pp}
    +
    \BK_{pp}^{\tau}
\)
is block diagonal with respect to the pressure cells, and
\cref{eq:stabilized_pressure_elimination_chi} can be evaluated pressure cell by pressure cell:
\begin{equation}
    \Delta p^{(E)}
    =
    \left[
    \left(
    \BK_{pp}
    +
    \BK_{pp}^{\tau}
    \right)^{(E)}
    \right]^{-1}
    \left[
    \left(\BF_p^{\tau}\right)^{(E)}
    -
    \left(
    \BK_{up}^{T}
    +
    \BK_{pu}^{\tau}
    \right)^{(E)}
    \Delta \Bu^{(E)}
    \right],
    \qquad
    \Omega_E^p\in\calT_h^p .
    \label{eq:local_stabilized_pressure_elimination_chi}
\end{equation}
Here, the superscript \((E)\) denotes the quantities associated with the
pressure cell \(\Omega_E^p\). This pressure-cell-wise inversion is the key
computational advantage of using discontinuous pressure spaces in the
stabilized formulation. The global system after condensation involves only
the displacement unknowns.

\revx{One consequence of the condensation deserves to be stated, because it bears on the motivation of solver reuse. For \(\chi=+1\) the stabilization form is symmetric, \(\calL_{+1}=-\calL\), and the condensed displacement operator is symmetric positive definite, so it can be handed to a displacement solver built on conjugate gradients or a Cholesky factorization without change. For \(\chi=-1\) the condensed operator is nonsymmetric, and such a solver must be replaced by a nonsymmetric Krylov method. The \(\chi\)-family therefore presents a trade: GLS preserves the symmetry that existing displacement solvers assume and pays with the conditional bound of \cref{eq:gls_tau_condition_local}, whereas VMS is unconditionally stable and pays with a nonsymmetric global operator. Which side of that trade to take depends on whether solver reuse or parameter robustness dominates in a given application; the numerical studies below use VMS.}
\subsection{Stability condition for discontinuous pressure spaces}
\label{subsec:discontinuous_stabilized_condition}

\noindent
The stability analysis in
\cref{subsec:residual_based_stabilization} shows that the residual-based
formulation introduces element-wise pressure-gradient control through terms
of the form
\begin{equation}
    \sum_{e\in\calT_h}
    \tau_e B_{\chi,e}
    \|\Bnabla p_h\|_{0,\Omega_e}^{2},
\end{equation}
in addition to the direct finite-compressibility contribution
$\lambda^{-1}\|p_h\|_{0,\Omega}^{2}$. For continuous pressure spaces, the
pressure-gradient contribution provides the additional pressure control
required by the stabilized formulation. For discontinuous pressure spaces,
however, it cannot control pressure modes that are constant within each
pressure cell. These modes therefore require separate consideration.

To isolate this component, we decompose the discontinuous pressure field
into its elementwise constant and zero-mean nonconstant components:
\begin{equation}
    p_h
    =
    \bar p_h
    +
    \widetilde p_h,
    \label{eq:p_decomposition}
\end{equation}
where, on each pressure cell $\Omega_E^p\in\calT_h^p$,
\begin{equation}
    \bar p_h|_{\Omega_E^p}
    =
    \frac{1}{|\Omega_E^p|}
    \int_{\Omega_E^p}
    p_h\,\dV,
    \qquad
    \widetilde p_h
    =
    p_h-\bar p_h .
    \label{eq:p_constant_nonconstant_decomposition}
\end{equation}
By construction,
\begin{equation}
    \int_{\Omega_E^p}
    \widetilde p_h\,\dV
    =
    0,
    \label{eq:tilde_pressure_zero_mean}
\end{equation}
and therefore the two components are $L^2$-orthogonal on every pressure
cell:
\begin{equation}
    \left(
        \bar p_h,\widetilde p_h
    \right)_{\Omega_E^p}
    =
    0.
    \label{eq:pressure_component_orthogonality}
\end{equation}
Consequently,
\begin{equation}
    \|p_h\|_{0,\Omega_E^p}^{2}
    =
    \|\bar p_h\|_{0,\Omega_E^p}^{2}
    +
    \|\widetilde p_h\|_{0,\Omega_E^p}^{2}.
    \label{eq:pressure_component_l2_decomposition}
\end{equation}
Since $\bar p_h$ is constant on each pressure cell,
\begin{equation}
    \Bnabla\bar p_h
    =
    \Bzero
    \qquad
    \text{in }\Omega_E^p,
    \label{eq:constant_pressure_gradient_zero}
\end{equation}
and hence
\begin{equation}
    \Bnabla p_h
    =
    \Bnabla\widetilde p_h
    \qquad
    \text{in }\Omega_E^p.
    \label{eq:pressure_gradient_nonconstant_only}
\end{equation}
Thus, the residual-based pressure-gradient stabilization acts only on the
nonconstant component $\widetilde p_h$, and provides
no control of the elementwise constant component $\bar p_h$.

For finite $\lambda$, the constant component is still controlled through
the direct compressibility contribution
$\lambda^{-1}\|p_h\|_{0,\Omega}^{2}$. In the incompressible limit, however,
\begin{equation}
    \frac{1}{\lambda}
    \|p_h\|_{0,\Omega}^{2}
    \rightarrow
    0,
\end{equation}
and the elementwise constant pressure component must therefore be
controlled through the displacement--pressure coupling.

To identify this remaining stability requirement, define the elementwise
constant pressure subspace
\begin{equation}
    \overline{\calP}_h
    :=
    \mathcal{P}(0,\mathcal{C}^{-1},nh).
    \label{eq:constant_pressure_subspace}
\end{equation}
If pressure is unique only up to a global constant in the exactly
incompressible problem, the corresponding zero-mean subspace is understood
in \cref{eq:constant_pressure_subspace}.

Uniform control of the elementwise constant component requires the discrete
inf--sup condition
\begin{equation}
    \inf_{\substack{
        \bar q_h\in\overline{\calP}_h\\
        \bar q_h\neq 0
    }}
    \sup_{\substack{
        \Bv_h\in\calU_{h,0}\\
        \Bv_h\neq\Bzero
    }}
    \frac{
        b(\Bv_h,\bar q_h)
    }{
        \|\Bv_h\|_{1,\Omega}
        \|\bar q_h\|_{0,\Omega}
    }
    \geq
    \bar{\beta}_0>0,
    \label{eq:constant_pressure_infsup}
\end{equation}
where $\bar{\beta}_0$ is independent of the mesh size.

Using Korn's inequality, this condition provides the mechanically scaled
pressure control
\begin{equation}
    \frac{1}{\sqrt{2\mu}}
    \|\bar q_h\|_{0,\Omega}
    \lesssim
    \frac{1}{\bar{\beta}_0}
    \sup_{\Bv_h\in\calU_{h,0}}
    \frac{
        |b(\Bv_h,\bar q_h)|
    }{
        \sqrt{2\mu}
        \|\Bepsilon(\Bv_h)\|_{0,\Omega}
    }.
    \label{eq:constant_pressure_scaled_control}
\end{equation}
Thus, the elementwise constant pressure component receives the same natural
$(2\mu)^{-1/2}$ pressure scaling as in the standard mixed formulation, but
its control now follows from the inf--sup stability of the reduced
constant-pressure subspace.

The two complementary pressure-control mechanisms motivate the following
mesh-dependent norm for the stabilized discontinuous-pressure formulation:
\begin{align}
    \left\|\!\left|\Bv_h,q_h\right|\!\right\|_{\mathrm{DG},\tau,\chi,\lambda}^{2}
    :={}&\sum_{e\in\calT_h}
    \biggl[
        2\mu A_{\chi,e}\|\Bepsilon(\Bv_h)\|_{0,\Omega_e}^{2}
    \nonumber\\
    &\quad+\tau_e B_{\chi,e}
        \|\Bnabla\widetilde q_h\|_{0,\Omega_e}^{2}\biggr]
        +\frac{1}{2\mu}\|\bar q_h\|_{0,\Omega}^{2}
        +\frac{1}{\lambda}\|q_h\|_{0,\Omega}^{2},
    \label{eq:dg_stabilized_mixed_norm}
\end{align}
where
$    q_h
    =
    \bar q_h+\widetilde q_h
$
is the elementwise decomposition defined above.

The individual pressure terms in
\cref{eq:dg_stabilized_mixed_norm} have distinct origins. The
pressure-gradient contribution controls the nonconstant component
$\widetilde q_h$, the term
$(2\mu)^{-1}\|\bar q_h\|_{0,\Omega}^{2}$ represents the constant-pressure
component controlled through
\cref{eq:constant_pressure_infsup}, and
$\lambda^{-1}\|q_h\|_{0,\Omega}^{2}$ represents the additional direct
pressure control provided by finite compressibility.

In the incompressible limit, the norm reduces to
\begin{equation}
    \left\|\!\left|
        \Bv_h,q_h
    \right|\!\right\|_{\mathrm{DG},\tau,\chi,\infty}^{2}
    =
    \sum_{e\in\calT_h}
    \left[
        2\mu A_{\chi,e}
        \|\Bepsilon(\Bv_h)\|_{0,\Omega_e}^{2}
        +
        \tau_e B_{\chi,e}
        \|\Bnabla\widetilde q_h\|_{0,\Omega_e}^{2}
    \right]
    +
    \frac{1}{2\mu}
    \|\bar q_h\|_{0,\Omega}^{2},
    \label{eq:dg_stabilized_mixed_norm_incompressible}
\end{equation}
which remains nondegenerate because the nonconstant and constant pressure
components are controlled by the stabilization and inf--sup mechanisms,
respectively.

\revx{The sufficiency of this complementary pressure control follows from the analysis of stabilized methods with discontinuous pressure by Franca, Hughes, and Stenberg~\cite{FrancaHughesStenberg1993}, together with the original analysis of the nonsymmetric form by Hughes, Franca, and Balestra~\cite{HughesFrancaBalestra1986}. The proof in~\cite{FrancaHughesStenberg1993} utilizes three ingredients: an elementwise inverse estimate for the displacement space, uniform inf--sup stability of the displacement space against the elementwise constant projection of the discontinuous pressure space, and a coercivity argument through Young's inequality. None of the three depends on the inter-element continuity of the displacement space. The inverse estimate is a statement about the polynomial restriction to each element, which a maximally continuous spline space possesses exactly as a \(\mathcal{C}^0\) space does, and it is precisely \cref{eq:element_inverse_estimate_stabilization}. The only spline-specific ingredient is the constant-mode inf--sup condition itself, \cref{eq:constant_pressure_infsup}, and that is supplied by the macroelement analysis of Antol\'in et al.~\cite{antolin2017isogeometric}. The two remaining differences from the setting of those papers are favorable rather than obstructive: the exact identity of \cref{eq:vms_energy_identity} makes the coercivity step for the VMS form simpler than for GLS, and the finite-compressibility term enters the coercivity estimate with a positive sign and the continuity estimate with a bounded one. The argument therefore applies to the present formulation as written, and no pressure-jump stabilization is required. In the present notation, the required projected pressure space is precisely \(\overline{\calP}_h\), and the error estimate in \cref{eq:method_dependent_error_estimate_incompressible} consequently extends to the discontinuous-pressure formulation with the continuous stabilized norm replaced by \cref{eq:dg_stabilized_mixed_norm_incompressible}. The quantitative step converting the pressure-gradient control in \cref{eq:dg_stabilized_mixed_norm} into \(L^2\) control of \(\tilde q_h\), with a constant explicit in the coarsening ratio \(n\), is recorded in the companion formalization and is not reproduced here.}

Now, it remains to determine when the constant-pressure inf--sup condition in
\cref{eq:constant_pressure_infsup} is satisfied by the spline spaces
considered here. For discontinuous pressure spaces, the spline
macroelement stability analysis developed by
Antol\'in et al.~\cite{antolin2017isogeometric} can be applied specifically
to the constant-pressure subspace $\overline{\calP}_h$. In the notation
adopted here, the corresponding \revx{sufficient} condition is
\begin{equation}
    n(k-r)>1.
    \label{eq:constant_pressure_inf_sup_condition}
\end{equation}

\begin{figure*}[!htb]
    \centering
    \includegraphics[width=\textwidth]
    {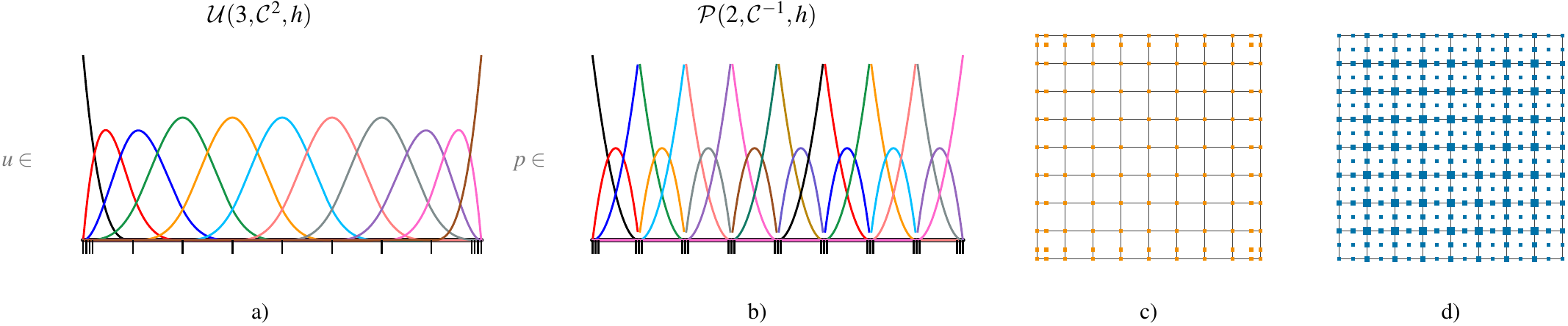}
    \caption{Same-mesh discontinuous-pressure pair for the representative
    cubic case,
    \unstablepair{\(\mathcal{U}(3,\mathcal{C}^2,h)/\mathcal{P}(2,\mathcal{C}^{-1},h)\)}.
    Panels show: (a) maximally continuous cubic displacement basis functions,
    (b) discontinuous quadratic pressure basis functions on the same mesh,
    (c) displacement mesh and control points, and
    (d) discontinuous pressure mesh and control points.}
    \label{fig:displacement_pressure_spaces_unstable_dg}
\end{figure*}

For a maximally continuous displacement space,
$r=k-1$, on the same pressure mesh, $n=1$, this gives
\begin{equation}
    1\left(k-(k-1)\right)
    =
    1
    \not>
    1.
    \label{eq:max_continuity_same_mesh_constant_pressure_fail}
\end{equation}
Hence, the same-mesh pair
\begin{equation}
   \unstablepair{ \mathcal{U}(k,\mathcal{C}^{k-1},h)/\mathcal{P}(0,\mathcal{C}^{-1},h)}
\end{equation}
does not satisfy the \revx{sufficient} constant-pressure macroelement condition\revx{, and its instability is confirmed by the discrete inf--sup tests of \Cref{subsubsec:kirsch_constant_pressure,subsubsec:cook_constant_pressure}}.
Because every higher-order discontinuous pressure space contains this
elementwise constant subspace, pressure-gradient stabilization alone cannot
establish the required uniform pressure control for the corresponding
same-mesh maximally continuous pair. This situation is illustrated in
\cref{fig:displacement_pressure_spaces_unstable_dg} for
$\unstablepair{\mathcal{U}(3,\mathcal{C}^2,h)/\allowbreak\mathcal{P}(2,\mathcal{C}^{-1},h)}$.

We now consider two natural spline constructions that satisfy the remaining
constant-pressure stability requirement.

\paragraph{Coarsen the pressure mesh}

Consider
\begin{equation}
    \stablepair{
        \mathcal{U}(k,\mathcal{C}^{k-1},h)/
        \mathcal{P}(k-1,\mathcal{C}^{-1},2h)
    }.
    \label{eq:remedy_pressure_coarsening}
\end{equation}
Although the complete pressure space has degree $k-1$, only its
elementwise constant subspace must satisfy the remaining inf--sup
condition. For $r=k-1$ and $n=2$,
\begin{equation}
    2\left(k-(k-1)\right)
    =
    2
    >
    1.
\end{equation}
Thus, the constant-pressure subspace satisfies the macroelement condition.
This construction preserves the maximum continuity of the displacement
space while using a pressure mesh twice as coarse as the displacement mesh.

\begin{figure*}[!htb]
    \centering
    \includegraphics[width=\textwidth]
    {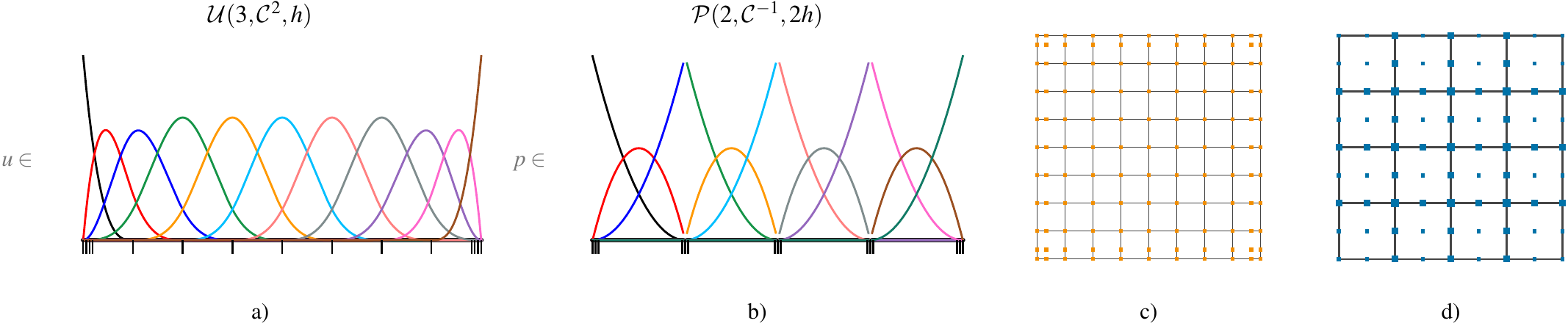}
    \caption{Coarsened discontinuous-pressure pair for the representative
    cubic case,
    \(\stablepair{\mathcal{U}(3,\mathcal{C}^2,h)/\mathcal{P}(2,\mathcal{C}^{-1},2h)}\).
    Panels show: (a) maximally continuous cubic displacement basis functions,
    (b) discontinuous quadratic pressure basis functions on a pressure mesh
    twice as coarse as the displacement mesh,
    (c) displacement mesh and control points, and
    (d) discontinuous pressure mesh and control points.}
    \label{fig:displacement_pressure_spaces_sub_dg}
\end{figure*}

\paragraph{Reduce the displacement continuity}

Alternatively, consider
\begin{equation}
    \stablepair{
        \mathcal{U}(k,\mathcal{C}^{k-2},h)/
        \mathcal{P}(k-1,\mathcal{C}^{-1},h)
    }.
    \label{eq:remedy_reduced_continuity}
\end{equation}
For $r=k-2$ and $n=1$, the constant-pressure condition gives
\begin{equation}
    1\left(k-(k-2)\right)
    =
    2
    >
    1.
\end{equation}
Thus, the elementwise constant pressure component satisfies the inf--sup
condition on the same pressure mesh. Reducing the displacement continuity
by one order therefore provides the additional displacement richness
required for constant-mode stability while preserving pressure-cell-local
elimination.

\begin{figure*}[!htb]
    \centering
    \includegraphics[width=\textwidth]
    {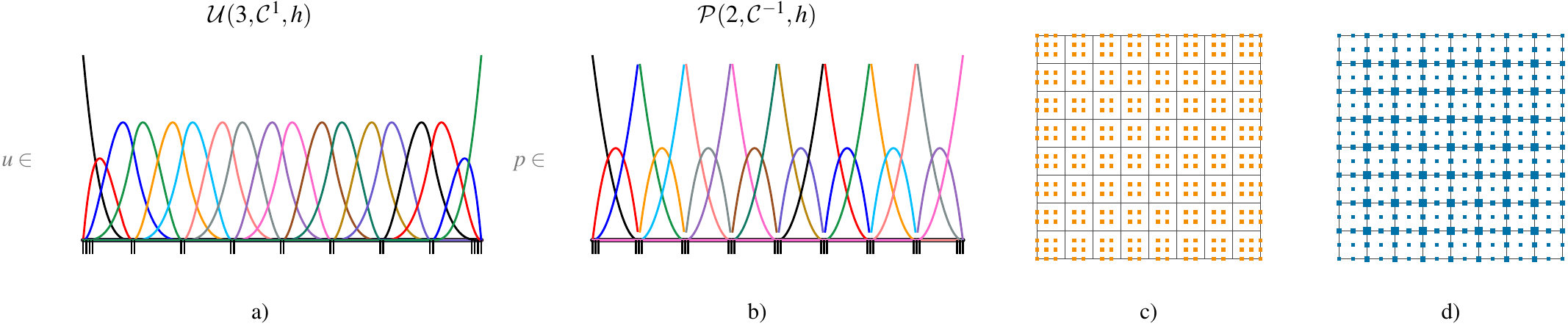}
    \caption{Reduced-continuity discontinuous-pressure pair for the
    representative cubic case,
    \(\stablepair{\mathcal{U}(3,\mathcal{C}^1,h)/\mathcal{P}(2,\mathcal{C}^{-1},h)}\).
    Panels show: (a) reduced-continuity cubic displacement basis functions,
    (b) discontinuous quadratic pressure basis functions on the same mesh,
    (c) displacement mesh and control points, and
    (d) discontinuous pressure mesh and control points.}
    \label{fig:displacement_pressure_spaces_reduced_dg}
\end{figure*}

\paragraph{Comparison with the unstabilized formulation}

The main consequence of the preceding analysis is the substantial reduction
in the pressure coarsening required for maximally continuous displacement
splines. For the unstabilized Galerkin formulation with
$k_p=k-1$ and $r=k-1$,
\cref{eq:discontinuous_max_continuity_condition} requires $n>k$, whereas
the stabilized formulation requires only the elementwise constant pressure
subspace to satisfy the inf--sup condition, for which
\cref{eq:constant_pressure_inf_sup_condition} gives $n>1$. Hence,
\begin{equation}
    \boxed{
        n_{\min}^{\mathrm{Gal}}
        =
        k+1,
        \qquad
        n_{\min}^{\mathrm{stab}}
        =
        2
    }.
    \label{eq:min_ratio_comparison_dg}
\end{equation}
Thus, residual stabilization replaces the degree-dependent pressure
coarsening required by the Galerkin formulation with a fixed factor of two,
independently of the displacement polynomial degree.

Among the two stable constructions considered above, pressure coarsening is
particularly attractive because it preserves the maximum continuity of the
B-spline displacement space while avoiding the additional displacement
degrees of freedom introduced by continuity reduction. It also reduces the
number of pressure unknowns while preserving pressure-cell-local static
condensation. Compared with the corresponding unstabilized Galerkin
construction, the fixed factor-two pressure mesh retains substantially
greater pressure resolution at higher polynomial degrees. The relative
accuracy and computational performance of the two alternatives are assessed
in the numerical examples.
\subsection{Immersed isogeometric discretization}
\label{subsec:immersed_iga_theory}

\noindent In the immersed isogeometric formulation, the physical domain
\(\Omega\) is represented independently of the analysis grid and is embedded
in a fixed Eulerian background mesh. The geometry of \(\Omega\) and its
boundary \(\Gamma\) are described by the physical NURBS discretization,
whereas the displacement and pressure basis functions are defined on the
background isogeometric grids. Consequently, the background mesh does not
need to conform to the physical boundary.

Although the approximation functions are defined over the background mesh,
the governing weak form is integrated only over the physical domain. For a
background cell intersected by the geometry, the volume integrals are
evaluated over the portion of the cell lying inside \(\Omega\), while the
boundary terms are evaluated over the corresponding portion of \(\Gamma\).
This separation allows the physical domain to be represented by a
boundary-conforming NURBS geometry while avoiding the construction of a
boundary-fitted background analysis mesh for the displacement and pressure
approximations.

For the linear small-deformation problems considered in this work, a complete
physical mesh and its associated interior control points are not strictly
required; only a sufficiently accurate geometric representation of
\(\Omega\) and its boundary \(\Gamma\) is needed to classify the background
cells and construct the cut-cell volume and surface quadrature. Nevertheless,
we retain a separate physical NURBS discretization as part of the
implementation. This choice anticipates future extensions to nonlinear
finite-deformation problems, in which the physical geometry and its control
points can evolve while the displacement and pressure approximations remain
defined on a fixed Eulerian background grid. Retaining this separation at the
present stage facilitates such extensions without altering the background
analysis-space construction.

The immersed setting requires three additional numerical ingredients. First,
essential boundary conditions cannot generally be imposed through the
background control variables because \(\Gamma_D\) does not coincide with the
background-grid boundary; they are therefore imposed weakly using Nitsche's
method \cite{nitsche1971variationsprinzip}. Second, background cells intersected by the physical boundary require
specialized volume and surface quadrature. Third, basis functions with only a
small portion of their support inside \(\Omega\) may lead to severe
ill-conditioning and must be treated using the extension and aggregation
procedures introduced below.

\begin{figure*}[!htb]
    \centering
    \includegraphics[width=\textwidth]
    {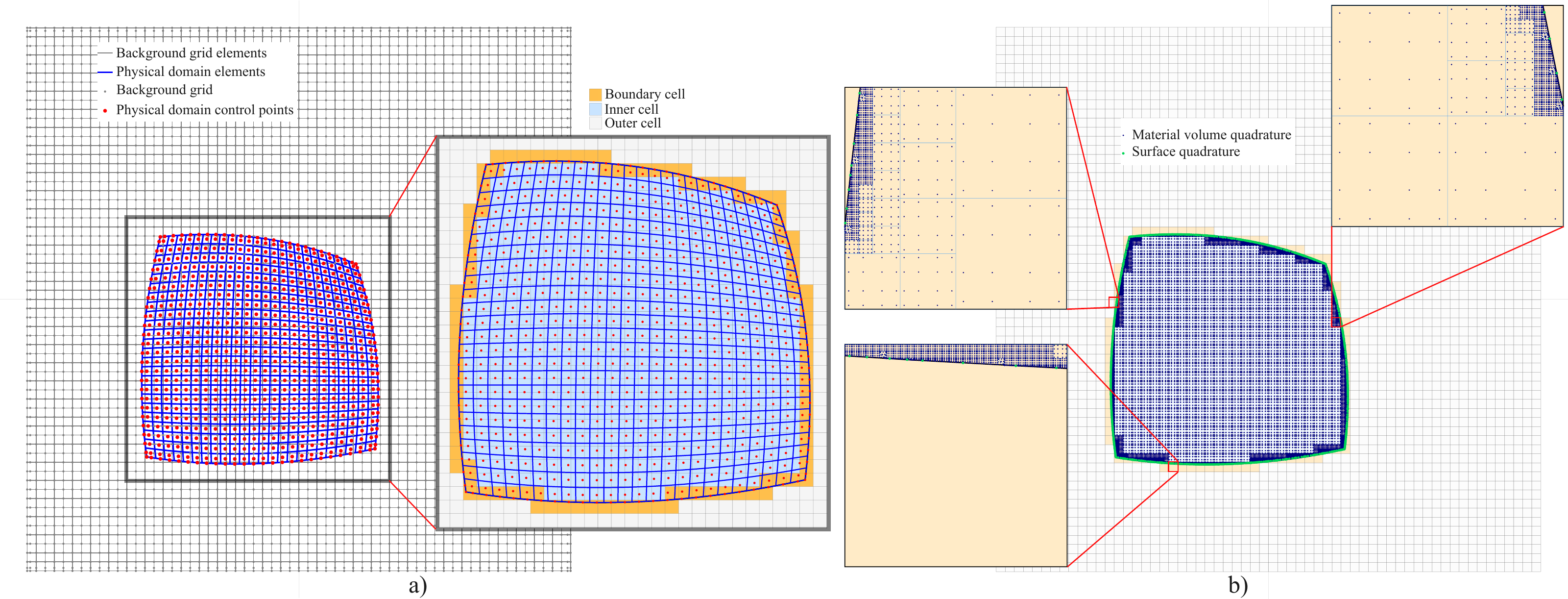}
\caption{Immersed isogeometric discretization and cut-cell integration.
(a) The physical NURBS discretization, shown by its physical elements and
control points, is embedded in an independent Eulerian background grid.
Background cells are classified as inner cells, which lie entirely inside
the physical domain; outer cells, which do not intersect the physical
domain; and boundary or cut cells, which are intersected by the physical
boundary. The enlarged view illustrates this classification near the
boundary.
(b) Numerical integration of the immersed domain. Inner cells are integrated
directly using standard tensor-product Gaussian quadrature, whereas boundary
cells are recursively subdivided. At the terminal subdivision depth, the remaining cut subcells are intersected with the physical domain, a polygonal approximation of their physical portions is constructed and triangulated, and Gaussian quadrature is applied on the resulting triangles. Material-volume quadrature points are retained only
inside the physical domain, while surface quadrature points are placed along
the physical boundary segment. The enlarged views show the local subdivision,
triangulation, and quadrature distributions near representative cut regions.}
    \label{fig:Mesh_Immersed}
\end{figure*}

The geometric embedding, cell classification, and cut-cell integration
procedure are summarized in \cref{fig:Mesh_Immersed}. Panel (a) illustrates
the physical NURBS discretization embedded in the background grid and the
resulting classification of background cells. Panel (b) shows the recursive
subdivision of boundary cells used to construct the material-volume and
surface quadrature rules.

\subsubsection{Active mesh and cell classification}
\label{subsubsec:cell_classification}

\noindent Let \(\calT_h^{\mathrm{bg}}\) denote the background grid. Each
background cell \(K\in\calT_h^{\mathrm{bg}}\) is classified according to its
intersection with the physical domain:
\begin{align}
    K \text{ is an inner cell}
    &\quad\Longleftrightarrow\quad
    K\subset\Omega,
    \label{eq:inner_cell_definition}
    \\
    K \text{ is an outer cell}
    &\quad\Longleftrightarrow\quad
    K\cap\Omega=\emptyset,
    \label{eq:outer_cell_definition}
    \\
    K \text{ is a boundary cell}
    &\quad\Longleftrightarrow\quad
    K\cap\Omega\neq\emptyset
    \quad\text{and}\quad
    K\not\subset\Omega.
    \label{eq:boundary_cell_definition}
\end{align}
Boundary cells are also referred to as cut cells because the physical
boundary passes through their interiors.

The active background mesh is the union of the inner and boundary cells:
\begin{equation}
    \calT_h^{\mathrm{act}}
    =
    \left\{
    K\in\calT_h^{\mathrm{bg}}
    :
    K\cap\Omega\neq\emptyset
    \right\}.
    \label{eq:active_background_mesh}
\end{equation}
Outer cells contain no physical material and are excluded from the
discretized problem. For every active cell, we define its physical volume
portion as
\begin{equation}
    \Omega_K
    :=
    K\cap\Omega.
    \label{eq:physical_cell_portion}
\end{equation}
For a boundary cell intersected by \(\Gamma\), the corresponding physical
boundary segment is
\begin{equation}
    \Gamma_K
    :=
    K\cap\Gamma.
    \label{eq:physical_boundary_segment}
\end{equation}

The classification is performed from the relative position of the background
cell and the physical NURBS geometry. Inner cells are completely covered by
the physical domain and retain their standard tensor-product integration
structure. Outer cells are inactive. Boundary cells contain only a portion
of the physical domain and require separate treatment for both volume and
surface integration.

The same cell classification also provides the geometric basis for the
cut-cell stabilization procedures introduced later. In particular, it
identifies the near-boundary regions in which basis functions may have
insufficient physical support and in which displacement extension or
pressure-cell aggregation is required.

\subsubsection{Subdivision quadrature for boundary cells}
\label{subsubsec:subdivision_quadrature}

\noindent Inner cells preserve the tensor-product structure of the
background grid and are integrated using standard Gaussian quadrature.
Boundary cells, however, contain non-tensor-product physical regions
\(\Omega_K\) whose boundaries are determined by the intersection of the
background cell with the physical NURBS geometry. These cells are integrated
using recursive subdivision followed by triangulation of the remaining cut
regions at the terminal subdivision level.

For each boundary cell \(K\), the procedure begins by subdividing the cell
into a set of smaller tensor-product subcells. Each subcell \(S\subset K\) is
then classified with respect to \(\Omega\):
\begin{align}
    S\subset\Omega
    &\quad\Longrightarrow\quad
    S \text{ is accepted as an inner subcell},
    \\
    S\cap\Omega=\emptyset
    &\quad\Longrightarrow\quad
    S \text{ is discarded},
    \\
    S\cap\Omega\neq\emptyset,\quad S\not\subset\Omega
    &\quad\Longrightarrow\quad
    S \text{ remains cut}.
\end{align}
A cut subcell is subdivided further if the prescribed maximum subdivision
depth has not yet been reached. At the terminal subdivision level, each
remaining cut subcell is intersected with the physical domain, and a polygonal approximation of the remaining physical region is constructed and triangulated. Standard Gaussian
quadrature is then applied on the resulting triangles. Thus, the physical
portion of a boundary cell is represented by a collection of accepted
tensor-product inner subcells together with triangles obtained from the
terminal cut subcells.

Standard tensor-product Gaussian quadrature is applied independently on each
accepted inner subcell, while triangular Gaussian quadrature is applied on
each triangle obtained at the terminal subdivision depth. If
\(\{\Bx_{K\alpha},w_{K\alpha}^{V}\}_{\alpha=1}^{n_K^V}\)
denotes the combined material-volume quadrature rule in \(K\), then a generic
volume integral is approximated as
\begin{equation}
    \int_{\Omega_K} f(\Bx)\dV
    \approx
    \sum_{\alpha=1}^{n_K^V}
    w_{K\alpha}^{V}
    f(\Bx_{K\alpha}).
    \label{eq:cut_cell_volume_quadrature}
\end{equation}
The quadrature set therefore contains contributions from both the accepted
tensor-product subcells and the triangulated terminal cut regions. The
quadrature weights include the mappings from the corresponding parent
quadrature domains to the subdivision subcells or physical triangles and,
where required, the mapping from the parametric representation to the
physical configuration.

Boundary cells also require integration over
\(\Gamma_K=K\cap\Gamma\). The physical boundary segment is obtained from the
NURBS geometry, and a one-dimensional surface quadrature rule is constructed
along this segment. If
\(\{\Bx_{K\beta}^{\Gamma},w_{K\beta}^{\Gamma}\}_{\beta=1}^{n_K^\Gamma}\)
denotes this rule, then
\begin{equation}
    \int_{\Gamma_K} g(\Bx)\dA
    \approx
    \sum_{\beta=1}^{n_K^\Gamma}
    w_{K\beta}^{\Gamma}
    g(\Bx_{K\beta}^{\Gamma}).
    \label{eq:cut_cell_surface_quadrature}
\end{equation}
The associated unit normals are evaluated from the physical NURBS boundary
at the surface quadrature points.

The material-volume quadrature is used for all cell-interior contributions,
including the Galerkin mixed form, the compressibility term, and the
residual-based stabilization. The surface quadrature is used for prescribed
tractions and for the consistency, adjoint-consistency, and penalty
contributions of the Nitsche formulation.

Recursive subdivision concentrates the integration resolution near the
physical boundary while retaining comparatively large tensor-product
quadrature subcells away from it, as illustrated in
\cref{fig:Mesh_Immersed}b. Triangulation at the terminal subdivision level
provides a geometric representation of the remaining non-tensor-product
physical regions and allows these regions to be integrated without generating
a globally boundary-fitted analysis mesh. Increasing the subdivision depth
reduces the size of the terminal cut regions and improves the geometric and
integration accuracy at the expense of additional quadrature cells and
points.
\subsubsection{Nitsche method for essential boundary conditions}
\label{subsubsec:nitsche_method}

\noindent
Because the physical boundary is not aligned with the background grid, the
essential boundary condition
\begin{equation}
    \Bu=\Bg
    \qquad
    \text{on }\Gamma_D
\end{equation}
cannot, in general, be imposed strongly through the background control
variables. Instead, it is imposed weakly using Nitsche's method. For boundaries on which only selected displacement components are prescribed, the same formulation is applied componentwise, or equivalently after projection onto the prescribed displacement directions. The full-vector notation used below is retained for compactness.

Since the Dirichlet condition is imposed weakly, the immersed displacement
trial and test functions are taken from the same unconstrained active
background space, denoted by
\(\calW_h\). In particular, no trace condition is built into
\(\calW_h\) on \(\Gamma_D\). Thus,
\begin{equation}
    \Bu_h,\Bv_h\in\calW_h .
    \label{eq:immersed_unconstrained_displacement_space}
\end{equation}
The extension procedure introduced below replaces \(\calW_h\) by the
corresponding extended displacement subspace without changing the Nitsche
formulation.

We define the mixed traction operator by
\begin{equation}
    \Bt(\Bu,p)
    :=
    \Bsigma(\Bu,p)\Bn
    =
    \left(
        2\mu\Bepsilon(\Bu)-p\BI
    \right)\Bn ,
    \label{eq:traction_operator}
\end{equation}
and introduce its displacement contribution
\begin{equation}
    \Bt_{\mu}(\Bu)
    :=
    2\mu\Bepsilon(\Bu)\Bn .
    \label{eq:displacement_traction_operator}
\end{equation}
Hence,
\begin{equation}
    \Bt(\Bu,p)
    =
    \Bt_{\mu}(\Bu)-p\Bn .
    \label{eq:mixed_traction_decomposition}
\end{equation}

Let \(\calB_{\tau}\) and \(\calF_{\tau}\) denote the Galerkin or
residual-based stabilized forms introduced previously, with
\(\tau_e=0\) understood for the unstabilized Galerkin formulation.
In the immersed setting, all volume integrals appearing in these forms are
evaluated over the physical portions of the active background cells.

The immersed formulation is: find
\begin{equation}
    (\Bu_h,p_h)
    \in
    \calW_h\times\calP_h
\end{equation}
such that
\begin{equation}
    \calB_h(\Bu_h,p_h;\Bv_h,q_h)
    =
    \calF_h(\Bv_h,q_h)
    \qquad
    \forall
    (\Bv_h,q_h)
    \in
    \calW_h\times\calP_h ,
    \label{eq:immersed_nitsche_problem}
\end{equation}
where
\begin{equation}
\begin{aligned}
    \calB_h(\Bu_h,p_h;\Bv_h,q_h)
    ={}&\calB_{\tau}(\Bu_h,p_h;\Bv_h,q_h)
       -\int_{\Gamma_D}\Bt(\Bu_h,p_h)\cdot\Bv_h\,\dA
    \\
    &-s_k\int_{\Gamma_D}\Bt_{\mu}(\Bv_h)\cdot\Bu_h\,\dA
       +\int_{\Gamma_D}q_h\,\Bn\cdot\Bu_h\,\dA
    \\
    &+\sum_{K\in\calT_h^{\Gamma_D}}
        \int_{\Gamma_D\cap K}
        \gamma_{N,K}\frac{\mu}{h_{\Gamma,K}}
        \Bu_h\cdot\Bv_h\,\dA ,
    \end{aligned}
    \label{eq:nitsche_bilinear}
\end{equation}
and
\begin{align}
    \calF_h(\Bv_h,q_h)
    ={}&\calF_{\tau}(\Bv_h,q_h)
       -s_k\int_{\Gamma_D}\Bt_{\mu}(\Bv_h)\cdot\Bg\,\dA
    \nonumber\\
    &+\int_{\Gamma_D}q_h\,\Bn\cdot\Bg\,\dA
       +\sum_{K\in\calT_h^{\Gamma_D}}
        \int_{\Gamma_D\cap K}
        \gamma_{N,K}\frac{\mu}{h_{\Gamma,K}}
        \Bg\cdot\Bv_h\,\dA .
    \label{eq:nitsche_linear}
\end{align}
Here \(\calT_h^{\Gamma_D}\) denotes the set of active background cells
intersected by \(\Gamma_D\), and \(h_{\Gamma,K}\) is the characteristic
background-grid size associated with the boundary segment
\(\Gamma_D\cap K\). The dimensionless parameter
\(\gamma_{N,K}\) controls the Nitsche penalty contribution.

The parameter \(s_k\) selects the adjoint-consistency variant:
\begin{equation}
    s_k
    =
    \begin{cases}
        +1,
        & \text{symmetric Nitsche formulation},\\[2mm]
        -1,
        & \text{skew-symmetric Nitsche formulation}.
    \end{cases}
    \label{eq:nitsche_variant_parameter}
\end{equation}

For \(s_k=+1\), the adjoint terms in
\cref{eq:nitsche_bilinear} can be combined as
\begin{equation}
    -
    \Bt_{\mu}(\Bv_h)\cdot\Bu_h
    +
    q_h\Bn\cdot\Bu_h
    =
    -
    \Bt(\Bv_h,q_h)\cdot\Bu_h ,
\end{equation}
which gives the symmetric mixed Nitsche form. For \(s_k=-1\),
\begin{equation}
    \Bt_{\mu}(\Bv_h)\cdot\Bu_h
    +
    q_h\Bn\cdot\Bu_h
    =
    \Bt(\Bv_h,-q_h)\cdot\Bu_h ,
\end{equation}
so that the skew-symmetric form is equivalently written using the adjoint
traction \(\Bt(\Bv_h,-q_h)\). This representation is consistent with the
mixed energy pairing
\begin{equation}
    (\Bv_h,q_h)
    =
    (\Bu_h,-p_h).
\end{equation}

\subsubsection{Nitsche stability estimate}
\label{subsubsec:nitsche_stability}

\noindent
We next examine the contribution of the weak boundary treatment to the
energy estimate. The pressure-control mechanisms established previously are
not replaced by the Nitsche argument. In particular, the
\((2\mu)^{-1}\|p_h\|_{0,\Omega}^{2}\) contribution appearing in the robust
mixed norm is controlled through the corresponding inf--sup mechanism, while
the residual stabilization controls the pressure-gradient contribution when
present.

For the present analysis, let
\begin{equation}
    \mathcal{E}_{\tau,h}^{2}(\Bu_h,p_h)
\end{equation}
denote the collection of nonnegative terms controlled directly by the
Galerkin or stabilized energy estimate derived previously. Thus,
\begin{equation}
    \calB_{\tau}(\Bu_h,p_h;\Bu_h,-p_h)
    \geq
    \mathcal{E}_{\tau,h}^{2}(\Bu_h,p_h).
    \label{eq:nitsche_volume_stability}
\end{equation}
For the Galerkin formulation,
\begin{equation}
    \mathcal{E}_{0,h}^{2}(\Bu_h,p_h)
    =
    2\mu
    \|\Bepsilon(\Bu_h)\|_{0,\Omega}^{2}
    +
    \frac{1}{\lambda}
    \|p_h\|_{0,\Omega}^{2},
    \label{eq:nitsche_galerkin_direct_energy}
\end{equation}
whereas for the residual-based formulations
\(\mathcal{E}_{\tau,h}\) is given by the nonnegative terms on the
right-hand side of the corresponding GLS, SUPG/PSPG, or VMS stability
estimate.

For each boundary cut cell
\(K\in\calT_h^{\Gamma_D}\), denote the displacement contribution to this
local energy by
\begin{equation}
    |\Bv_h|_{u,h,K}^{2}
    :=
    2\mu A_{\chi,K}
    \|\Bepsilon(\Bv_h)\|_{0,\Omega_K}^{2},
    \label{eq:nitsche_local_displacement_energy}
\end{equation}
where \(A_{\chi,K}\) is the displacement weight appearing in the
corresponding stabilized estimate. For the Galerkin formulation,
\(A_{\chi,K}=1\).

We use the scaled boundary inverse estimate
\begin{equation}
    \frac{h_{\Gamma,K}}{\mu}
    \left\|
        \Bt_{\mu}(\Bv_h)
    \right\|_{0,\Gamma_D\cap K}^{2}
    \leq
    \lambda_{N,K}
    |\Bv_h|_{u,h,K}^{2},
    \qquad
    K\in\calT_h^{\Gamma_D},
    \label{eq:nitsche_boundary_inverse}
\end{equation}
where \(\lambda_{N,K}\) is the dimensionless local boundary inverse
constant.

Testing the Nitsche contribution with
\begin{equation}
    (\Bv_h,q_h)
    =
    (\Bu_h,-p_h)
\end{equation}
gives
\begin{align}
    \calN_h(\Bu_h,p_h;\Bu_h,-p_h)
    ={}&-\int_{\Gamma_D}
        \left[\Bt_{\mu}(\Bu_h)-p_h\Bn\right]
        \cdot\Bu_h\,\dA
    \nonumber\\
    &-s_k\int_{\Gamma_D}\Bt_{\mu}(\Bu_h)\cdot\Bu_h\,\dA
       -\int_{\Gamma_D}p_h\Bn\cdot\Bu_h\,\dA
    \nonumber\\
    &+\sum_{K\in\calT_h^{\Gamma_D}}
        \gamma_{N,K}\frac{\mu}{h_{\Gamma,K}}
        \|\Bu_h\|_{0,\Gamma_D\cap K}^{2}.
    \label{eq:nitsche_energy_contribution}
\end{align}
The two pressure boundary contributions cancel exactly. Therefore,
\begin{equation}
    \begin{aligned}
    \calN_h(\Bu_h,p_h;\Bu_h,-p_h)
    ={}&-(1+s_k)
    \int_{\Gamma_D}\Bt_{\mu}(\Bu_h)\cdot\Bu_h\,\dA
    \\
    &+\sum_{K\in\calT_h^{\Gamma_D}}
    \gamma_{N,K}\frac{\mu}{h_{\Gamma,K}}
    \|\Bu_h\|_{0,\Gamma_D\cap K}^{2}.
    \end{aligned}
    \label{eq:nitsche_energy_combined}
\end{equation}

The potentially negative consistency contribution is bounded cellwise using
\cref{eq:nitsche_boundary_inverse} and Young's inequality:
\begin{align}
    &|1+s_k|\|\Bt_{\mu}(\Bu_h)\|_{0,\Gamma_D\cap K}
        \|\Bu_h\|_{0,\Gamma_D\cap K}
    \nonumber\\
    &\qquad\leq\eta|\Bu_h|_{u,h,K}^{2}
       +\frac{(1+s_k)^2\lambda_{N,K}}{4\eta}
        \frac{\mu}{h_{\Gamma,K}}
        \|\Bu_h\|_{0,\Gamma_D\cap K}^{2},\quad 0<\eta<1 .
    \label{eq:nitsche_young}
\end{align}

Combining
\cref{eq:nitsche_volume_stability,eq:nitsche_energy_combined,eq:nitsche_young}
and using the fact that the sum of the local displacement-energy
contributions over the boundary cells is bounded by the global direct energy
gives
\begin{align}
    \calB_h(\Bu_h,p_h;\Bu_h,-p_h)
    \geq{}&(1-\eta)\mathcal{E}_{\tau,h}^{2}(\Bu_h,p_h)
    \nonumber\\
    &+\sum_{K\in\calT_h^{\Gamma_D}}
        \left(\gamma_{N,K}
        -\frac{(1+s_k)^2\lambda_{N,K}}{4\eta}\right)
        \frac{\mu}{h_{\Gamma,K}}
        \|\Bu_h\|_{0,\Gamma_D\cap K}^{2}.
    \label{eq:nitsche_final_stability_bound}
\end{align}

Consequently, for any fixed \(0<\eta<1\), positivity of the boundary
contribution is guaranteed if
\begin{equation}
    \gamma_{N,K}
    >
    \frac{(1+s_k)^2}{4\eta}
    \lambda_{N,K},
    \qquad
    K\in\calT_h^{\Gamma_D}.
    \label{eq:nitsche_penalty_condition_eta}
\end{equation}
Since \(\eta\) may be chosen arbitrarily close to one, such an
\(\eta\in(0,1)\) exists whenever
\begin{equation}
    \gamma_{N,K}
    >
    \frac{(1+s_k)^2}{4}
    \lambda_{N,K}.
    \label{eq:nitsche_penalty_condition}
\end{equation}

The corresponding sufficient conditions are therefore
\begin{align}
    s_k=+1
    &:
    &
    \gamma_{N,K}
    &>
    \lambda_{N,K},
    &&
    \text{symmetric Nitsche formulation},
    \label{eq:nitsche_symmetric_condition}
    \\
    s_k=-1
    &:
    &
    \gamma_{N,K}
    &>
    0,
    &&
    \text{skew-symmetric Nitsche formulation}.
    \label{eq:nitsche_skew_condition}
\end{align}

For the skew-symmetric choice \(s_k=-1\), the Nitsche consistency
contributions cancel exactly in the mixed energy pairing. Consequently, the
energy estimate imposes no positive lower bound on
\(\gamma_{N,K}\) relative to the boundary inverse constant
\(\lambda_{N,K}\). Any \(\gamma_{N,K}>0\) provides a positive
boundary-trace contribution. The pressure stability remains governed by the
inf--sup and residual-stabilization mechanisms established previously.

In the computations, we use
\begin{equation}
    s_k=-1,
    \qquad
    \gamma_{N,K}
    =
    0.01\lambda_{N,K},
    \label{eq:skew_nitsche_gamma}
\end{equation}
which provides mild weak control of the boundary displacement without
requiring a cut-dependent coercivity threshold. We assume that
\(\Gamma_D\) has sufficient measure to eliminate rigid-body motions once the
positive boundary contribution is included.

\paragraph{Computation of the boundary inverse constant.}

The boundary inverse constant involves only the displacement traction
\(\Bt_{\mu}\), because the pressure contributions cancel in the mixed energy
pairing. Let \(\Bd_K\) denote the vector of local displacement coefficients
associated with the active displacement basis functions on the boundary cut
cell \(K\). Define the local boundary matrix \(\BA_{N,K}\) by
\begin{equation}
    \Bd_K^{T}
    \BA_{N,K}
    \Bd_K
    :=
    \frac{h_{\Gamma,K}}{\mu}
    \left\|
        \Bt_{\mu}(\Bu_h)
    \right\|_{0,\Gamma_D\cap K}^{2}.
    \label{eq:nitsche_boundary_matrix_quadratic}
\end{equation}
Equivalently,
\begin{equation}
    \left[\BA_{N,K}\right]_{AB}
    =
    \frac{h_{\Gamma,K}}{\mu}
    \int_{\Gamma_D\cap K}
    \Bt_{\mu}(\Bvarphi_A^u)
    \cdot
    \Bt_{\mu}(\Bvarphi_B^u)
    \dA ,
    \label{eq:nitsche_boundary_matrix}
\end{equation}
where \(\Bvarphi_A^u\) denotes the \(A\)-th local vector-valued
displacement basis function.

Similarly, define the local displacement-energy matrix
\(\BM_{N,K}\) through
\begin{equation}
    \Bd_K^{T}
    \BM_{N,K}
    \Bd_K
    :=
    |\Bu_h|_{u,h,K}^{2}.
    \label{eq:nitsche_energy_matrix}
\end{equation}
For the Galerkin formulation, this matrix corresponds to
\begin{equation}
    |\Bu_h|_{u,h,K}^{2}
    =
    2\mu
    \|\Bepsilon(\Bu_h)\|_{0,\Omega_K}^{2},
\end{equation}
whereas for the GLS, SUPG/PSPG, or VMS formulations the corresponding
coefficient \(A_{\chi,K}\) is inherited from the method-dependent stability
estimate introduced previously.

The sharp local boundary inverse constant is the largest finite generalized
Rayleigh quotient
\begin{equation}
    \lambda_{N,K}
    =
    \max_{\Bd_K\neq\Bzero}
    \frac{
        \Bd_K^{T}\BA_{N,K}\Bd_K
    }{
        \Bd_K^{T}\BM_{N,K}\Bd_K
    },
    \label{eq:nitsche_rayleigh_quotient}
\end{equation}
or, equivalently, the largest finite generalized eigenvalue satisfying
\begin{equation}
    \BA_{N,K}\Bd_K
    =
    \lambda_{N,K}
    \BM_{N,K}\Bd_K .
    \label{eq:nitsche_generalized_eigenproblem}
\end{equation}
The generalized eigenproblem is evaluated on the non-null subspace of
\(\BM_{N,K}\). The resulting value accounts for the local spline degree,
continuity, cut-cell geometry, physical boundary segment, and quadrature
rule. When the extended B-spline construction is used, the matrices are
formed using the actual extended displacement basis employed in the immersed
assembly. The resulting \(\lambda_{N,K}\) is then used directly in
\cref{eq:skew_nitsche_gamma}.
\subsubsection{Cut-cell instability for continuous displacement fields}
\label{subsubsec:extended_bspline}

\noindent In an immersed discretization, a B-spline basis function may have only a very small
portion of its support inside the physical domain. The corresponding control variable is then
only weakly constrained by the discrete bilinear form, which can produce very small matrix
entries and severe deterioration of the conditioning of the immersed system. For the
continuous displacement field, this cut-cell instability is treated using an extended
B-spline construction
\cite{hollig2001weighted,yamaguchi2021extended}.

The construction is based on the polynomial-reproduction property of B-splines. In one
dimension, Marsden's identity \cite{marsden1970identity,hollig2013approximation} states that a polynomial of degree \(k\) can be represented
exactly in a degree-\(k\) B-spline basis:
\begin{equation}
    (x-t)^k
    =
    \sum_{i\in\mathbb{Z}}
    \Psi_{i,k}^{h}(t)\,
    N_{i,k}^{h}(x),
    \label{eq:marsden_identity_1d}
\end{equation}
where, for a uniform knot spacing \(h\),
\begin{equation}
    \Psi_{i,k}^{h}(t)
    =
    h^k
    \prod_{r=1}^{k}
    \left(
    i+r-\frac{t}{h}
    \right).
    \label{eq:marsden_coefficient_1d}
\end{equation}
Differentiation of \cref{eq:marsden_identity_1d} with respect to \(t\) shows that every
polynomial of degree at most \(k\) can be represented by the same B-spline basis. The
corresponding tensor-product construction extends this property to multiple dimensions.

Let \(\mathcal I\), \(\mathcal B\), and \(\mathcal O\) denote the sets of inner, boundary,
and outer displacement control variables, respectively. A control variable is classified
according to the physical support of its associated basis function:
\begin{align}
    A\in\mathcal I
    &\quad\Longleftrightarrow\quad
    \exists\,K\in\calT_h^{\mathrm{inner}}
    \ \text{such that}\
    \operatorname{supp}(N_A^u)\cap K\neq\emptyset,
    \\
    A\in\mathcal B
    &\quad\Longleftrightarrow\quad
    A\notin\mathcal I
    \quad\text{and}\quad
    \exists\,K\in\calT_h^{\mathrm{boundary}}
    \ \text{such that}\
    \operatorname{supp}(N_A^u)\cap K\neq\emptyset,
    \\
    A\in\mathcal O
    &\quad\Longleftrightarrow\quad
    A\notin\mathcal I\cup\mathcal B .
\end{align}

For each boundary index \(\boldsymbol{j}\in\mathcal B\), a neighboring tensor-product block
\(\mathcal I_{\boldsymbol{j}}\subset\mathcal I\) of
\((k+1)^d\) inner indices is selected. If
\(\boldsymbol{l}=(l_1,\ldots,l_d)\) is the lower-corner index of this block, then
\begin{equation}
    \mathcal I_{\boldsymbol{j}}
    =
    \left\{
    \boldsymbol{i}\in\mathcal I:
    l_\alpha\leq i_\alpha\leq l_\alpha+k,
    \quad
    \alpha=1,\ldots,d
    \right\}.
    \label{eq:interior_extension_block}
\end{equation}
The boundary displacement coefficient is extrapolated from this block according to
\begin{equation}
    \Bd_{\boldsymbol{j}}
    =
    \sum_{\boldsymbol{i}\in\mathcal I_{\boldsymbol{j}}}
    E_{\boldsymbol{i}\boldsymbol{j}}
    \Bd_{\boldsymbol{i}},
    \qquad
    \boldsymbol{j}\in\mathcal B.
    \label{eq:boundary_cp_extrapolation}
\end{equation}
The extrapolation coefficients are the tensor-product Lagrange coefficients
\begin{equation}
    E_{\boldsymbol{i}\boldsymbol{j}}
    =
    \prod_{\alpha=1}^{d}
    \left[
    \prod_{\substack{s_\alpha=l_\alpha\\s_\alpha\neq i_\alpha}}^{l_\alpha+k}
    \frac{j_\alpha-s_\alpha}
         {i_\alpha-s_\alpha}
    \right],
    \qquad
    \boldsymbol{i}\in\mathcal I_{\boldsymbol{j}}.
    \label{eq:extended_bspline_coefficients}
\end{equation}
Because these coefficients reproduce polynomials of degree \(k\), replacing the boundary
coefficients by \cref{eq:boundary_cp_extrapolation} preserves the polynomial-completeness
properties of the original spline space.

To express the resulting approximation directly in terms of inner coefficients, define the
inverse incidence set
\begin{equation}
    \mathcal I_{\boldsymbol{i}}^{-1}
    :=
    \left\{
    \boldsymbol{j}\in\mathcal B:
    \boldsymbol{i}\in\mathcal I_{\boldsymbol{j}}
    \right\}.
    \label{eq:inverse_incidence_set}
\end{equation}
Substitution of \cref{eq:boundary_cp_extrapolation} into the displacement approximation gives
\begin{equation}
    \Bu_h(\Bx)
    =
    \sum_{\boldsymbol{i}\in\mathcal I}
    \widetilde{N}_{\boldsymbol{i},k}^{u}(\Bx)
    \Bd_{\boldsymbol{i}},
    \label{eq:extended_displacement_approximation}
\end{equation}
where the extended displacement basis functions are
\begin{equation}
    \widetilde{N}_{\boldsymbol{i},k}^{u}(\Bx)
    =
    N_{\boldsymbol{i},k}^{u}(\Bx)
    +
    \sum_{\boldsymbol{j}\in\mathcal I_{\boldsymbol{i}}^{-1}}
    E_{\boldsymbol{i}\boldsymbol{j}}
    N_{\boldsymbol{j},k}^{u}(\Bx).
    \label{eq:extended_displacement_basis}
\end{equation}
Thus, the contribution of each eliminated boundary basis function is redistributed among the
basis functions associated with its neighboring inner block.

\begin{figure*}[!htb]
    \centering
    \includegraphics[width=\textwidth]
    {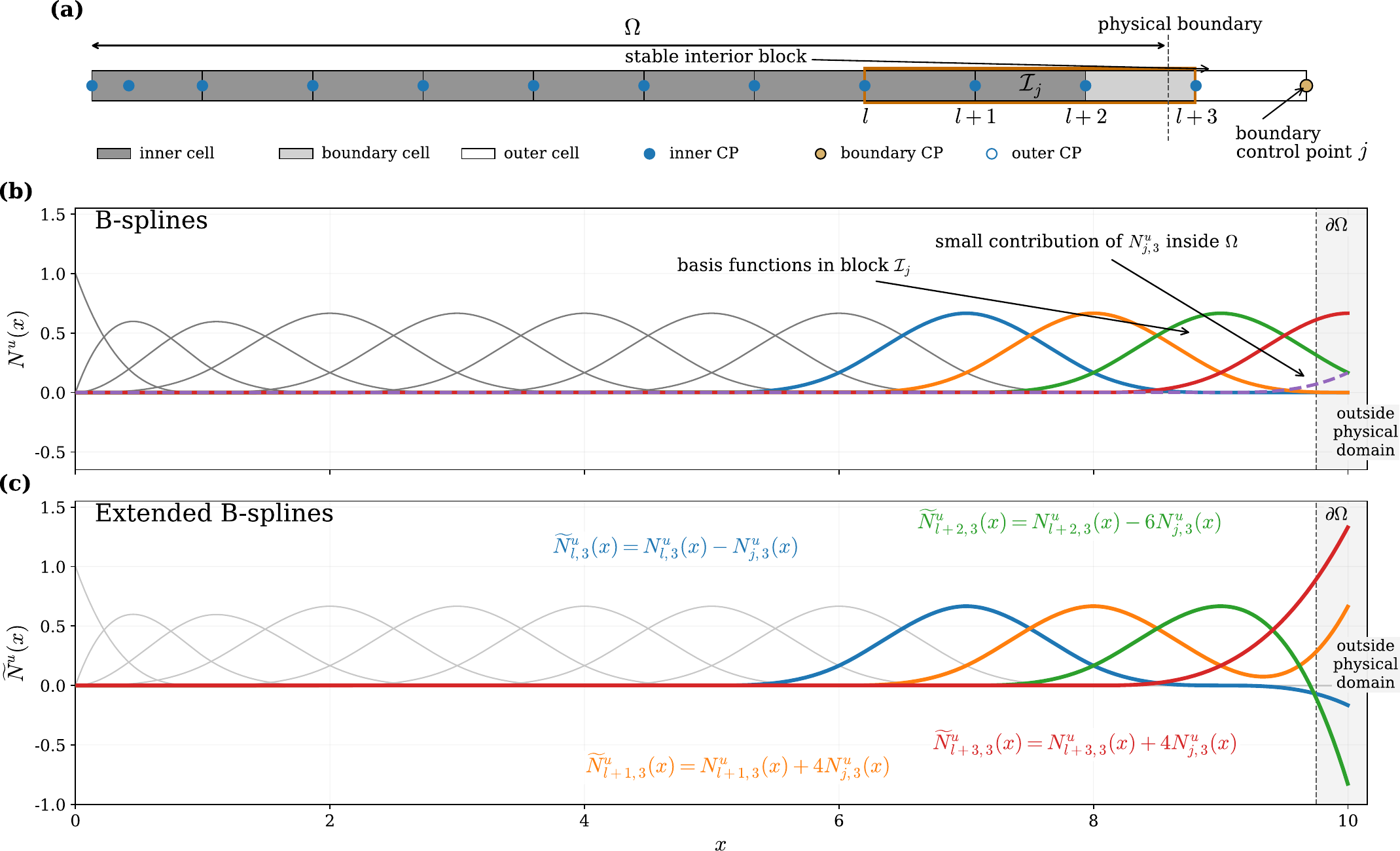}
    \caption{One-dimensional cubic extended B-spline construction for the continuous
    displacement field.
    \textbf{(a)} Classification of background cells and displacement control points near the
    physical boundary. The boundary control point \(j\), whose basis function has only a
    small contribution inside \(\Omega\), is associated with the neighboring inner block
    \(\mathcal I_j=\{l,l+1,l+2,l+3\}\).
    \textbf{(b)} Original cubic B-spline basis functions. The dashed basis
    \(N_{j,3}^{u}\) has small physical support and is removed as an independent basis
    function.
    \textbf{(c)} Extended cubic B-spline basis functions obtained using the coefficients
    \([-1,4,-6,4]\). The equations are colored consistently with their corresponding
    modified basis functions.}
    \label{fig:Extended_B-spline}
\end{figure*}

The one-dimensional cubic construction is illustrated in
\cref{fig:Extended_B-spline}. For the configuration shown, the boundary control point \(j\)
is associated with the four-point inner block
\begin{equation}
    \mathcal I_j
    =
    \{l,l+1,l+2,l+3\}.
\end{equation}
Because the block lies immediately to the left of \(j\), evaluation of
\cref{eq:extended_bspline_coefficients} gives
\begin{equation}
    \left[
    E_{lj},
    E_{(l+1)j},
    E_{(l+2)j},
    E_{(l+3)j}
    \right]
    =
    [-1,\,4,\,-6,\,4].
    \label{eq:cubic_extension_coefficients}
\end{equation}
Consequently,
\begin{align}
    \widetilde{N}_{l,3}^{u}
    &=
    N_{l,3}^{u}-N_{j,3}^{u},
    \\
    \widetilde{N}_{l+1,3}^{u}
    &=
    N_{l+1,3}^{u}+4N_{j,3}^{u},
    \\
    \widetilde{N}_{l+2,3}^{u}
    &=
    N_{l+2,3}^{u}-6N_{j,3}^{u},
    \\
    \widetilde{N}_{l+3,3}^{u}
    &=
    N_{l+3,3}^{u}+4N_{j,3}^{u}.
    \label{eq:cubic_extended_basis_example}
\end{align}
The small physical contribution of \(N_{j,3}^{u}\) is therefore transferred to four
well-supported interior basis functions, and \(N_{j,3}^{u}\) is removed as an independent
basis function.

\begin{figure*}[!htb]
    \centering
    \includegraphics[width=\textwidth]
    {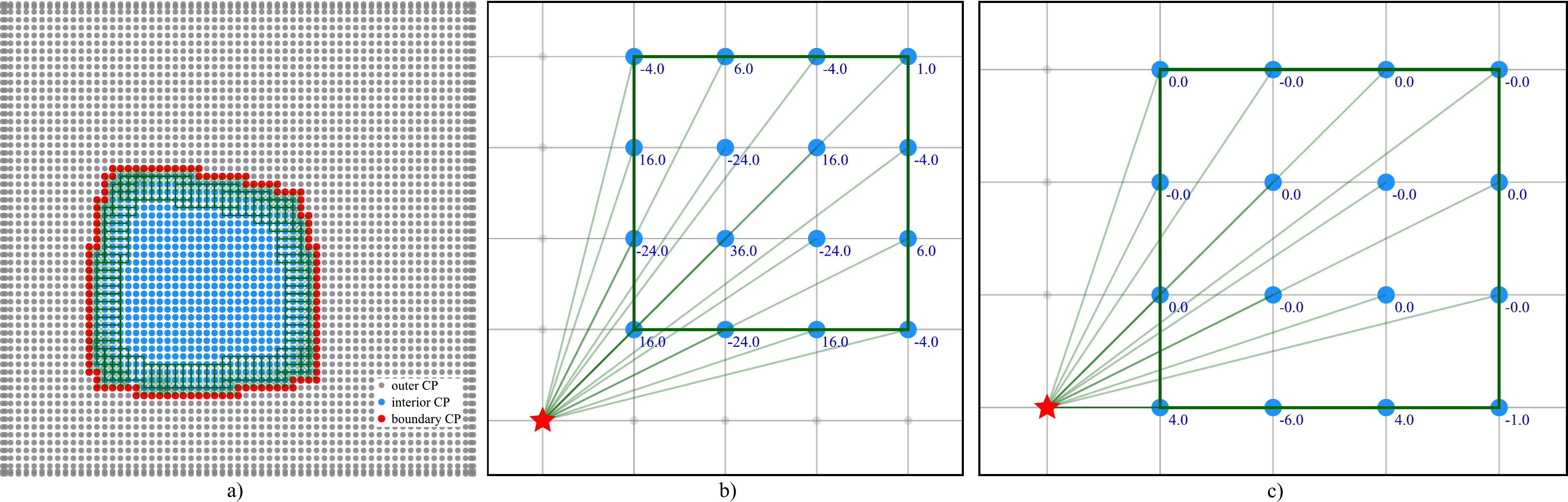}
    \caption{Two-dimensional cubic extended B-spline construction.
    \textbf{(a)} Classification of displacement control points for an immersed geometry.
    Inner, boundary, and outer control points are shown in blue, red, and gray,
    respectively; each boundary point is assigned a neighboring \(4\times4\) block of inner
    control points.
    \textbf{(b)} Tensor-product extension coefficients for a boundary control point located
    diagonally relative to its associated inner block. Both coordinate directions contribute
    nontrivial cubic extrapolation vectors, producing nonzero coefficients throughout the
    \(4\times4\) block.
    \textbf{(c)} Coefficient pattern for an aligned boundary configuration. One coordinate
    direction contributes an interpolation selector, so the nonzero extrapolation
    coefficients are restricted to a single row or column of the inner block.}
    \label{fig:Extended_B-spline_2D}
\end{figure*}

In multiple dimensions, both the B-spline basis and the extension coefficients are obtained
by tensor products. For a boundary index
\(\boldsymbol{j}=(j_1,\ldots,j_d)\), the coefficient associated with
\(\boldsymbol{i}=(i_1,\ldots,i_d)\) is
\begin{equation}
    E_{\boldsymbol{i}\boldsymbol{j}}
    =
    \prod_{\alpha=1}^{d}
    E_{i_\alpha j_\alpha}^{(\alpha)}.
    \label{eq:tensor_product_extension_coefficient}
\end{equation}
For example, in two dimensions, if the one-dimensional extrapolation vectors in the two
coordinate directions are denoted by \(\Be^{(1)}\) and \(\Be^{(2)}\), the coefficient array is
\begin{equation}
    \BE_{\boldsymbol{j}}
    =
    \Be^{(1)}
    \left(\Be^{(2)}\right)^T.
    \label{eq:two_dimensional_extension_matrix}
\end{equation}
The resulting coefficient pattern depends on the position of the boundary control point
relative to the selected \(4\times4\) cubic inner block. A diagonal configuration produces
nonzero coefficients throughout the block, whereas a boundary point aligned with one row or
column of the block produces nonzero coefficients only along that aligned direction. These
representative configurations are shown in
\cref{fig:Extended_B-spline_2D}.

In the examples employing continuous pressure approximations, the same extended B-spline construction is applied to the pressure space. For discontinuous pressure approximations, which are the main focus of the paper, the cellwise aggregation procedure introduced in the following subsection is used instead.
\subsubsection{Aggregation of discontinuous pressure degrees of freedom}
\label{subsubsec:pressure_aggregation_theory}

\noindent The pressure approximation considered in this work may be discontinuous across
pressure-cell interfaces. Consequently, a boundary pressure cell possesses its own
element-local coefficient vector, and the continuous extended B-spline construction cannot
be applied directly. Instead, boundary pressure cells are aggregated with neighboring inner
pressure cells by transferring the donor-cell polynomial to the boundary cell.

The pressure grid is first divided into inner, boundary, and outer cells. An inner pressure
cell lies entirely within the physical domain and retains an independent local coefficient
vector. A boundary pressure cell intersects both the physical and fictitious domains and is
treated as dependent. \revx{Every cut pressure cell is therefore dependent; no cut cell retains independent pressure unknowns.} Each boundary pressure cell is assigned to the nearest inner pressure
cell based on the distance between their cell centers. Outer pressure cells contain no
physical domain and are removed.

Let \(X\) denote a boundary pressure cell and \(A\) its assigned inner donor cell. The local
pressure approximations on these two cells are written as
\begin{equation}
    p_h|_X
    =
    \BN_X^{p}\Bp_X,
    \qquad
    p_h|_A
    =
    \BN_A^{p}\Bp_A,
    \label{eq:local_pressure_polynomials}
\end{equation}
where \(\BN_X^{p}\) and \(\BN_A^{p}\) denote the vectors of element-local pressure basis
functions associated with the boundary cell \(X\) and donor cell \(A\), respectively, and
\(\Bp_X\) and \(\Bp_A\) are the corresponding local pressure coefficient vectors.

The donor-cell polynomial is continued onto the boundary pressure cell \(X\). Here,
\(\BN_A^{p}\Bp_A\) evaluated on \(X\) denotes the polynomial continuation of the
pressure polynomial represented on \(A\), rather than the restriction of globally supported
donor basis functions. The dependent boundary-cell coefficients \(\Bp_X\) are then obtained
from the \(L^2\)-projection problem
\begin{equation}
    \Bp_X
    =
    \arg\min_{\Bq_X}
    \int_{X}
    \left|
    \BN_X^{p}\Bq_X
    -
    \BN_A^{p}\Bp_A
    \right|^2
    \dV.
    \label{eq:l2_projection_problem_p}
\end{equation}
Stationarity with respect to \(\Bq_X\) gives the local normal equations
\begin{equation}
    \BM_{XX}\Bp_X
    =
    \BM_{XA}\Bp_A,
    \label{eq:pressure_projection_normal_equations}
\end{equation}
where
\begin{align}
    \BM_{XX}
    &=
    \int_{X}
    \left(\BN_X^{p}\right)^T
    \BN_X^{p}
    \dV,
    \label{eq:pressure_boundary_mass_matrix}
    \\
    \BM_{XA}
    &=
    \int_{X}
    \left(\BN_X^{p}\right)^T
    \BN_A^{p}
    \dV.
    \label{eq:pressure_cross_projection_matrix}
\end{align}
Provided that \(\BM_{XX}\) is nonsingular on the local pressure space, the boundary-cell
coefficient vector is expressed in terms of the donor-cell coefficients as
\begin{equation}
    \Bp_X
    =
    \BT_{XA}\Bp_A,
    \qquad
    \BT_{XA}
    =
    \BM_{XX}^{-1}\BM_{XA}.
    \label{eq:pressure_aggregation_matrix}
\end{equation}
Thus, the entries of \(\BT_{XA}\) are not prescribed extrapolation coefficients. They are
computed from the \(L^2\) projection of the continued donor polynomial over the complete
boundary pressure cell \(X\). Using the complete cell rather than only its physical portion
avoids forming a local mass matrix over an arbitrarily small cut region, while preserving
the polynomial transfer between the donor and boundary pressure spaces.

Note that for tensor-product Bernstein polynomial pressure bases on rectangular pressure
cells, the Gramian matrix \(\BM_{XX}\) and its inverse admit explicit analytical expressions.
Consequently, the local projection can be evaluated without numerically inverting the
Bernstein mass matrix; see, e.g., the B\'ezier projection framework of
Thomas et al.~\cite{THOMAS201555}.

\begin{figure*}[!htb]
    \centering
    \includegraphics[width=\textwidth]
    {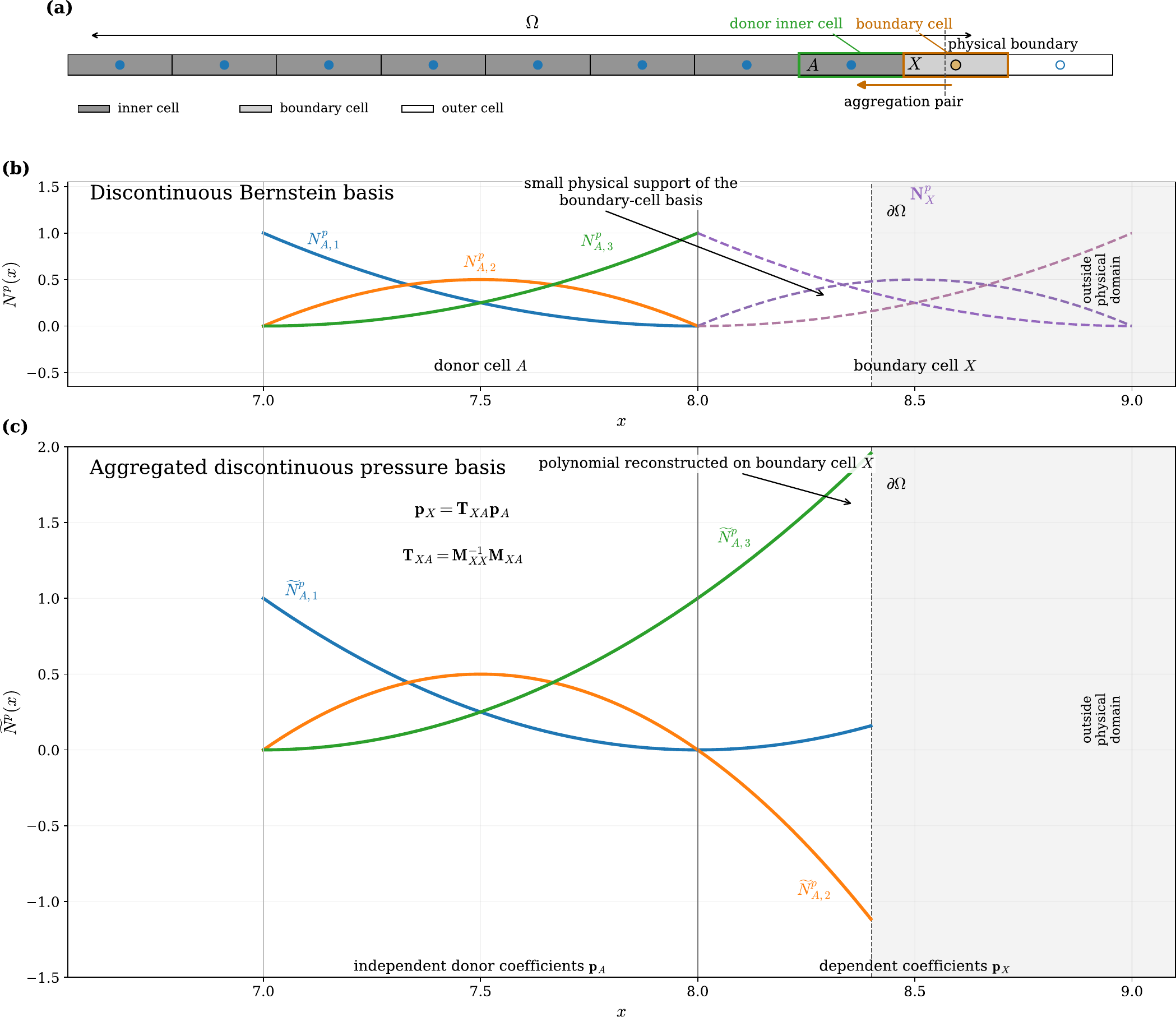}
    \caption{One-dimensional aggregation of discontinuous pressure degrees of freedom.
    \textbf{(a)} A boundary pressure cell \(X\) is assigned to the neighboring inner donor
    cell \(A\).
    \textbf{(b)} Original element-local quadratic Bernstein pressure basis functions on
    \(A\) and \(X\). The boundary-cell basis vector \(\BN_X^{p}\) has only a portion
    of its support inside the physical domain.
    \textbf{(c)} Aggregated pressure basis obtained from the \(L^2\)-projection relation
    \(\Bp_X=\BT_{XA}\Bp_A\), with
    \(\BT_{XA}=\BM_{XX}^{-1}\BM_{XA}\). The donor coefficients remain independent, whereas
    the boundary-cell coefficients are reconstructed as dependent coefficients.}
    \label{fig:Discontinuous_Pressure_SHape_Functions}
\end{figure*}

The one-dimensional construction is illustrated in
\cref{fig:Discontinuous_Pressure_SHape_Functions}. In the representative example, quadratic
Bernstein pressure basis functions are used on the donor cell \(A\) and the boundary cell
\(X\). The original basis functions \(\BN_X^{p}\) have only a portion of their support inside
the physical domain because the physical boundary intersects \(X\). After applying
\cref{eq:pressure_aggregation_matrix}, the polynomial represented by
\(\BN_A^{p}\Bp_A\) is reconstructed on \(X\), and therefore also on its physical portion,
while the local coefficient vector \(\Bp_X\) becomes dependent on the independent donor
coefficients \(\Bp_A\). The resulting
aggregated pressure basis therefore preserves the local polynomial approximation while
removing the boundary-cell coefficients as independent unknowns.

\begin{figure*}[!htb]
    \centering
    \includegraphics[width=\textwidth]
    {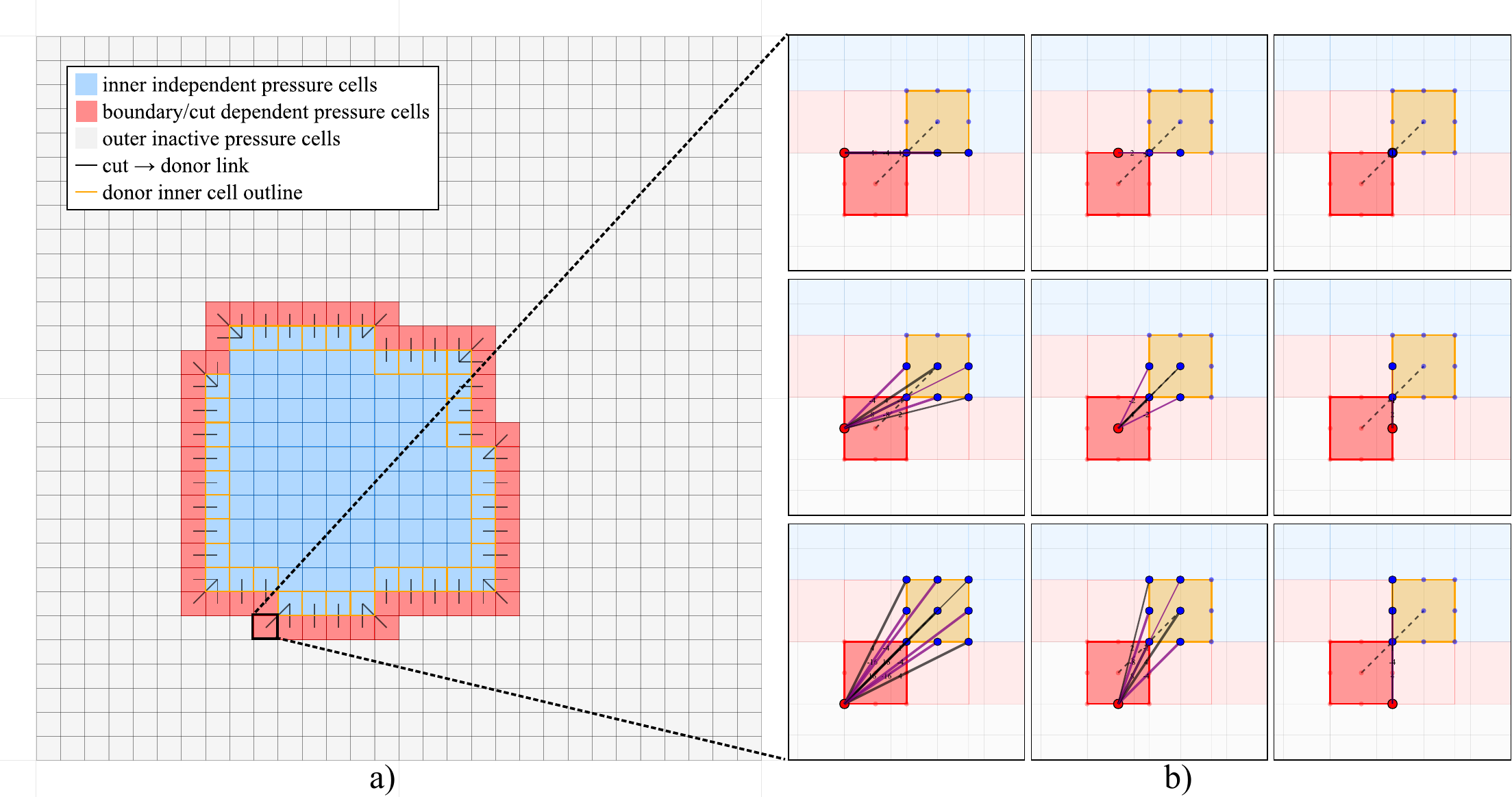}
    \caption{Selection of donor cells for discontinuous-pressure aggregation.
    \textbf{(a)} Classification of the pressure grid into independent inner cells, dependent
    boundary cells, and inactive outer cells. Each boundary cell is assigned to the nearest
    inner donor cell based on the distance between their cell centers.
    \textbf{(b)} Representative donor configurations for boundary cells located near
    straight edges, corners, and diagonal cut regions. The red cell denotes the dependent
    boundary cell, the yellow-outlined cell denotes the selected inner donor cell, and the
    connecting lines illustrate the dependence of the boundary-cell pressure coefficients on
    the donor-cell coefficients through the local \(L^2\)-projection matrix
    \(\BT_{XA}\).}
    \label{fig:Donor}
\end{figure*}

In two dimensions, each boundary pressure cell is assigned to the nearest inner donor cell
based on the distance between their cell centers. Because the pressure approximation is
discontinuous, the aggregation is performed independently for each boundary cell and does
not impose continuity across neighboring pressure-cell interfaces. The donor-cell
assignments used for representative edge, corner, and diagonal cut configurations are shown
in \cref{fig:Donor}.

For each aggregation pair \((X,A)\), the vectors \(\BN_X^{p}\) and \(\BN_A^{p}\) contain
the tensor-product element-local pressure basis functions on the boundary and donor cells,
respectively. The local transfer matrix \(\BT_{XA}\) is again computed from
\cref{eq:pressure_aggregation_matrix}, with all projection integrals evaluated over the
complete boundary pressure cell \(X\). The connecting lines shown in the local configurations
illustrate the dependence of the boundary-cell pressure coefficients on the selected
donor-cell coefficients; they do not represent direct pointwise interpolation.

Equivalently, at the algebraic level, the displacement extension and pressure extension or
aggregation can be represented collectively by
\begin{equation}
    \begin{bmatrix}
        \Bd\\
        \Bp
    \end{bmatrix}
    =
    \underbrace{
    \begin{bmatrix}
        \BE_u & \Bzero\\
        \Bzero & \BE_p
    \end{bmatrix}}_{\BE}
    \begin{bmatrix}
        \Bd_I\\
        \Bp_I
    \end{bmatrix},
    \label{eq:immersed_extension_matrix}
\end{equation}
where \(\BE_u\) denotes the displacement extended B-spline transformation,
while \(\BE_p\) denotes either the corresponding extended B-spline
transformation for continuous pressure approximations or the cellwise
aggregation transformation for discontinuous pressure approximations.
Accordingly, the complete mixed system can be represented as
\begin{equation}
    \BK^{\mathrm{imm}}
    =
    \BE^T\BK\BE,
    \qquad
    \BR^{\mathrm{imm}}
    =
    \BE^T\BR.
    \label{eq:immersed_system_transformation}
\end{equation}
The global transformation \(\BE\) need not be formed explicitly. In the implementation,
the corresponding local extension and aggregation maps are applied directly to the basis
functions during assembly and are subsequently used to reconstruct the dependent
coefficients.
\section{Kirsch problem}
\label{subsec:kirsch_problem}


\noindent
The Kirsch problem is considered first to assess the accuracy, stability, and
convergence properties of the mixed formulations introduced in the preceding
sections. The problem consists of an infinite elastic plate containing a
circular hole and subjected to a uniform far-field tensile stress. It is a
particularly useful benchmark for nearly incompressible elasticity because a
smooth analytical solution is available, while the stress concentration around
the hole produces strongly nonuniform stress and pressure fields. As the
incompressible limit is approached, a displacement-only formulation is
susceptible to volumetric locking, whereas an inf-sup-unstable mixed
formulation may produce spurious oscillations in the pressure and stress
fields. The problem therefore provides a controlled setting for examining the
accuracy and inf-sup stability of the mixed discretizations and the
effectiveness of the residual-based stabilization methods.

\begin{figure*}[!htb]
    \centering
    \includegraphics[width=\textwidth]
    {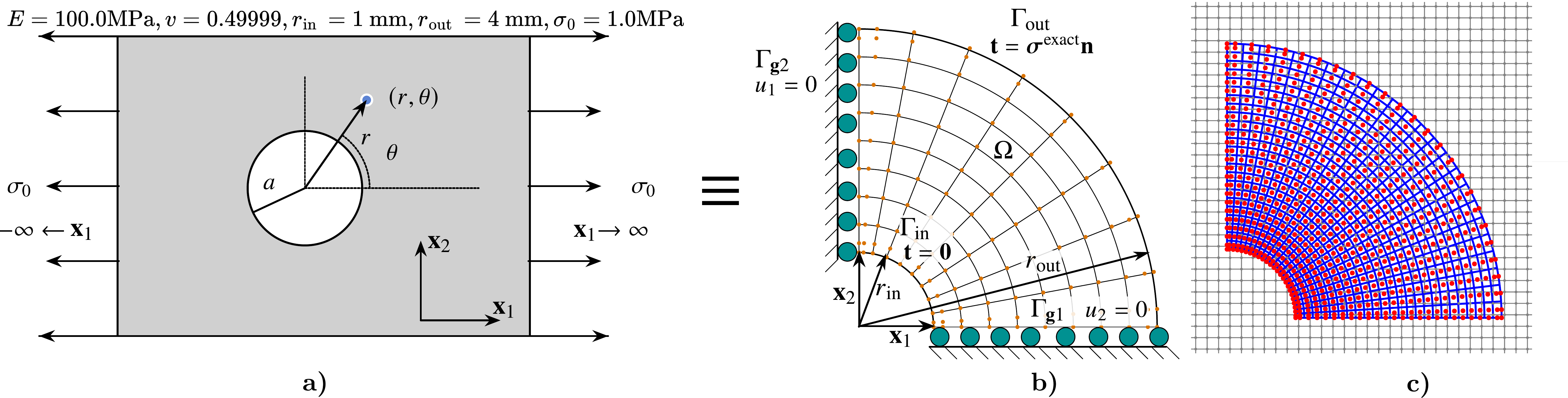}
    \caption{
    Schematic representation and discretization of the Kirsch problem.
    \textbf{(a)} Infinite plate containing a circular hole of radius
    \(a=r_{\mathrm{in}}=1~\mathrm{mm}\) subjected to a uniform far-field
    tensile stress \(\sigma_0=1~\mathrm{MPa}\), with
    \(E=100~\mathrm{MPa}\) and \(\nu=0.49999\).
    \textbf{(b)} Owing to symmetry, the problem is solved on a quarter-annular
    domain with \(r_{\mathrm{out}}=4~\mathrm{mm}\). The symmetry conditions
    \(u_1=0\) and \(u_2=0\) are imposed on
    \(\Gamma_{g2}\) and \(\Gamma_{g1}\), respectively. The inner boundary
    \(\Gamma_{\mathrm{in}}\) is traction free, while the exact Kirsch traction
    \(\Bt=\Bsigma^{\mathrm{exact}}\Bn\) is prescribed on
    \(\Gamma_{\mathrm{out}}\).
    \textbf{(c)} Representative immersed discretization. The quarter-annular
    physical domain is embedded in an independent
    \(10~\mathrm{mm}\times10~\mathrm{mm}\) Cartesian background grid, with
    the center of the corresponding circle located at
    \((x_1,x_2)=(3.12321321,\,3.12321321)~\mathrm{mm}\).
    The physical discretization and its control points are nonconforming with
    respect to the background grid.
    }
    \label{fig:KirschScheme}
\end{figure*}

The geometry and boundary conditions are illustrated in
\cref{fig:KirschScheme}. An infinite plate containing a circular hole of radius
\(a=r_{\mathrm{in}}=1~\mathrm{mm}\) is subjected to a uniform tensile stress
\(\sigma_0=1~\mathrm{MPa}\) in the \(x_1\)-direction. The material parameters
are
$E=100~\mathrm{MPa}, \
\nu=0.49999$,
so that the material response is very close to the incompressible limit. The
calculations are performed under plane-strain conditions, for which the Lamé
parameters are
\begin{equation}
    \mu
    =
    \frac{E}{2(1+\nu)},
    \qquad
    \lambda
    =
    \frac{E\nu}{(1+\nu)(1-2\nu)}.
    \label{eq:kirsch_lame_parameters}
\end{equation}

Owing to symmetry, only one quarter of the plate is analyzed. The infinite
domain is truncated at
\(r_{\mathrm{out}}=4~\mathrm{mm}\), giving the quarter-annular computational
domain
\begin{equation}
    \Omega
    =
    \left\{
        (r,\theta):
        r_{\mathrm{in}}\leq r\leq r_{\mathrm{out}},
        \quad
        0\leq\theta\leq\frac{\pi}{2}
    \right\}.
    \label{eq:kirsch_domain}
\end{equation}
Here, \(r\) and \(\theta\) are polar coordinates centered at the hole, with
\(\theta=0\) measured from the \(x_1\)-axis. The symmetry conditions are
\begin{equation}
    u_2=0
    \quad\text{on}\quad
    \Gamma_{g1},
    \qquad
    u_1=0
    \quad\text{on}\quad
    \Gamma_{g2}.
    \label{eq:kirsch_symmetry_bc}
\end{equation}
The inner circular boundary \(\Gamma_{\mathrm{in}}\) is traction free, whereas
the exact Kirsch traction is prescribed on the artificial outer boundary
\(\Gamma_{\mathrm{out}}\):
\begin{equation}
    \Bt
    =
    \Bsigma^{\mathrm{exact}}\Bn
    \qquad
    \text{on}\quad
    \Gamma_{\mathrm{out}}.
    \label{eq:kirsch_exact_traction}
\end{equation}
Prescribing the exact traction on the truncated outer boundary makes the
restriction of the infinite-domain Kirsch solution the exact solution of the
finite quarter-annular problem.

For an infinite plate subjected to uniaxial tension in the \(x_1\)-direction,
the analytical Kirsch in-plane stress components in polar coordinates are
\begin{align}
    \sigma_{rr}^{\mathrm{exact}}
    &=
    \frac{\sigma_0}{2}
    \left(
        1-\frac{a^2}{r^2}
    \right)
    +
    \frac{\sigma_0}{2}
    \left(
        1
        -4\frac{a^2}{r^2}
        +3\frac{a^4}{r^4}
    \right)
    \cos(2\theta),
    \label{eq:kirsch_sigma_rr}
    \\
    \sigma_{\theta\theta}^{\mathrm{exact}}
    &=
    \frac{\sigma_0}{2}
    \left(
        1+\frac{a^2}{r^2}
    \right)
    -
    \frac{\sigma_0}{2}
    \left(
        1+3\frac{a^4}{r^4}
    \right)
    \cos(2\theta),
    \label{eq:kirsch_sigma_tt}
    \\
    \sigma_{r\theta}^{\mathrm{exact}}
    &=
    -
    \frac{\sigma_0}{2}
    \left(
        1
        +2\frac{a^2}{r^2}
        -3\frac{a^4}{r^4}
    \right)
    \sin(2\theta).
    \label{eq:kirsch_sigma_rt}
\end{align}
These in-plane stress components are independent of whether plane-stress or
plane-strain conditions are assumed. The corresponding displacement field,
however, depends on this choice. For plane strain, introduce the Kolosov
constant
\begin{equation}
    \kappa_{\mathrm{K}}
    =
    3-4\nu.
    \label{eq:kirsch_kolosov_constant}
\end{equation}
The exact radial and circumferential displacement components are
\begin{align}
    u_r^{\mathrm{exact}}
    &=
    \frac{\sigma_0}{8\mu}
    \left[
        (\kappa_{\mathrm{K}}-1)r
        +2\frac{a^2}{r}
        +
        \left(
            2r
            +2(\kappa_{\mathrm{K}}+1)\frac{a^2}{r}
            -2\frac{a^4}{r^3}
        \right)
        \cos(2\theta)
    \right],
    \label{eq:kirsch_u_r}
    \\
    u_\theta^{\mathrm{exact}}
    &=
    -
    \frac{\sigma_0}{8\mu}
    \left[
        2r
        +2(\kappa_{\mathrm{K}}-1)\frac{a^2}{r}
        +2\frac{a^4}{r^3}
    \right]
    \sin(2\theta).
    \label{eq:kirsch_u_theta}
\end{align}
The corresponding Cartesian displacement components are obtained from
\begin{equation}
    u_1^{\mathrm{exact}}
    =
    u_r^{\mathrm{exact}}\cos\theta
    -
    u_\theta^{\mathrm{exact}}\sin\theta,
    \qquad
    u_2^{\mathrm{exact}}
    =
    u_r^{\mathrm{exact}}\sin\theta
    +
    u_\theta^{\mathrm{exact}}\cos\theta.
    \label{eq:kirsch_cartesian_displacement}
\end{equation}

The Cartesian in-plane stress components are obtained through the standard
tensor transformation:
\begin{align}
    \sigma_{11}^{\mathrm{exact}}
    &=
    \sigma_{rr}^{\mathrm{exact}}\cos^2\theta
    +
    \sigma_{\theta\theta}^{\mathrm{exact}}\sin^2\theta
    -
    2\sigma_{r\theta}^{\mathrm{exact}}
    \sin\theta\cos\theta,
    \\
    \sigma_{22}^{\mathrm{exact}}
    &=
    \sigma_{rr}^{\mathrm{exact}}\sin^2\theta
    +
    \sigma_{\theta\theta}^{\mathrm{exact}}\cos^2\theta
    +
    2\sigma_{r\theta}^{\mathrm{exact}}
    \sin\theta\cos\theta,
    \\
    \sigma_{12}^{\mathrm{exact}}
    &=
    \left(
        \sigma_{rr}^{\mathrm{exact}}
        -
        \sigma_{\theta\theta}^{\mathrm{exact}}
    \right)
    \sin\theta\cos\theta
    +
    \sigma_{r\theta}^{\mathrm{exact}}
    \left(
        \cos^2\theta-\sin^2\theta
    \right).
    \label{eq:kirsch_cartesian_stress}
\end{align}

Under plane-strain conditions, \(\epsilon_{33}=0\), and the out-of-plane
normal stress is
\begin{equation}
    \sigma_{33}^{\mathrm{exact}}
    =
    \nu
    \left(
        \sigma_{11}^{\mathrm{exact}}
        +
        \sigma_{22}^{\mathrm{exact}}
    \right)
    =
    \nu
    \left(
        \sigma_{rr}^{\mathrm{exact}}
        +
        \sigma_{\theta\theta}^{\mathrm{exact}}
    \right).
    \label{eq:kirsch_sigma_33}
\end{equation}
According to the mixed-pressure convention adopted in
\cref{eq:p_volumetric_relation}, namely
\(p=-\lambda\Bnabla\cdot\Bu\), the exact pressure is therefore
\begin{align}
    p^{\mathrm{exact}}
    &=
    -\lambda\Bnabla\cdot\Bu^{\mathrm{exact}}
    =
    -\sigma_{33}^{\mathrm{exact}} =
    -\nu\sigma_0
    \left[
        1
        -2\frac{a^2}{r^2}\cos(2\theta)
    \right].
    \label{eq:kirsch_exact_pressure}
\end{align}
The analytical displacement, pressure, and stress fields provide
direct references for evaluating the corresponding numerical errors and for
identifying spurious oscillations in the nearly incompressible limit.

Both mesh-conforming and immersed discretizations are considered. In the
mesh-conforming calculations, the quarter-annular geometry is represented
exactly by the analysis mesh and is uniformly refined using
\begin{equation}
    n_r
    \in
    \{2,4,8,16,32,64,128\},
    \qquad
    h
    =
    \frac{r_{\mathrm{out}}-r_{\mathrm{in}}}{n_r}
    =
    \frac{3}{n_r}~\mathrm{mm},
    \label{eq:kirsch_conforming_mesh_sequence}
\end{equation}
where \(n_r\) denotes the number of elements in the radial direction.

In the immersed calculations, the same physical quarter-annular geometry is
embedded in the independent
\(10~\mathrm{mm}\times10~\mathrm{mm}\) Cartesian background domain shown in
\cref{fig:KirschScheme}c. The center of the corresponding circle is placed at
$
    (x_1^c,x_2^c)
    =
(3.12321321, \allowbreak 3.12321321)~\mathrm{mm}$,
so that the physical boundaries remain non-aligned with the background grid.
Uniform Cartesian meshes with
\begin{equation}
    n_x=n_y
    \in
    \{40,60,80,120,160,240,320,480,640\},
    \qquad
    h
    =
    \frac{10}{n_x}~\mathrm{mm},
    \label{eq:kirsch_immersed_mesh_sequence}
\end{equation}
are used. For the immersed calculations, the physical portions of the cut
cells are integrated using the recursive-subdivision
\((n_{\textrm{cut}}=7)\) and terminal-triangulation procedure described in
\cref{subsubsec:subdivision_quadrature}. Essential boundary conditions are
imposed weakly using the Nitsche formulation, with
$
    \gamma_{N,K}
    =
    0.01\lambda_{N,K}.
$
Poorly supported degrees of freedom are treated using the displacement
extension and pressure extension or aggregation procedures introduced
previously.

Three classes of pressure approximation are considered in the numerical
study. First, continuous pressure spaces are examined to assess the classical
Galerkin stability properties and the effect of residual-based stabilization.
Second, discontinuous piecewise-constant pressure spaces are considered to
isolate the elementwise constant pressure mode and its associated inf-sup
requirement. Third, higher-order discontinuous pressure spaces are studied to assess the stabilized local-pressure formulation and the effects of using
same-mesh and coarsened pressure grids. These three classes are considered
consistently for both the mesh-conforming and immersed discretizations. Finally, the elementwise inverse constants are examined on representative
fixed meshes to quantify the dependence of the local stabilization scaling on
the polynomial degree and cut-cell geometry.

\subsection{Continuous Pressure}

\noindent We first consider continuous pressure approximations. This set of numerical experiments has two objectives. The first is to verify the inf-sup behavior predicted in \cref{subsec:inf_sup_galerkin_IGA} for the Galerkin pairs \unstablepair{$ \mathcal{U}(k,\mathcal{C}^{k-1},h)/\allowbreak\mathcal{P}(k-1,\mathcal{C}^{k-2},h)$}, and \stablepair{$\allowbreak \mathcal{U}(k,\mathcal{C}^{k-1},h)/\allowbreak\mathcal{P}(k-1,\mathcal{C}^{k-2},2h)$}. As discussed previously, maximum continuity of the displacement space makes the same-mesh pair fail the \revx{sufficient} isogeometric inf-sup \revx{criterion}, whereas coarsening the pressure grid by a factor of two satisfies the corresponding stability condition while preserving the maximum continuity of the displacement field. The second objective is to examine whether residual-based stabilization can recover stable and accurate solutions for the equal-order family \unstablepair{$ \mathcal{U}(k,\mathcal{C}^{k-1},h)/\allowbreak\mathcal{P}(k,\mathcal{C}^{k-1},h)$}, which does not satisfy the Galerkin inf-sup condition. Both questions are examined first in the mesh-conforming setting and subsequently for the immersed formulation.

\subsubsection{Mesh-Conforming}

\begin{figure*}[!htb]
    \centering
    \includegraphics[width=\textwidth]
    {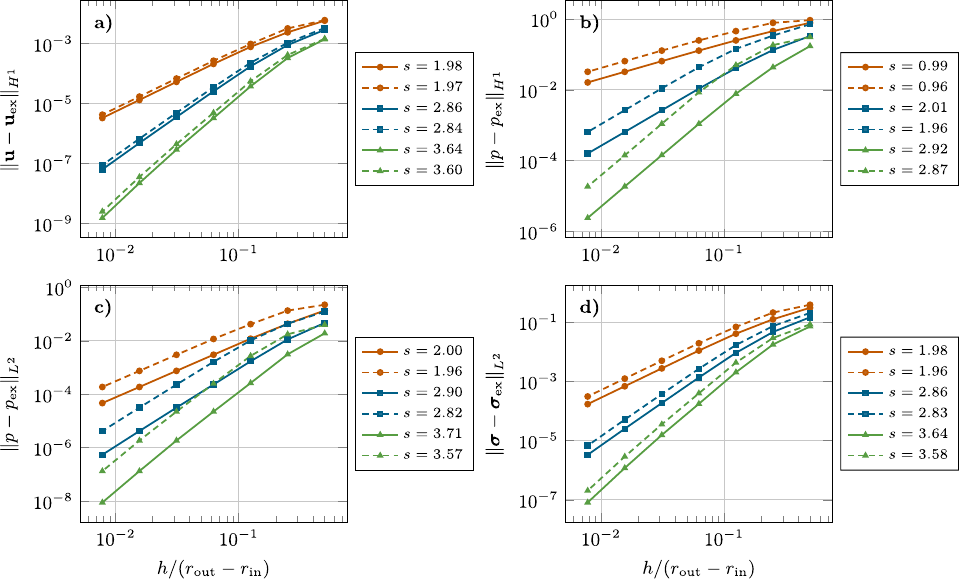}
    \caption{Mesh-conforming Galerkin convergence results.
    The \ktwo, \kthree, and \kfour{} curves correspond to the indicated
    polynomial orders. Solid curves use
    \unstablepair{$\mathcal{U}(k,\mathcal{C}^{k-1},h)/\mathcal{P}(k-1,\mathcal{C}^{k-2},h)$}, whereas dashed curves use
    $\stablepair{\mathcal{U}(k,\mathcal{C}^{k-1},h)/\mathcal{P}(k-1,\mathcal{C}^{k-2},2h)}$. Although the $\unstablepair{\mathcal{U}(k,\mathcal{C}^{k-1},h)/\mathcal{P}(k-1,\mathcal{C}^{k-2},h)}$ case \revx{is not certified by the sufficient inf-sup criterion}, both families exhibit optimal rates of convergence.}
    \label{fig:Gal_Cont_Conforming}
\end{figure*}

\noindent The mesh-conforming Galerkin convergence results for the pressure degree-reduced Taylor--Hood-type families \unstablepair{\( \allowbreak \mathcal{U}(k,\mathcal{C}^{k-1},h)/ \mathcal{P}(k-1,\mathcal{C}^{k-2},h)\) } and \stablepair{\(\mathcal{U}(k,\mathcal{C}^{k-1},h)/\allowbreak \mathcal{P}(k-1,\mathcal{C}^{k-2},2h)\)} are shown in \cref{fig:Gal_Cont_Conforming}. The four panels report the displacement \(H^1\)-error, pressure \(H^1\)-error, pressure \(L^2\)-error, and stress \(L^2\)-error, respectively. As discussed in the stability analysis, the equal-mesh pair \unstablepair{\(\mathcal{U}(k,\mathcal{C}^{k-1},h)/\allowbreak\mathcal{P}(k-1,\mathcal{C}^{k-2},h)\)} \revx{fails the sufficient isogeometric inf-sup criterion}. In contrast, coarsening the pressure grid by a factor of two gives \stablepair{\(\mathcal{U}(k,\mathcal{C}^{k-1},h)/\allowbreak\mathcal{P}(k-1,\mathcal{C}^{k-2},2h)\)}, which satisfies the corresponding inf-sup condition while preserving maximum continuity of the displacement space.

Despite this difference in theoretical stability, both families exhibit the expected asymptotic convergence rates for all error measures considered in \cref{fig:Gal_Cont_Conforming}. This illustrates an important distinction between convergence observed for a particular benchmark and uniform mixed stability. The Kirsch problem has a smooth analytical solution over the computational domain, and the instability of the same-mesh pair is therefore not necessarily apparent from the convergence rates alone. Consequently, the observed optimal convergence of \unstablepair{\(\mathcal{U}(k,\mathcal{C}^{k-1},h)/\allowbreak\mathcal{P}(k-1,\mathcal{C}^{k-2},h)\)} in this example should not be interpreted as evidence of uniform inf-sup stability.

\begin{figure*}[!htb]
    \centering
    \includegraphics[width=\textwidth]
    {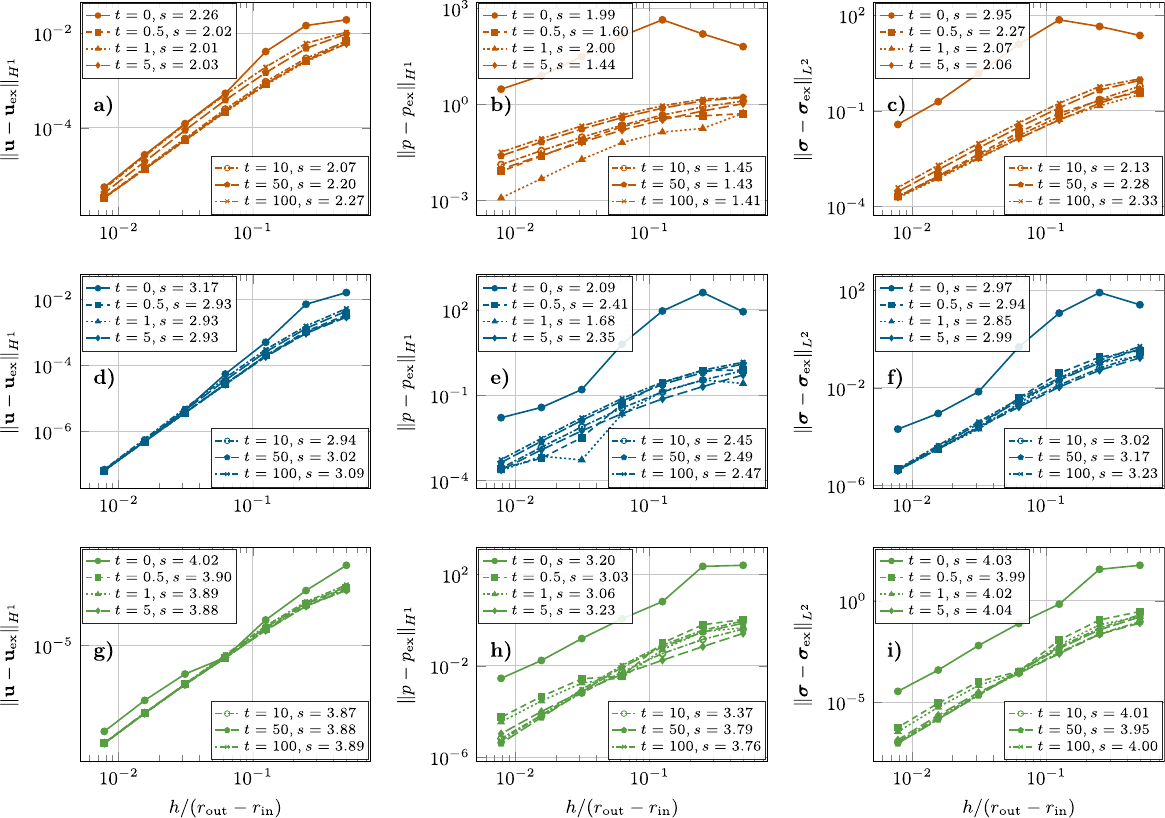}
    \caption{Mesh-conforming \stablepair{VMS} convergence results for the
    stabilization-parameter study. The \ktwo, \kthree, and \kfour{}
    rows correspond to the indicated polynomial orders. All cases use
    the equal-order \stablepair{VMS-stabilized pair}
    \stablepair{$\mathcal{U}(k,\mathcal{C}^{k-1},h)/\mathcal{P}(k,\mathcal{C}^{k-1},h)$}. The \unstablepair{Galerkin case \(t=0\)} exhibits large errors on coarse meshes due to the
lack of inf-sup stability, whereas the \stablepair{stabilized cases
\(t=0.5,1,5,10,50,100\)} exhibit substantially smaller errors and recover the
expected asymptotic convergence rates.}
    \label{fig:VMS_Cont_Conforming}
\end{figure*}

Next, to assess the effect of residual-based stabilization more directly, we consider the equal-order approximation \unstablepair{\(\mathcal{U}(k,\mathcal{C}^{k-1},h)/\allowbreak\mathcal{P}(k,\mathcal{C}^{k-1},h)\)}. This provides a more demanding test of the stabilization than the pressure degree-reduced family \unstablepair{\(\mathcal{U}(k,\mathcal{C}^{k-1},h)/\allowbreak\mathcal{P}(k-1,\mathcal{C}^{k-2},h)\)}, for which even the \revx{uncertified} same-mesh pair exhibits optimal convergence for the present smooth benchmark. The corresponding mesh-conforming results are shown in \cref{fig:VMS_Cont_Conforming}. The case \(t=0\) corresponds to the unstabilized Galerkin formulation, whereas \(t=\tau_e\Lambda_e\in\{0.5,1,5,10,50,100\}\) denotes the VMS-stabilized cases. Here, \(\Lambda_e\) is the local inverse constant introduced in \cref{subsec:tau_parameter}, such that \(t\) defines a nondimensional stabilization level. 
The effect of the inf-sup instability is considerably more apparent for the equal-order Galerkin discretization, particularly in the pressure- and stress-based error measures on the coarser meshes and for the lower polynomial orders. Once VMS stabilization is introduced, these errors are substantially reduced and the refinement curves recover regular asymptotic behavior with the expected convergence rates. Moreover, the stabilized solutions obtained over a broad range of positive \(t\) values remain relatively close to one another. This behavior is consistent with the VMS stability estimate derived previously: the VMS formulation does not require an upper stability bound on \(t\), although the value of \(t\) continues to affect the accuracy of the discrete solution. The sensitivity to \(t\) is also more pronounced for some lower-order approximations. For example, the pressure \(H^1\)-error for \(k=2\) exhibits a stronger dependence on the stabilization parameter than the corresponding \(k=3\) and \(k=4\) cases.

\begin{figure*}[!htb]
    \centering
    \includegraphics[width=\textwidth]
    {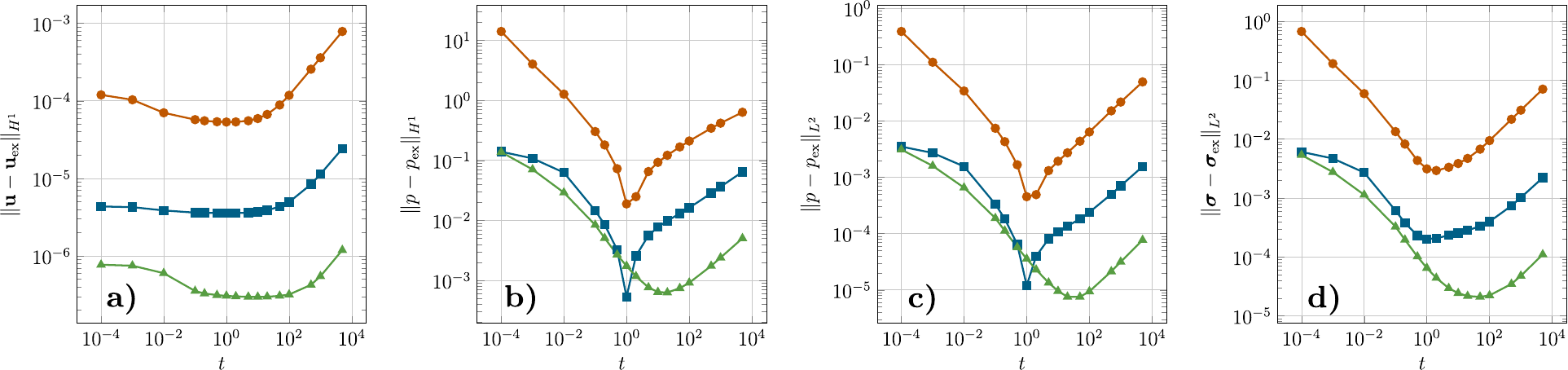}
    \caption{Effect of {stabilization} parameter \(t=\tau_e\Lambda_e\) on a fixed \(32\times32\)
    displacement mesh for
    \stablepair{VMS-stablized pair \(\mathcal{U}(k,\mathcal{C}^{k-1},h)/\mathcal{P}(k,\mathcal{C}^{k-1},h)\)}.
    From left to right, the panels show the displacement \(H^1\)-error,
    pressure \(H^1\)-error, pressure \(L^2\)-error, and stress \(L^2\)-error.
    Colors denote
    \(\textcolor{kTwoColor}{k=2}\),
    \(\textcolor{kThreeColor}{k=3}\), and
    \(\textcolor{kFourColor}{k=4}\).}
    \label{fig:conforming_vms_Continous}
\end{figure*}

The dependence of the numerical error on the stabilization magnitude is examined more directly in \cref{fig:conforming_vms_Continous}, where the displacement mesh is fixed at \(32\times32\) elements and \(t=\tau_e\Lambda_e\) is varied over several orders of magnitude. The four panels show the displacement \(H^1\)-error, pressure \(H^1\)-error, pressure \(L^2\)-error, and stress \(L^2\)-error. For all three polynomial orders, very small values of \(t\) provide insufficient stabilization, and the solution retains the influence of the underlying inf-sup instability. Increasing \(t\) initially reduces the pressure and stress errors substantially, and an intermediate stabilization range produces the smallest errors. For sufficiently large \(t\), however, the errors increase again as the formulation becomes over-stabilized.
The numerical character of the under- and over-stabilized regimes is different. For small \(t\), insufficient stabilization permits spurious pressure oscillations associated with the unstable pressure modes. For excessively large \(t\), the pressure field remains smooth but becomes increasingly inaccurate and over-diffusive as the stabilization contribution becomes too strong. Thus, the absence of an upper stability restriction on \(t\) for the VMS formulation should not be interpreted as parameter-independent accuracy. Rather, \(t\) should be selected within an intermediate range that sufficiently controls the unstable pressure modes without introducing excessive numerical diffusion. 

\begin{figure*}[!htbp]
    \centering
    \includegraphics[width=\linewidth]{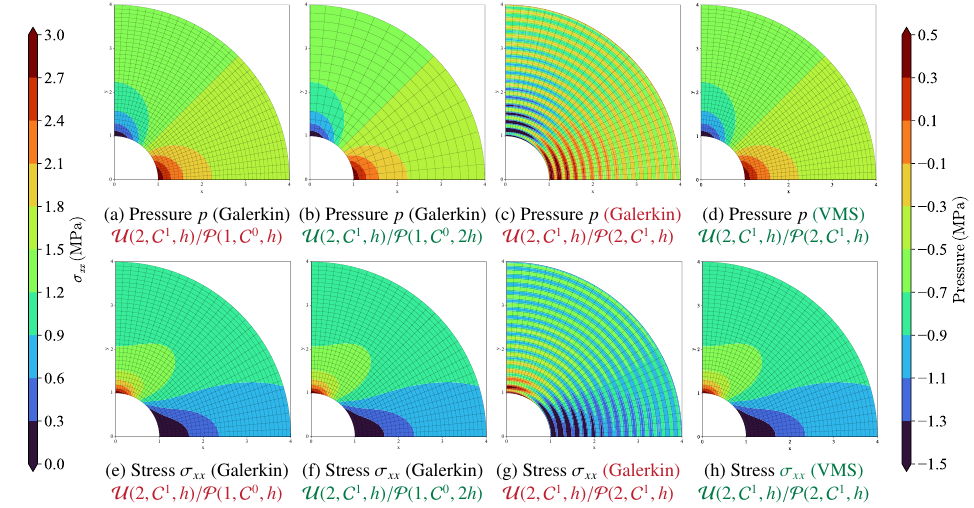}
    \caption{Pressure $p$ and normal stress $\sigma_{xx}$ obtained using
    different displacement--pressure discretizations. Each column presents
    the pressure and stress fields corresponding to the same discretization.
    The shared stress colorbar is shown on the left, and the shared pressure
    colorbar is shown on the right.}
    \label{fig:pressure_stress_comparison}

\end{figure*}

Representative pressure and stress fields are shown in \cref{fig:pressure_stress_comparison} for \(k=2, \ n_r=32\). The first two columns show the Galerkin approximations \unstablepair{\(\mathcal{U}(2,\mathcal{C}^1,h)/\allowbreak\mathcal{P}(1,\mathcal{C}^0,h)\)} and \stablepair{\(\mathcal{U}(2,\mathcal{C}^1,h)/\allowbreak\mathcal{P}(1,\mathcal{C}^0,2h)\)}, respectively, while the third and fourth columns show the unstable \unstablepair{Galerkin} and \stablepair{VMS-stabilized} equal-order approximation \unstablepair{\(\mathcal{U}(2,\mathcal{C}^1,h)/\allowbreak\mathcal{P}(2,\mathcal{C}^1,h)\)}. The pressure field is shown in the upper row and the corresponding Cauchy normal stress \(\sigma_{xx}\) in the lower row. Both \unstablepair{\(\mathcal{U}(2,\mathcal{C}^1,h)/\allowbreak\mathcal{P}(1,\mathcal{C}^0,h)\)}, and \stablepair{\(\mathcal{U}(2,\mathcal{C}^1,h)/\allowbreak\mathcal{P}(1,\mathcal{C}^0,2h)\)} Galerkin spaces produce smooth pressure and stress distributions for this benchmark, despite the different theoretical inf-sup properties of the \(h\) and \(2h\) pressure grids. This observation is consistent with the convergence results in \cref{fig:Gal_Cont_Conforming}, where the smooth Kirsch solution does not strongly expose the instability of the same-mesh Taylor--Hood pair. In contrast, the equal-order \unstablepair{\(\mathcal{U}(2,\mathcal{C}^1,h)/\allowbreak\mathcal{P}(2,\mathcal{C}^1,h)\)} Galerkin solution reveals the mixed instability much more clearly through pronounced oscillations in the pressure and stress fields. Introducing VMS stabilization suppresses these spurious modes and recovers smooth fields comparable to those obtained using the pressure degree-reduced discretizations. The field comparisons therefore complement the convergence studies and demonstrate directly the role of residual-based stabilization in controlling the unstable continuous pressure modes.

\subsubsection{Immersed discretization}

\noindent The same continuous-pressure families are next examined using the immersed formulation. Consequently, the comparison with the mesh-conforming results provides a direct assessment of whether the additional immersed treatments alter the stability and convergence behavior established above.

\begin{figure*}[!htb]
    \centering
    \includegraphics[width=\textwidth]
    {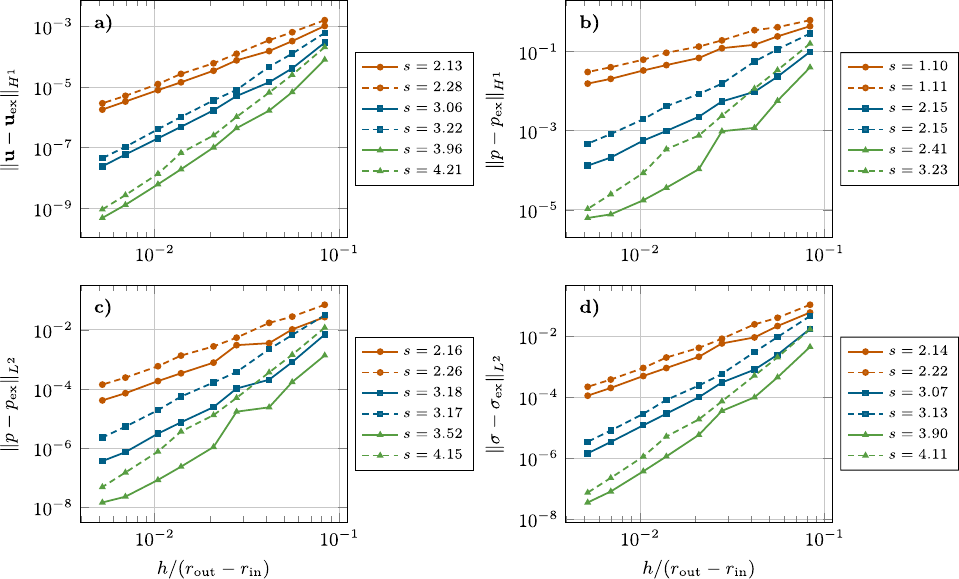}
    \caption{Immersed Galerkin convergence results.
    The \ktwo, \kthree, and \kfour{} curves correspond to the indicated
    polynomial orders. Solid curves use
    \unstablepair{$\mathcal{U}(k,\mathcal{C}^{k-1},h)/\mathcal{P}(k-1,\mathcal{C}^{k-2},h)$}, whereas dashed curves use
    \stablepair{$\mathcal{U}(k,\mathcal{C}^{k-1},h)/\mathcal{P}(k-1,\mathcal{C}^{k-2},2h)$}. Although the \unstablepair{$\mathcal{U}(k,\mathcal{C}^{k-1},h)/\mathcal{P}(k-1,\mathcal{C}^{k-2},h)$} case \revx{is not certified by the sufficient inf-sup criterion}, both families exhibit optimal rates of convergence.}
    \label{fig:Gal_Cont_Immersed}
\end{figure*}

The immersed Galerkin convergence results for
\unstablepair{\(\mathcal{U}(k,\mathcal{C}^{k-1},h)/\allowbreak \mathcal{P}(k-1,\mathcal{C}^{k-2},h)\)} and
\stablepair{\(\mathcal{U}(k,\mathcal{C}^{k-1},h)/\allowbreak \mathcal{P}(k-1,\mathcal{C}^{k-2},2h)\)} are shown in
\cref{fig:Gal_Cont_Immersed}. As in the mesh-conforming case, the solid
curves correspond to the same-mesh pressure approximation, whereas the
dashed curves correspond to the \(2h\) pressure grid. The displacement,
pressure, and stress errors decrease asymptotically as the background mesh
is refined, and both families retain their expected optimal convergence
rates. In particular, the immersed geometry, cut-cell integration, and
extended B-spline treatment do not introduce a loss of asymptotic accuracy.
As observed for the mesh-conforming discretization, the same-mesh family
\unstablepair{\(\mathcal{U}(k,\mathcal{C}^{k-1},h)/\allowbreak \mathcal{P}(k-1,\mathcal{C}^{k-2},h)\) } exhibits optimal
convergence despite \revx{failing the sufficient isogeometric inf-sup criterion} for
the maximally continuous displacement space. It also generally produces
smaller errors than
\stablepair{\(\mathcal{U}(k,\mathcal{C}^{k-1},h)/\allowbreak \mathcal{P}(k-1,\mathcal{C}^{k-2},2h)\)} for this benchmark.
\revx{This difference is expected from the two-mesh error estimate of
\cref{eq:method_dependent_error_estimate}: with \(h_p=nh_u\), the
pressure-approximation terms scale as \(n^{2(k_p+1)}\) and \(n^{2k_p}\),
respectively, relative to the same-mesh case. Thus, pressure-grid coarsening
increases the error constant while preserving the asymptotic convergence rate
for fixed \(n\).} The smaller error of the same-mesh discretization,
however, should not be interpreted as evidence of inf-sup stability. The
\(2h\) family retains the theoretical stability established previously,
whereas the same-mesh family does not.

Compared with the nearly monotonic convergence curves obtained for the
mesh-conforming discretization, the immersed results exhibit more noticeable
non-monotone variations about the asymptotic convergence trends. These
variations become more pronounced for the higher-order cases \(k=3\) and
\(k=4\). In an immersed discretization, successive mesh refinements do not
produce geometrically nested cut configurations: the physical boundary
intersects the background cells at different relative locations, resulting
in different cut-cell volume fractions and different physical supports of
the near-boundary basis functions. Consequently, the extended B-spline
construction and the associated cut-cell integration vary with refinement,
which can produce local fluctuations in the measured errors. The effect is
more noticeable for higher-order B-splines because of their wider supports
and the correspondingly larger number of basis functions influenced by the
cut region. Nevertheless, these variations remain superimposed on the
expected asymptotic behavior, and the overall convergence rates remain
optimal for all cases considered.

\begin{figure*}[!htb]
    \centering
    \includegraphics[width=\textwidth]
    {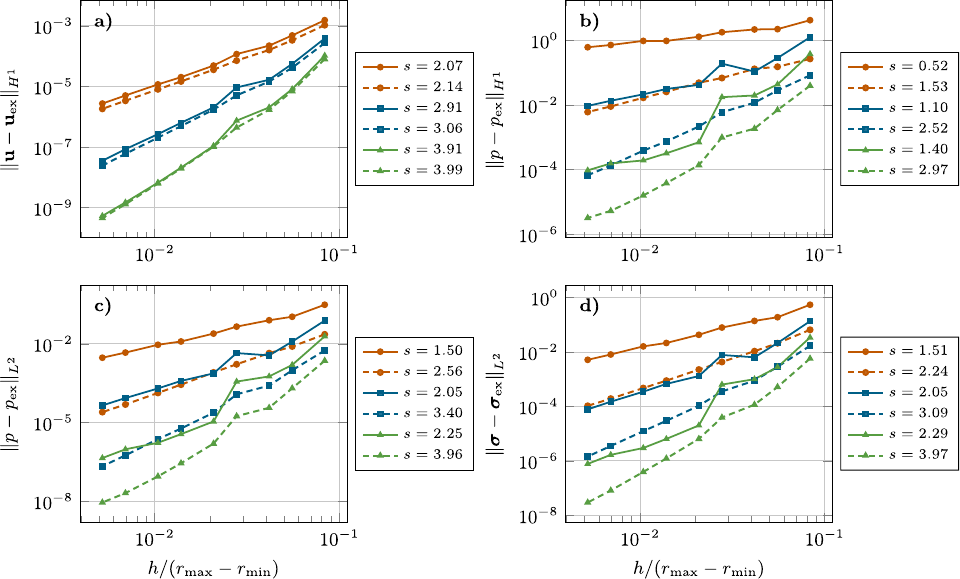}
    \caption{Immersed equal-order \unstablepair{Galerkin}--\stablepair{VMS} comparison.
    The \ktwo, \kthree, and \kfour{} curves correspond to the indicated
    polynomial orders. Solid curves denote Galerkin results, whereas
    dashed curves denote \stablepair{VMS} results; both use
    \unstablepair{$\mathcal{U}(k,\mathcal{C}^{k-1},h)/\mathcal{P}(k,\mathcal{C}^{k-1},h)$}. The \stablepair{VMS} curves perform optimally in terms of convergence rate, whereas the \unstablepair{Galerkin} results exhibit large errors, especially for lower-order coarse meshes.}
    \label{fig:VMS_Cont_Immersed}
\end{figure*}

Next, we consider a comparison between \stablepair{VMS-stabilized} and \unstablepair{Galerkin} discretization of the equal order pair \unstablepair{$ \mathcal{U}(k,\mathcal{C}^{k-1},h)/\allowbreak\mathcal{P}(k,\mathcal{C}^{k-1},h)$}.
The corresponding convergence curves are shown in \cref{fig:VMS_Cont_Immersed}. The unstable Galerkin approximation again exhibits significantly larger errors, particularly for the lower polynomial orders and coarser background meshes. The VMS formulation uses $t=5$ and markedly reduces these errors and restores the regular convergence behavior observed in the mesh-conforming calculations. Thus, the residual-based stabilization remains effective when the physical boundary cuts through the analysis grid, indicating that the volume stabilization, cut-cell quadrature, displacement extension, and Nitsche boundary treatment operate consistently within the immersed formulation.

\begin{figure*}[!htbp]
    \centering
    \includegraphics[width=\linewidth]{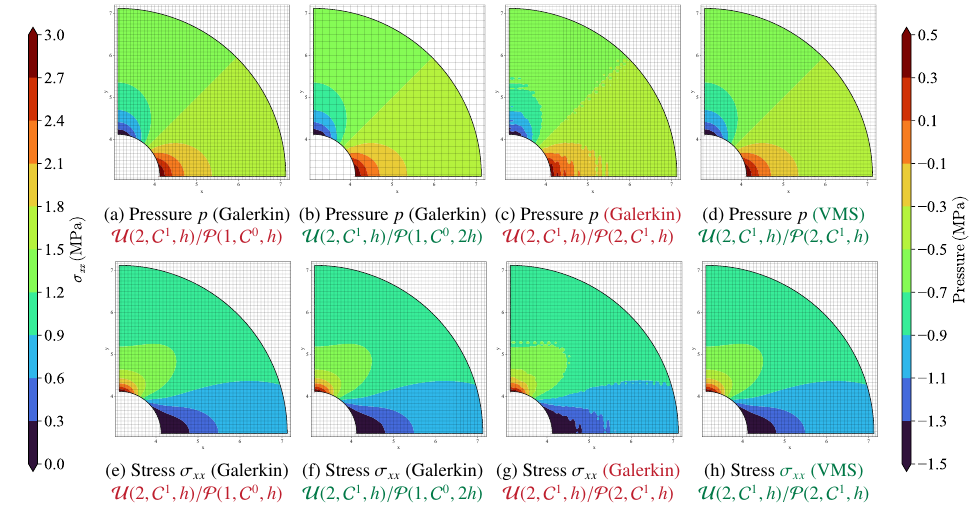}
    \caption{Pressure $p$ and normal stress $\sigma_{xx}$ obtained using
    different displacement--pressure discretizations in the immersed
    formulation. Each column presents the pressure and stress fields
    corresponding to the same discretization. The shared stress colorbar
    is shown on the left, and the shared pressure colorbar is shown on the
    right.}
    \label{fig:kirsch_immersed_pressure_stress_comparison}

\end{figure*}

Finally, the corresponding immersed pressure and stress fields are shown in
\cref{fig:kirsch_immersed_pressure_stress_comparison}. The same four
discretizations considered in the mesh-conforming comparison are examined
here using a Cartesian background grid with \(n_x=n_y=120\). Despite the nonconforming intersection of the quarter-annular physical domain
with the background grid, the
\stablepair{\(\mathcal{U}(k,\mathcal{C}^{k-1},h)/\allowbreak\mathcal{P}(k-1,\mathcal{C}^{k-2},2h)\)}
Galerkin approximation produces smooth pressure and stress fields. As in the mesh-conforming case,
the equal-order Galerkin approximation \unstablepair{$\mathcal{U}(k,\mathcal{C}^{k-1},h)/\allowbreak\mathcal{P}(k,\mathcal{C}^{k-1},h)$} exhibits pronounced pressure and stress
oscillations associated with the lack of inf-sup stability. The introduction
of VMS stabilization effectively suppresses these spurious modes and recovers
smooth pressure and stress distributions. The close qualitative agreement
between the stabilized immersed solution and its mesh-conforming counterpart
demonstrates that the combination of cut-cell integration and the immersed
boundary treatment preserves the stabilizing effect of the residual-based
VMS formulation.

\subsection{Constant Pressure}
\label{subsubsec:kirsch_constant_pressure}

\noindent For discontinuous pressure spaces, the residual-based stabilization
controls the nonconstant elementwise pressure modes, whereas stability of the
elementwise constant pressure mode must be provided by the underlying mixed
discretization. It is therefore important to examine the constant-pressure
case separately before considering higher-order discontinuous pressure
approximations. To isolate this contribution, we consider the Galerkin
families \unstablepair{\(\allowbreak \mathcal{U}(k,\mathcal{C}^{k-1},h)/\allowbreak \mathcal{P}(0,\mathcal{C}^{-1},h)\)} and
\stablepair{\(\allowbreak \mathcal{U}(k,\mathcal{C}^{k-1},h)/\allowbreak\mathcal{P}(0,\mathcal{C}^{-1},2h)\)}, with \(k=2,3,4\), for both
mesh-conforming and immersed discretizations.

\subsubsection{Mesh-Conforming}

\begin{figure*}[!htb]
    \centering
    \includegraphics[width=\textwidth]
    {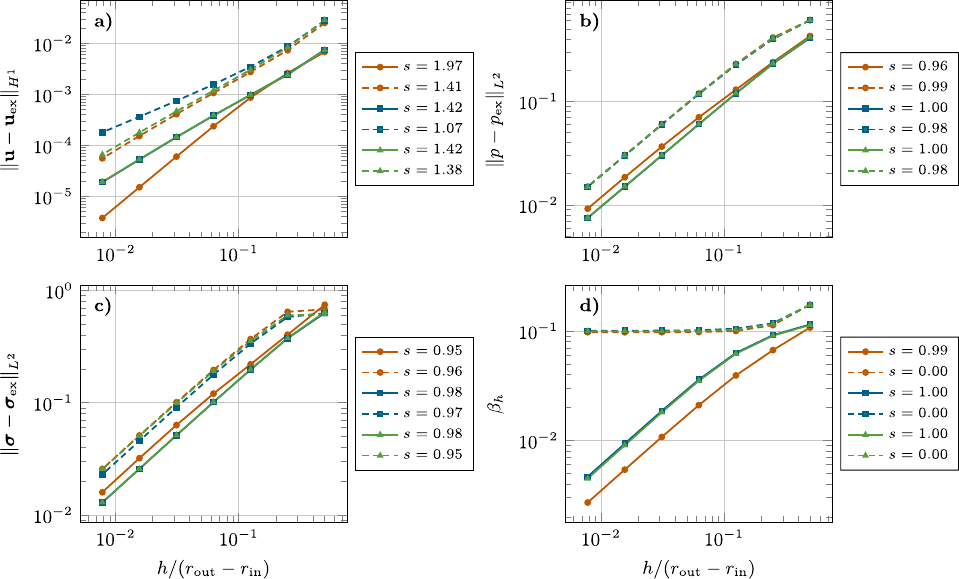}

    \caption{Mesh-conforming Galerkin convergence results for the
    discontinuous constant-pressure discretization. The \ktwo,
    \kthree, and \kfour{} curves correspond to the indicated
    displacement orders. Solid curves use
    \unstablepair{$\mathcal{U}(k,\mathcal{C}^{k-1},h)/\mathcal{P}(0,\mathcal{C}^{-1},h)$}, whereas dashed curves use
    \stablepair{$\mathcal{U}(k,\mathcal{C}^{k-1},h)/\mathcal{P}(0,\mathcal{C}^{-1},2h)$}.}

    \label{fig:DGconstP_conforming_convergence}
\end{figure*}

\noindent The mesh-conforming Galerkin convergence results are shown in
\cref{fig:DGconstP_conforming_convergence}. The four panels report the
displacement \(H^1\)-error, pressure \(L^2\)-error, stress \(L^2\)-error,
and discrete inf-sup constant \(\beta_h\), respectively. Solid curves
correspond to \unstablepair{\(\mathcal{U}(k,\mathcal{C}^{k-1},h)/\allowbreak\mathcal{P}(0,\mathcal{C}^{-1},h)\)}, whereas dashed curves
correspond to \stablepair{\(\mathcal{U}(k,\mathcal{C}^{k-1},h)/\allowbreak\mathcal{P}(0,\mathcal{C}^{-1},2h)\)}.
Since the pressure approximation remains piecewise constant for all
displacement orders, increasing \(k\) does not lead to the higher-order
convergence observed for the continuous pressure families. In particular,
the pressure and stress errors converge at approximately first order for
all \(k=2,3,4\). The displacement \(H^1\)-error is also limited by the
low-order pressure approximation and, apart from the same-mesh \(k=2\)
case, exhibits rates substantially below those expected from the
higher-order displacement spaces. Thus, enriching the displacement space
alone does not improve the overall convergence order when it is coupled
to a piecewise-constant pressure approximation.

The discrete inf--sup results shown in panel~(d) of
\cref{fig:DGconstP_conforming_convergence}, computed using the generalized
eigenvalue procedure described~\ref{app:numerical_inf_sup_test}, reveal a
clear distinction between the two pressure-grid resolutions. For the
same-mesh family
\unstablepair{\(\mathcal{U}(k,\mathcal{C}^{k-1},h)/\allowbreak\mathcal{P}(0,\mathcal{C}^{-1},h)\)}, \(\beta_h\) decreases
systematically under mesh refinement for all \(k\), indicating that the
discrete inf--sup constant is not bounded away from zero as
\(h\rightarrow0\). The same-mesh constant-pressure family therefore does not
satisfy a uniform inf--sup condition. In contrast, for
\stablepair{\(\mathcal{U}(k,\mathcal{C}^{k-1},h)/\allowbreak\mathcal{P}(0,\mathcal{C}^{-1},2h)\)}, \(\beta_h\) remains
approximately constant under refinement for all displacement orders.
Coarsening the pressure grid reduces the number of independent
elementwise-constant pressure modes relative to the displacement space,
allowing the displacement divergence field to control the pressure space
uniformly. The \(2h\) pressure-grid family therefore satisfies the discrete
inf--sup requirement for the elementwise constant pressure mode.

\begin{figure*}[!htbp]
    \centering
    \includegraphics[width=\linewidth]{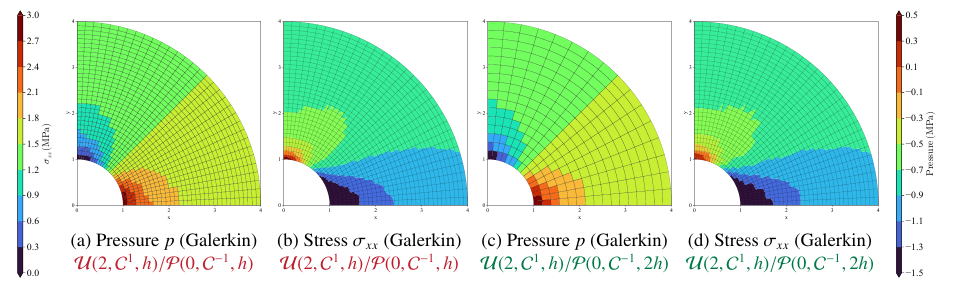}
    \caption{Pressure $p$ and normal stress $\sigma_{xx}$ obtained using
    mesh-conforming quadratic-displacement and discontinuous
    constant-pressure discretizations. The first two panels use
    \unstablepair{$\mathcal{U}(2,\mathcal{C}^1,h)/\mathcal{P}(0,\mathcal{C}^{-1},h)$}, while the last two panels use \stablepair{$\mathcal{U}(2,\mathcal{C}^1,h)/\mathcal{P}(0,\mathcal{C}^{-1},2h)$}. The shared stress colorbar is shown
    on the left, and the shared pressure colorbar is shown on the right.}
    \label{fig:kirsch_conforming_Q2Q0_pressure_stress}

\end{figure*}

Representative pressure and normal-stress fields on the
\(32\times32\) displacement mesh are shown in
\cref{fig:kirsch_conforming_Q2Q0_pressure_stress} for the quadratic
displacement space. The first two panels correspond to
\unstablepair{\(\mathcal{U}(2,\mathcal{C}^1,h)/\allowbreak\mathcal{P}(0,\mathcal{C}^{-1},h)\)}, whereas the last two use
\stablepair{\(\mathcal{U}(2,\mathcal{C}^1,h)/\allowbreak\mathcal{P}(0,\mathcal{C}^{-1},2h)\)}. Although the discrete inf-sup results
demonstrate that the same-mesh pair is not uniformly inf-sup stable,
pronounced spurious oscillations are not observed in its pressure field
for this smooth benchmark. This illustrates that failure of the discrete
inf-sup condition does not necessarily produce visually apparent pressure
oscillations for a particular problem and mesh resolution.

The stress fields exhibit a different behavior. Since the Cauchy stress
contains contributions from the displacement-derived strain field and the
discontinuous piecewise-constant pressure field, discontinuities in the
pressure approximation are inherited by the computed stress. Consequently,
elementwise variations are visible in the stress contours for both the
\(h\) and \(2h\) pressure grids. These features should therefore be
distinguished from the spurious pressure oscillations associated with
inf-sup instability; they arise naturally from combining the smooth
displacement approximation with a discontinuous constant-pressure
approximation.

\subsubsection{Immersed discretization}

\begin{figure*}[!htb]
    \centering
    \includegraphics[width=\textwidth]
    {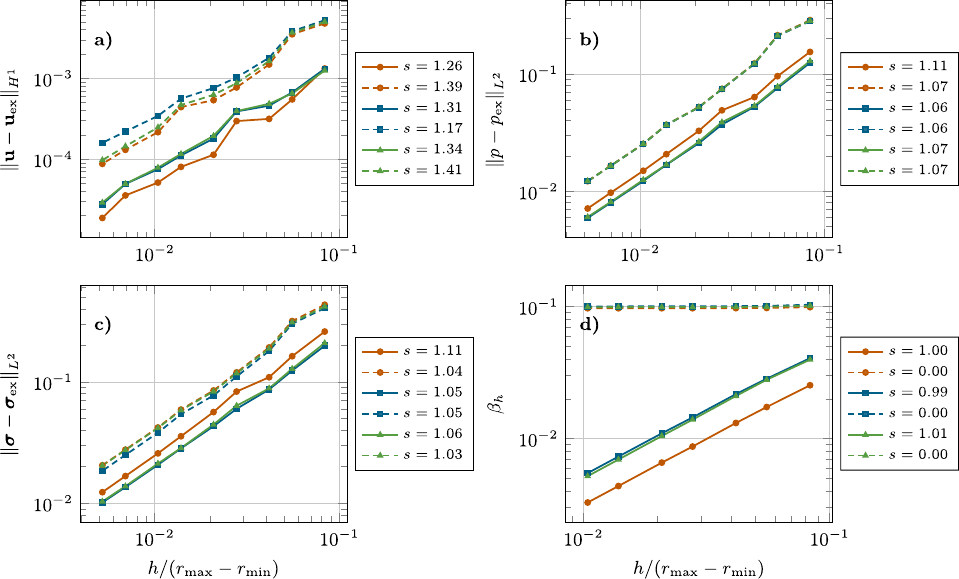}

    \caption{Immersed Galerkin convergence results for the
    discontinuous constant-pressure discretization. The \ktwo,
    \kthree, and \kfour{} curves correspond to the indicated
    displacement orders. Solid curves use
    \unstablepair{$\mathcal{U}(k,\mathcal{C}^{k-1},h)/\mathcal{P}(0,\mathcal{C}^{-1},h)$}, whereas dashed curves use
    \stablepair{$\mathcal{U}(k,\mathcal{C}^{k-1},h)/\mathcal{P}(0,\mathcal{C}^{-1},2h)$}.}

    \label{fig:DGconstP_immersed_convergence}
\end{figure*}

\noindent
The corresponding immersed Galerkin convergence results are shown in
\cref{fig:DGconstP_immersed_convergence}. The same two families,
\unstablepair{\(\mathcal{U}(k,\mathcal{C}^{k-1},h)\allowbreak /\allowbreak\mathcal{P}(0,\mathcal{C}^{-1},h)\)} and
\stablepair{\(\mathcal{U}(k,\mathcal{C}^{k-1},h)/\allowbreak\mathcal{P}(0,\mathcal{C}^{-1},2h)\)}, are considered on the nonconforming
Cartesian background grids. As in the mesh-conforming case, the use of a
piecewise-constant pressure space limits the asymptotic accuracy of the
coupled approximation. The pressure and stress errors converge at
approximately first order for all \(k=2,3,4\), and increasing the
displacement polynomial degree does not recover the higher-order rates that
would otherwise be expected from the displacement space alone. The
displacement \(H^1\)-error is also affected by the low-order pressure
approximation, although the immersed refinement curves exhibit additional
nonmonotone variations associated with changes in the cut-cell configuration
under refinement.

The discrete inf--sup results shown in panel~(d) of
\cref{fig:DGconstP_immersed_convergence}, computed using the immersed
generalization of the eigenvalue procedure described in~\ref{app:numerical_inf_sup_test}, exhibit the same qualitative
behavior as the mesh-conforming results. For the same-mesh family
\unstablepair{\(\mathcal{U}(k,\mathcal{C}^{k-1},h)/\allowbreak\mathcal{P}(0,\mathcal{C}^{-1},h)\)}, \(\beta_h\) decreases under
mesh refinement, indicating that the discrete inf--sup constant is not
bounded away from zero. In contrast, for
\stablepair{\(\mathcal{U}(k,\mathcal{C}^{k-1},h)/\allowbreak\mathcal{P}(0,\mathcal{C}^{-1},2h)\)}, \(\beta_h\) remains
approximately constant for all displacement orders. Thus, coarsening the
pressure grid by a factor of two again provides a uniformly stable pairing
for the elementwise constant pressure mode. These results additionally
demonstrate that the displacement-extension, pressure-cell aggregation, and
cut-cell integration procedures preserve the constant-mode inf--sup
stability of the underlying \(2h\) displacement--pressure pair.

\begin{figure*}[!htbp]
    \centering
    \includegraphics[width=\linewidth]{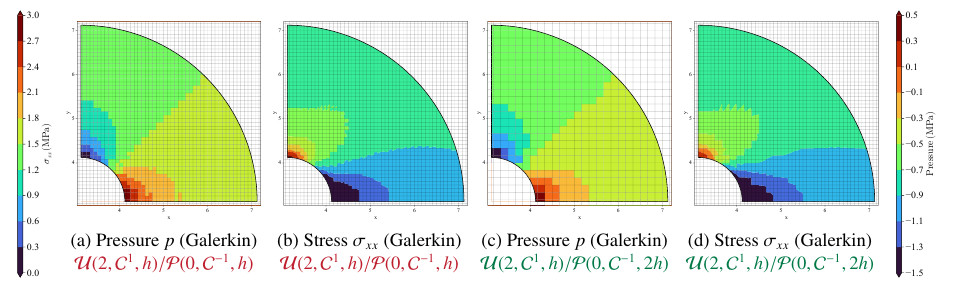}
    \caption{Pressure $p$ and normal stress $\sigma_{xx}$ obtained using
    immersed quadratic-displacement and discontinuous constant-pressure
    discretizations. The first two panels use
    \unstablepair{$\mathcal{U}(2,\mathcal{C}^1,h)/\mathcal{P}(0,\mathcal{C}^{-1},h)$}, while the last two use
    \stablepair{$\mathcal{U}(2,\mathcal{C}^1,h)/\mathcal{P}(0,\mathcal{C}^{-1},2h)$}. The shared stress colorbar is shown on
    the left, and the shared pressure colorbar is shown on the right.}
    \label{fig:kirsch_immersed_Q2Q0_pressure_stress}

\end{figure*}

Representative immersed pressure and stress fields are shown in
\cref{fig:kirsch_immersed_Q2Q0_pressure_stress} for the quadratic
displacement space with $n_x=n_y=120$. The first two panels correspond to
\unstablepair{\(\mathcal{U}(2,\mathcal{C}^1,h)/\allowbreak\mathcal{P}(0,\mathcal{C}^{-1},h)\)}, whereas the last two use
\stablepair{\(\mathcal{U}(2,\mathcal{C}^1,h)/\allowbreak \mathcal{P}(0,\mathcal{C}^{-1},2h)\)}. Unlike the mesh-conforming results, the
unstable same-mesh discretization exhibits mild spurious pressure
oscillations for the smooth Kirsch benchmark at the resolution considered.
The \(2h\) case, in contrast, produces a regular piecewise-constant pressure
field, consistent with its stable \(\beta_h\) behavior. In both cases, the
stress field exhibits elementwise variations because it combines
contributions from the smooth displacement approximation and the
discontinuous piecewise-constant pressure field.

\subsection{Higher-Order Discontinuous Pressure}
\label{subsubsec:kirsch_discontinuous_pressure}

\noindent
We finally consider the higher-order {VMS stabilized} discontinuous pressure families
\unstablepair{\(\mathcal{U}(k,\mathcal{C}^{k-1},h)/\allowbreak\mathcal{P}(k-1,\mathcal{C}^{-1},h)\)} and
\stablepair{\(\mathcal{U}(k,\mathcal{C}^{k-1},h)/\allowbreak \mathcal{P}(k-1,\mathcal{C}^{-1},2h)\)}, with \(k=2,3,4\).
As established in the stability analysis, the residual-based VMS term
controls the nonconstant elementwise pressure modes through the pressure
gradient, but it does not act on the elementwise constant component.
Consequently, stability of the constant pressure mode must still be supplied
by the underlying displacement--pressure pair.

The constant-pressure results of the preceding section are therefore directly
relevant here. For the maximally continuous displacement space,
\unstablepair{\(\mathcal{U}(k,\mathcal{C}^{k-1},h)/\allowbreak\mathcal{P}(k-1,\mathcal{C}^{-1},h)\)} does not satisfy the required
constant-mode inf-sup condition, whereas coarsening the pressure grid gives
\stablepair{\(\mathcal{U}(k,\mathcal{C}^{k-1},h)/\allowbreak\mathcal{P}(k-1,\mathcal{C}^{-1},2h)\)}, which satisfies this condition.
The following studies examine how this distinction appears in practice once
the remaining discontinuous pressure modes are controlled by VMS
stabilization.

\subsubsection{Mesh-Conforming}

\begin{figure*}[!htb]
    \centering
    \includegraphics[width=\textwidth]
    {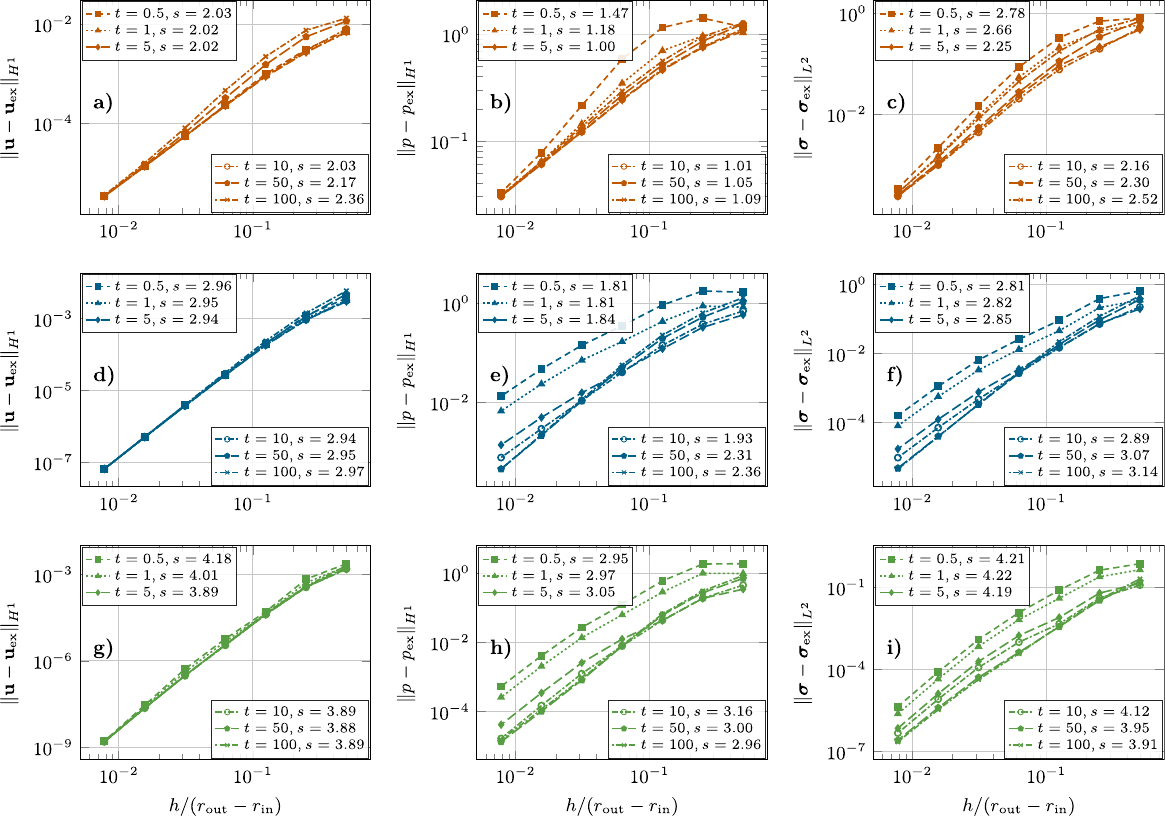}
    \caption{Mesh-conforming VMS convergence study for
    \unstablepair{\(\mathcal{U}(k,\mathcal{C}^{k-1},h)/\mathcal{P}(k-1,\mathcal{C}^{-1},h)\)}.
    Rows correspond to
    \(\textcolor{kTwoColor}{k=2}\),
    \(\textcolor{kThreeColor}{k=3}\), and
    \(\textcolor{kFourColor}{k=4}\);
    columns show the displacement \(H^1\)-error, pressure \(H^1\)-error, and
    stress \(L^2\)-error. The curves compare
    \(t=0.5,1,5,10,50,\) and \(100\), with the reported slopes computed from
    the asymptotic refinement range.}
    \label{fig:conforming_vms_h}
\end{figure*}

\noindent
We first consider the same-mesh family
\unstablepair{\(\mathcal{U}(k,\mathcal{C}^{k-1},h)/\allowbreak\mathcal{P}(k-1,\mathcal{C}^{-1},h)\)}. The corresponding VMS convergence
results are shown in \cref{fig:conforming_vms_h} for
\(t=\tau_e\Lambda_e\in\{0.5,1,5,10,50,100\}\). The rows correspond to
\(k=2,3,\) and \(4\), while the columns report the displacement
\(H^1\)-error, pressure \(H^1\)-error, and stress \(L^2\)-error.
Although this family does not satisfy the constant-mode inf-sup condition,
the displacement \(H^1\)-error, pressure \(H^1\)-error, and stress
\(L^2\)-error all exhibit regular asymptotic convergence over a broad range
of stabilization parameters.

The good practical behavior of the same-mesh family for this example does
not contradict the stability analysis. The Kirsch problem is smooth, and the
elementwise constant unstable mode is not strongly excited by this benchmark.
VMS stabilization simultaneously controls the nonconstant pressure modes,
allowing the discretization to exhibit accurate and nearly optimal behavior
over a wide range of \(t\), despite the absence of a uniform inf-sup bound
for the constant component.

\begin{figure*}[!htb]
    \centering
    \includegraphics[width=\textwidth]
    {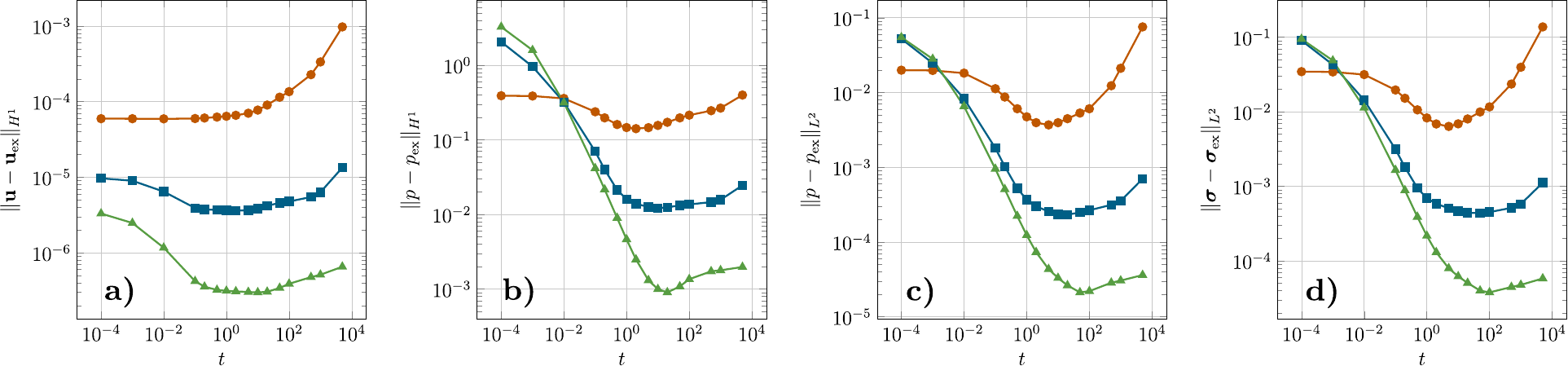}
    \caption{Effect of \(t=\tau_e\Lambda_e\) on a fixed \(32\times32\)
    displacement mesh for
    \unstablepair{\(\mathcal{U}(k,\mathcal{C}^{k-1},h)/\mathcal{P}(k-1,\mathcal{C}^{-1},h)\)}.
    From left to right, the panels show the displacement \(H^1\)-error,
    pressure \(H^1\)-error, pressure \(L^2\)-error, and stress \(L^2\)-error.
    Colors denote
    \(\textcolor{kTwoColor}{k=2}\),
    \(\textcolor{kThreeColor}{k=3}\), and
    \(\textcolor{kFourColor}{k=4}\).}
    \label{fig:tau_sweep_h}
\end{figure*}

The influence of the stabilization magnitude is examined more directly in
\cref{fig:tau_sweep_h} on a fixed \(32\times32\) displacement mesh. For
very small \(t\), the stabilization is insufficient and the solution remains
affected by unstable pressure modes. Increasing \(t\) substantially reduces
the pressure and stress errors, producing an intermediate range in which the
errors remain small. Compared with the continuous-pressure discretization,
the discontinuous formulation maintains this low-error behavior over a
considerably wider range of stabilization parameters. This reduced sensitivity to \(t\) is consistent with the element-local
character of the discontinuous pressure approximation, for which the
residual-based stabilization acts locally on the nonconstant pressure modes
without introducing the same interelement pressure coupling present in the
continuous pressure space.
For sufficiently large \(t\), however, the errors eventually increase due to
over-stabilization. The two limiting regimes have different numerical
characteristics: insufficient stabilization produces oscillatory pressure
fields, whereas excessive stabilization produces smooth but increasingly
inaccurate fields. The location of the minimum error also shifts toward
larger values of \(t\) as the polynomial order increases. Thus, although the
VMS formulation remains effective over a broad range of stabilization levels,
the value that minimizes the discretization error depends on the
approximation order.

\begin{figure*}[!htb]
    \centering
    \includegraphics[width=\textwidth]
    {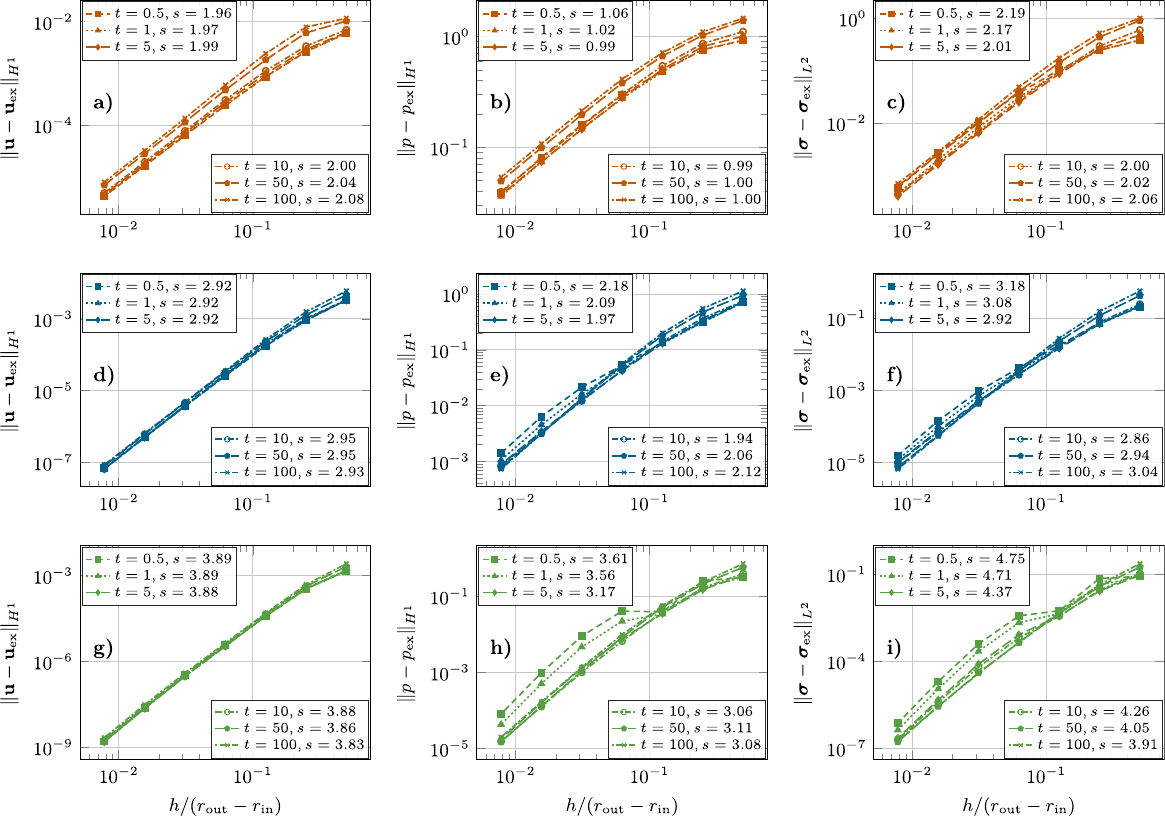}
    \caption{Mesh-conforming VMS convergence study for
   \stablepair{\(\mathcal{U}(k,\mathcal{C}^{k-1},h)/\mathcal{P}(k-1,\mathcal{C}^{-1},2h)\)}.
    Rows correspond to
    \(\textcolor{kTwoColor}{k=2}\),
    \(\textcolor{kThreeColor}{k=3}\), and
    \(\textcolor{kFourColor}{k=4}\);
    columns show the displacement \(H^1\)-error, pressure \(H^1\)-error, and
    stress \(L^2\)-error. The curves compare
    \(t=0.5,1,5,10,50,\) and \(100\), with the reported slopes computed from
    the asymptotic refinement range.}
    \label{fig:conforming_vms_2h}
\end{figure*}

\begin{figure*}[!htb]
    \centering
    \includegraphics[width=\textwidth]
    {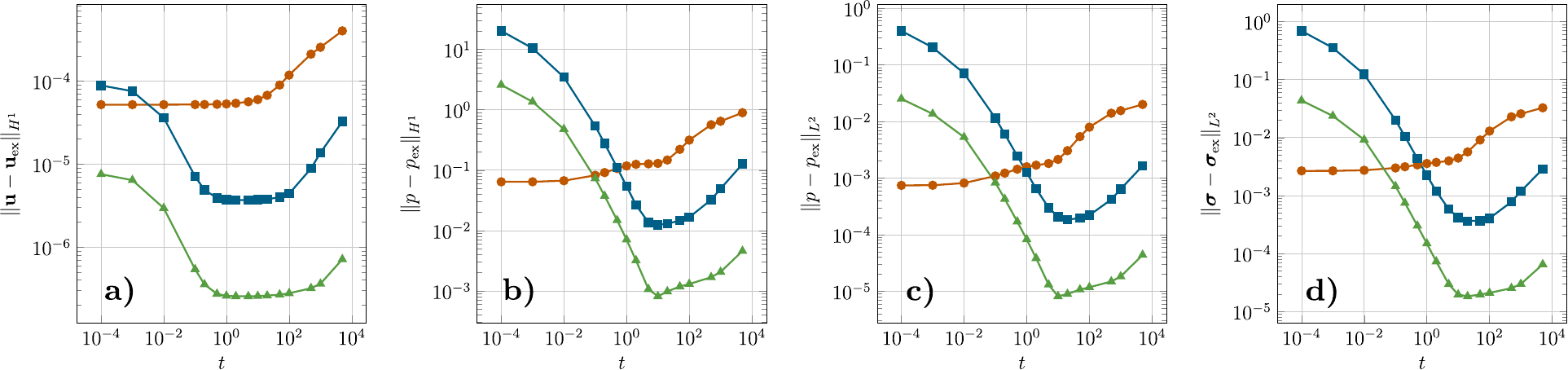}
    \caption{Effect of \(t=\tau_e\Lambda_e\) on a fixed \(32\times32\)
    displacement mesh for
    \stablepair{\(\mathcal{U}(k,\mathcal{C}^{k-1},h)/\mathcal{P}(k-1,\mathcal{C}^{-1},2h)\)}.
    The error measures and degree colors are the same as in
    \cref{fig:tau_sweep_h}. The curves again exhibit an intermediate
    stabilization range that balances insufficient stabilization and
    over-stabilization.}
    \label{fig:tau_sweep_2h}
\end{figure*}

We next consider the coarsened-pressure family
\stablepair{\(\mathcal{U}(k,\mathcal{C}^{k-1},h)/\allowbreak\mathcal{P}(k-1,\mathcal{C}^{-1},2h)\)}. Unlike the same-mesh family, this
discretization satisfies the constant-mode inf-sup requirement identified in
the preceding section. The VMS convergence results are shown in
\cref{fig:conforming_vms_2h}. The displacement, pressure, and stress errors
again converge regularly under mesh refinement, and the expected asymptotic
behavior is retained for all three polynomial orders. The corresponding stabilization-parameter study is shown in
\cref{fig:tau_sweep_2h}. The same general behavior is observed as for the
\(h\) pressure grid: insufficient stabilization at small \(t\), an
intermediate low-error range, and increasing error at large \(t\) due to
over-stabilization. The minimum again shifts toward larger stabilization
levels as \(k\) increases. Compared with the same-mesh discretization, the
low-error range is somewhat more localized; nevertheless, a clear
intermediate range exists in which the stable \(2h\) formulation provides
accurate solutions.

\begin{figure*}[!htbp]
    \centering
    \includegraphics[width=\linewidth]{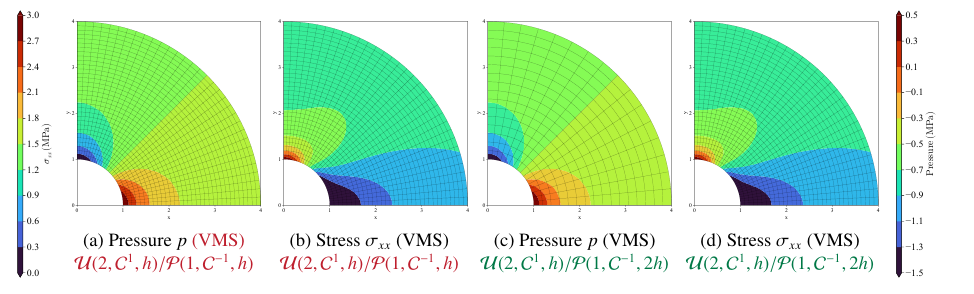}
    \caption{Pressure $p$ and normal stress $\sigma_{xx}$ obtained using
    mesh-conforming quadratic-displacement and discontinuous linear-pressure
    discretizations. The first two panels use
    \unstablepair{$\mathcal{U}(2,\mathcal{C}^1,h)/\mathcal{P}(1,\mathcal{C}^{-1},h)$}, while the last two panels use
    \stablepair{$\mathcal{U}(2,\mathcal{C}^1,h)/\mathcal{P}(1,\mathcal{C}^{-1},2h)$}. The shared stress colorbar is shown
    on the left, and the shared pressure colorbar is shown on the right.}
    \label{fig:kirsch_DG_conforming_Q2Q1_pressure_stress}

\end{figure*}

Representative mesh-conforming pressure and stress fields for \(k=2, \ n_r=32\) are
shown in \cref{fig:kirsch_DG_conforming_Q2Q1_pressure_stress}. The first two
panels use \unstablepair{\(\mathcal{U}(2,\mathcal{C}^1,h)/\allowbreak\mathcal{P}(1,\mathcal{C}^{-1},h)\)}, while the last two use
\stablepair{\(\mathcal{U}(2,\mathcal{C}^1,h)/\allowbreak\mathcal{P}(1,\mathcal{C}^{-1},2h)\)}. In both cases, VMS stabilization produces
smooth pressure and stress distributions without any oscillations observed
for the unstable \unstablepair{\(\mathcal{U}(2,\mathcal{C}^1,h)/\allowbreak\mathcal{P}(1,\mathcal{C}^{-1},h)\)} case. The fields are also
qualitatively comparable to those obtained using the continuous pressure
spaces, showing that the discontinuous representation can retain high
accuracy while preserving the local pressure structure required for
elementwise elimination.

\subsubsection{Immersed discretization}

\begin{figure*}[!htb]
    \centering
    \includegraphics[width=1.0\textwidth]
    {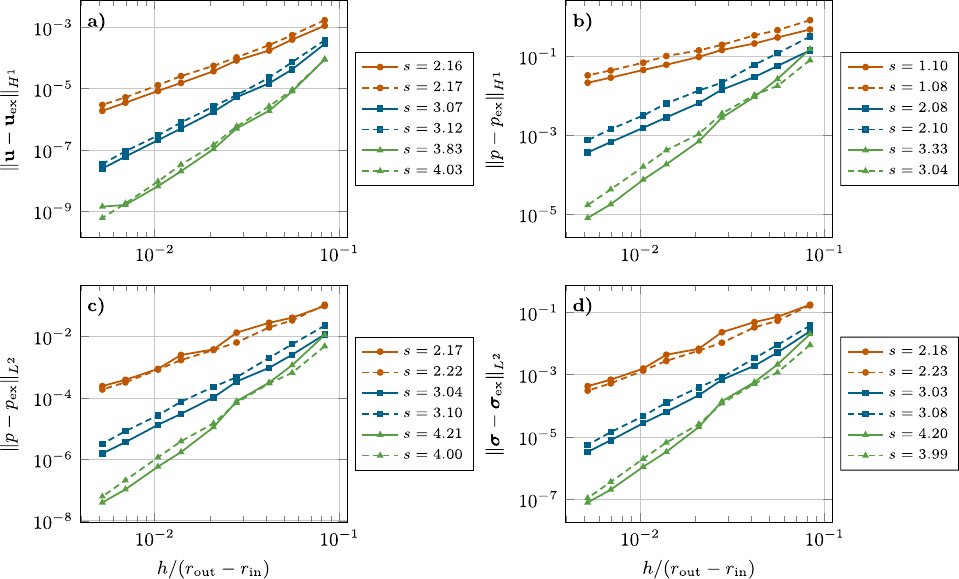}
    \caption{Immersed \stablepair{VMS} convergence comparison for discontinuous pressure spaces.
    Solid curves correspond to
    \unstablepair{\(\mathcal{U}(k,\mathcal{C}^{k-1},h)/\mathcal{P}(k-1,\mathcal{C}^{-1},h)\)}, whereas dashed curves correspond to
    \stablepair{\(\mathcal{U}(k,\mathcal{C}^{k-1},h)/\mathcal{P}(k-1,\mathcal{C}^{-1},2h)\)}.
    The panels report the displacement \(H^1\)-error, pressure \(H^1\)-error,
    pressure \(L^2\)-error, and stress \(L^2\)-error.
    Colors indicate
    \(\textcolor{kTwoColor}{k=2}\),
    \(\textcolor{kThreeColor}{k=3}\), and
    \(\textcolor{kFourColor}{k=4}\).}
    \label{fig:immersed_galerkin_ratio_comparison}
\end{figure*}

\noindent
The discontinuous pressure formulation is next examined using the immersed
discretization. The convergence comparison between
\unstablepair{\(\mathcal{U}(k,\mathcal{C}^{k-1},h)/\allowbreak\mathcal{P}(k-1,\mathcal{C}^{-1},h)\)} and
\stablepair{\(\mathcal{U}(k,\mathcal{C}^{k-1},h)/\allowbreak\mathcal{P}(k-1,\mathcal{C}^{-1},2h)\)} is shown in
\cref{fig:immersed_galerkin_ratio_comparison}, using a fixed nondimensional
stabilization parameter \(t=\tau_e\Lambda_e=5\) throughout the refinement
study. As in the previous immersed studies, the convergence curves exhibit
small nonmonotonic variations as the background grid is refined. These
variations arise from changes in the cut-cell configuration and, consequently,
in the displacement extension and pressure-cell aggregation associated with
successive immersed meshes. Nevertheless, both families retain the expected
overall convergence behavior.

The same-mesh family generally produces slightly smaller errors for this
smooth benchmark, owing to the greater approximation capacity provided by
the finer \(h\) pressure grid. This improvement in accuracy should, however,
be distinguished from stability. The pair
\unstablepair{\(\mathcal{U}(k,\mathcal{C}^{k-1},h)/\allowbreak\mathcal{P}(k-1,\mathcal{C}^{-1},h)\)} does not satisfy the inf-sup requirement
for the elementwise constant pressure mode, whereas the coarsened-pressure
pair \stablepair{\(\mathcal{U}(k,\mathcal{C}^{k-1},h)/\allowbreak\mathcal{P}(k-1,\mathcal{C}^{-1},2h)\)} satisfies this condition. Thus, the
slightly smaller errors observed for the same-mesh family in the Kirsch
problem do not alter the stability distinction between the two
discretizations.

\begin{figure*}[!htbp]
    \centering
    \includegraphics[width=\linewidth]{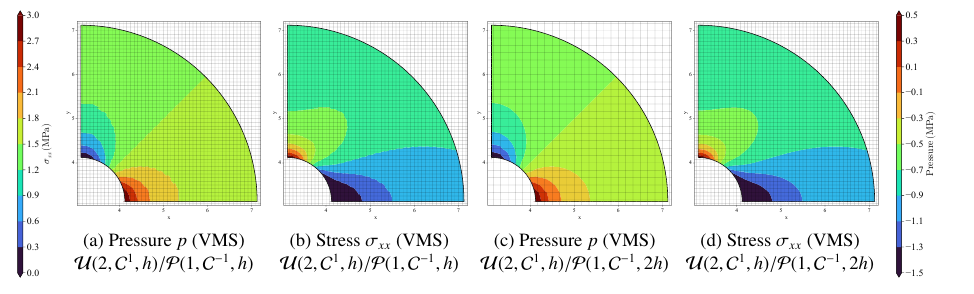}
    \caption{Pressure $p$ and normal stress $\sigma_{xx}$ obtained using
    immersed quadratic-displacement and discontinuous linear-pressure
    discretizations. The first two panels use
    $\mathcal{U}(2,\mathcal{C}^1,h)/\mathcal{P}(1,\mathcal{C}^{-1},h)$, while the last two panels use
    $\mathcal{U}(2,\mathcal{C}^1,h)/\mathcal{P}(1,\mathcal{C}^{-1},2h)$. The shared stress colorbar is shown
    on the left, and the shared pressure colorbar is shown on the right.}
    \label{fig:kirsch_DG_immersed_Q2Q1_pressure_stress}

\end{figure*}

Representative immersed pressure and stress fields for \(k=2, \ n_x=n_y=120\) are shown in
\cref{fig:kirsch_DG_immersed_Q2Q1_pressure_stress}. Despite the discontinuous
pressure approximation and the nonconforming intersection of the physical
domain with the background grid, the computed pressure and stress
distributions remain smooth at the macroscopic scale and free of spurious
oscillations. Both the \unstablepair{\(\mathcal{U}(k,\mathcal{C}^{k-1},h)/\allowbreak\mathcal{P}(k-1,\mathcal{C}^{-1},h)\)} and \stablepair{\(\mathcal{U}(k,\mathcal{C}^{k-1},h)/\allowbreak\mathcal{P}(k-1,\mathcal{C}^{-1},2h)\)} discretizations produce fields that are
qualitatively consistent with their mesh-conforming counterparts, indicating
that the VMS stabilization and pressure-cell aggregation remain effective in
the presence of cut cells.

Overall, the Kirsch results support the stability mechanisms established in
the preceding analysis and demonstrate that the proposed residual-based
formulation remains accurate and robust for higher-order discontinuous
pressure approximations.


\subsection{Local stabilization-parameter scaling}

\noindent
Finally, before closing this section, we quantify the degree and geometry dependence of the local inverse constant.
We define the dimensionless factor
\begin{equation}
    \alpha_K
    =
    \frac{2\mu}{\Lambda_K h_K^2},
    \qquad
    \tau_K
    =
    t\,\alpha_K\frac{h_K^2}{2\mu},
    \label{eq:kirsch_dimensionless_stabilization_factor}
\end{equation}
where \(h_K^2=|\Omega_K|\). For a cut background cell,
\(\Omega_K=K\cap\Omega\), so that \(h_K^2\) is the physical cut-cell area.
The minimum, mean, and maximum values of \(\alpha_K\) are reported in
\cref{tab:kirsch_stabilization_parameter} for maximum-continuity displacement
spaces on fixed mesh-conforming and immersed grids.

\begin{table*}[!t]
    \centering
    \caption{
    Elementwise dimensionless stabilization factor
    \(\alpha_K=2\mu/(\Lambda_Kh_K^2)\) for
    \(\mathcal{U}(k,\mathcal{C}^{k-1},h)\). For the immersed discretization,
    \(h_K^2=|K\cap\Omega|\), and the statistics are evaluated over all active
    physical cells. \revx{\(\Lambda_K\) is computed from the generalized eigenproblem of \Cref{subsec:tau_parameter} restricted to the complement of the rigid-body modes.}
    }
    \label{tab:kirsch_stabilization_parameter}
    \begin{tabular}{llcccc}
        \toprule
        Discretization
        & Mesh
        & \(k\)
        & \(\min_K\alpha_K\)
        & \(\operatorname{mean}_K\alpha_K\)
        & \(\max_K\alpha_K\)
        \\
        \midrule
        Mesh-conforming & \(32\times32\)  & 2 & 0.0298 & 0.0368 & 0.0407 \\
        Mesh-conforming & \(32\times32\)  & 3 & 0.0065 & 0.0085 & 0.0098 \\
        Mesh-conforming & \(32\times32\)  & 4 & 0.0023 & 0.0031 & 0.0036 \\
        \midrule
        Immersed        & \(120\times120\) & 2 & 0.0023 & 0.0392 & 0.0409 \\
        Immersed        & \(120\times120\) & 3 & 0.0008 & 0.0094 & 0.0108 \\
        Immersed        & \(120\times120\) & 4 & 0.0005 & 0.0035 & 0.0092 \\
        \bottomrule
    \end{tabular}
\end{table*}

The characteristic magnitude of \(\alpha_K\) decreases with increasing
polynomial degree. The immersed mean values remain close to their
mesh-conforming counterparts, whereas the wider immersed ranges reflect
cell-to-cell variations in the physical support and local inverse constants
of the cut cells. These variations are accounted for by computing
\(\Lambda_K\), and hence \(\tau_K=t/\Lambda_K\), separately on each active
cell.

\revx{The small immersed minima in
\cref{tab:kirsch_stabilization_parameter} originate from highly slender
physical portions of cut cells. Although the characteristic measure in
\cref{eq:kirsch_dimensionless_stabilization_factor} accounts for the
immersed geometry through $h_K^2=|K\cap\Omega|$, it represents only the
area of the physical portion and does not account for its aspect ratio.
The local inverse constant, by contrast, is governed by the smallest
geometric length scale of the physical support. To illustrate this, consider
a strip-like cut region of dimensions $H\times\delta$, with $\delta\ll H$.
Then
$h_K^2 = |K\cap\Omega| \sim H\delta$,
whereas the corresponding inverse estimate scales as
$\Lambda_K
    \sim
    {2\mu}/{\delta^2}$.
Consequently,
\begin{equation}
    \alpha_K
    =
    \frac{2\mu}{\Lambda_K h_K^2}
    \sim
    \frac{\delta}{H}.
\end{equation}
Thus, the area-based normalization accounts for the reduction in physical
cell size but does not eliminate the effect of a deteriorating aspect ratio.
As the physical portion becomes increasingly slender, $\alpha_K$ therefore
decreases, while
$\tau_K=t/\Lambda_K=O(\delta^2/(2\mu))$ becomes correspondingly small.}


\section{Cook's Membrane}
\label{subsec:cooks_membrane}
\noindent
The Cook's membrane problem is considered next as a more challenging
benchmark with singularities for assessing the stability of the mixed formulations in the
nearly incompressible regime. In contrast to the Kirsch problem, for which
a smooth analytical solution is available, Cook's membrane develops
strongly nonuniform pressure and stress fields under the applied shear
loading and provides a particularly useful test for exposing spurious
pressure modes. Consequently, this example is used primarily to examine
the qualitative and discrete inf-sup behavior of the different
displacement--pressure pairs and to assess the effectiveness of the
residual-based stabilization formulations.

\begin{figure*}[!htb]
    \centering
    \includegraphics[width=\textwidth]
    {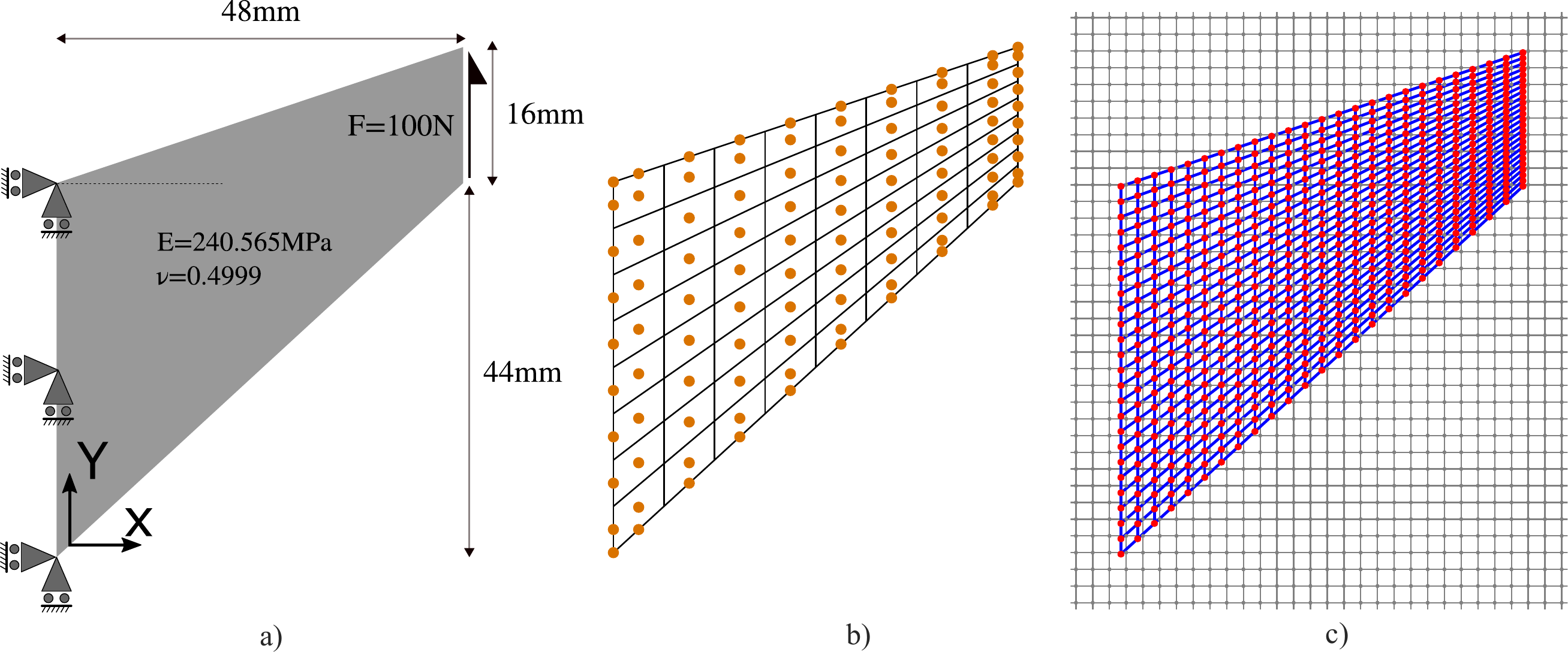}
    \caption{
    Schematic representation and discretization of Cook's membrane.
    \textbf{(a)} Geometry, loading, and boundary conditions for the
    nearly incompressible membrane with
    \(E=240.565~\mathrm{MPa}\) and \(\nu=0.4999\).
    The membrane has a width of \(48~\mathrm{mm}\), with right-edge
    heights of \(44~\mathrm{mm}\) and \(60~\mathrm{mm}\) at the lower
    and upper corners, respectively. The left boundary is fixed, and a
    uniform vertical traction of \(6.25~\mathrm{N/mm}\), corresponding
    to a resultant force of \(100~\mathrm{N}\), is applied along the
    \(16~\mathrm{mm}\)-long right edge.
    \textbf{(b)} Representative mesh-conforming isogeometric
    discretization of the physical Cook domain and its associated
    control points.
    \textbf{(c)} Representative immersed discretization, in which the
    physical Cook domain is embedded in an independent
    \(120~\mathrm{mm}\times120~\mathrm{mm}\) Cartesian background grid.
    The reference point of the Cook geometry is placed at
    \((x_0,y_0)=(61.37,\,55.82)~\mathrm{mm}\), deliberately offset from
    the background grid so that the physical boundary cuts through the
    background elements.
    }
    \label{fig:CookScheme}
\end{figure*}

The geometry, loading, and representative discretizations are shown in
\cref{fig:CookScheme}. The membrane has a width of \(48~\mathrm{mm}\),
with the lower and upper corners of the right boundary located at heights
of \(44~\mathrm{mm}\) and \(60~\mathrm{mm}\), respectively. The left
boundary is fixed, while a uniform vertical traction of
\(6.25~\mathrm{N/mm}\) is applied along the \(16~\mathrm{mm}\)-long right
boundary, giving a resultant vertical force of \(100~\mathrm{N}\). The material parameters are
\(E=240.565~\mathrm{MPa}\) and \(\nu=0.4999\), placing the material response
close to the incompressible limit. As in the preceding example, plane-strain
conditions are assumed, such that \(\epsilon_{33}=0\). In this nearly
incompressible regime, an inf-sup-unstable displacement--pressure pair may
admit pronounced spurious pressure modes. These nonphysical oscillations are
subsequently transmitted to the Cauchy stress through the pressure contribution
to the constitutive relation.

Both mesh-conforming and immersed discretizations are considered, as
illustrated in panels (b) and (c) of \cref{fig:CookScheme}. In the
mesh-conforming case, the spline mesh follows the boundaries of the Cook
domain. In the immersed case, the same physical domain is embedded in an
independent \(120~\mathrm{mm}\times120~\mathrm{mm}\) Cartesian background
grid. The reference point of the Cook geometry is placed at
\((x_0,y_0)=(61.37,55.82)~\mathrm{mm}\), deliberately offset from the
background grid so that the physical boundary cuts through the background
elements. In all immersed calculations, the physical Cook geometry is represented by
a fixed \(24\times24\) bilinear \(\mathcal{C}^0\) NURBS mesh with
\(25\times25\) physical control points. This physical discretization is held
fixed during background-grid refinement and is used only to represent the
domain and its boundary and to construct the cut-cell volume and surface
quadrature. The displacement and pressure unknowns remain defined exclusively
on the independent background grids. The physical portions of the cut cells are integrated using the
subdivision quadrature described previously, with
\(n_{\textrm{cut}}=7\). Essential boundary conditions on the immersed
boundary are imposed weakly using Nitsche's method, with the stabilization
parameter
$\gamma_{N,K}=0.01\lambda_{N,K}$.
Poorly supported displacement and pressure basis functions are treated using
the displacement-extension and pressure-extension/aggregation procedures,
respectively, introduced in the immersed formulation.

For the mesh-refinement studies, the mesh-conforming Cook domain is
discretized using
\begin{equation}
    n_x=n_y
    \in
    \{2,4,8,16,32,64\},
    \label{eq:cook_conforming_mesh_sequence}
\end{equation}
where \(n_x\) and \(n_y\) denote the numbers of displacement elements in
the two parametric directions. For the immersed formulation, the independent
\(120~\mathrm{mm}\times120~\mathrm{mm}\) Cartesian background domain is
discretized using
\begin{equation}
    n_x=n_y
    \in
    \{40,60,80,120,160,240,320\},
    \qquad
    h=\frac{120}{n_x}~\mathrm{mm}.
    \label{eq:cook_immersed_mesh_sequence}
\end{equation}
These refinement sequences are used in the discrete inf-sup studies for the
continuous and discontinuous pressure spaces. The mesh size reported in the
inf-sup plots is normalized by
\(L_{\mathrm{ref}}=60~\mathrm{mm}\), corresponding to the upper height of
Cook's membrane.

As in the Kirsch problem, three classes of pressure approximation are
examined. We first consider continuous pressure spaces to study the
Galerkin inf-sup behavior and the effect of residual-based stabilization.
We then consider the discontinuous constant-pressure space to isolate the
elementwise constant pressure mode and directly assess its inf-sup stability.
Finally, higher-order discontinuous pressure spaces are examined to
demonstrate the combined roles of residual-based stabilization and inf-sup
control of the elementwise constant pressure mode. The same questions are
considered for both the mesh-conforming and immersed formulations.


\subsection{Continuous Pressure}

\noindent
We first consider continuous pressure approximations. The Cook's membrane
problem is particularly useful for this purpose because the instability of
the continuous Galerkin pairs is considerably more visible in the pressure
and stress fields than for the smooth Kirsch benchmark. We first compare
the pressure degree-reduced families
\unstablepair{\(\mathcal{U}(k,\mathcal{C}^{k-1},h)/\allowbreak\mathcal{P}(k-1,\mathcal{C}^{k-2},h)\)} and
\stablepair{\(\mathcal{U}(k,\mathcal{C}^{k-1},h)/\allowbreak\mathcal{P}(k-1,\mathcal{C}^{k-2},2h)\)}. According to the isogeometric
inf-sup condition derived previously, the same-mesh pair does not satisfy
the required stability condition, whereas coarsening the pressure grid by
a factor of two provides an inf-sup-stable pair while preserving maximum
continuity of the displacement space. We then investigate the
effectiveness of residual-based stabilization by comparing the GLS,
SUPG/PSPG-type, and VMS formulations. Finally, we perform
numerical inf-sup tests for these two families and for the equal-order pair
\unstablepair{\(\mathcal{U}(k,\mathcal{C}^{k-1},h)/\allowbreak\mathcal{P}(k,\mathcal{C}^{k-1},h)\)}.

\subsubsection{Mesh-Conforming}

\begin{figure*}[!htbp]
    \centering
    \includegraphics[width=\linewidth]{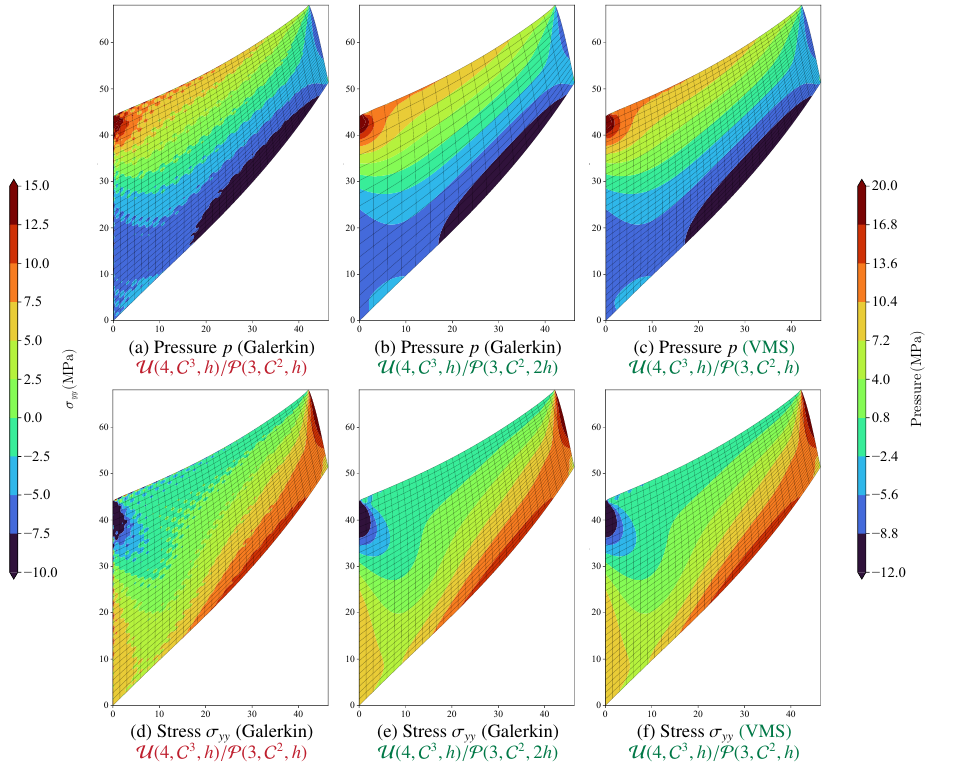}
    \caption{
    Pressure \(p\) and Cauchy stress \(\sigma_{yy}\) for Cook's membrane
    obtained using quartic displacement and cubic pressure spaces.
    Each column presents the pressure and stress fields associated with
    the same discretization. The first column uses the \unstablepair{Galerkin pair
    \(\mathcal{U}(4,\mathcal{C}^3,h)/\mathcal{P}(3,\mathcal{C}^2,h)\)}, the second uses the coarsened-pressure
    \stablepair{Galerkin pair \(\mathcal{U}(4,\mathcal{C}^3,h)/\mathcal{P}(3,\mathcal{C}^2,2h)\)}, and the third uses the
    \stablepair{VMS-stabilized pair \(\mathcal{U}(4,\mathcal{C}^3,h)/\mathcal{P}(3,\mathcal{C}^2,h)\)}.
    The shared stress colorbar is shown on the left, and the shared
    pressure colorbar is shown on the right.
    }
    \label{fig:cook_Q4Q3_pressure_stress_comparison}

\end{figure*}

\noindent
Representative mesh-conforming pressure and stress fields for a $32\times32$ displacement mesh are shown in
\cref{fig:cook_Q4Q3_pressure_stress_comparison}. The first column uses
the \unstablepair{Galerkin pair \(\mathcal{U}(4,\mathcal{C}^3,h)/\allowbreak\mathcal{P}(3,\mathcal{C}^2,h)\)}, the second uses
\stablepair{Galerkin pair \(\mathcal{U}(4,\mathcal{C}^3,h)/\allowbreak\mathcal{P}(3,\mathcal{C}^2,2h)\)}, and the third uses the \stablepair{VMS-stabilized pair \(\mathcal{U}(4,\mathcal{C}^3,h)/\allowbreak\mathcal{P}(3,\mathcal{C}^2,h)\)}. The difference between the
two Galerkin discretizations is clearly visible. The theoretically
unstable same-mesh pair develops oscillations in both the pressure and
the corresponding Cauchy stress, whereas coarsening the pressure grid
produces smooth fields. The latter observation is consistent with the
inf-sup condition established previously: the \(2h\) pressure grid provides a sufficiently rich displacement space relative to the pressure space.
The third column demonstrates the alternative stabilization mechanism.
Although the underlying same-mesh pair is not Galerkin inf-sup stable,
the residual-based VMS term controls the unstable continuous pressure
modes and recovers smooth pressure and stress fields. Cook's membrane
therefore makes the distinction between Galerkin inf-sup stability and
residual-based stabilization considerably more apparent than the Kirsch
problem, where the smooth solution can mask the instability of
some Galerkin pairs.

\begin{figure*}[!htb]
    \centering
    \includegraphics[width=\textwidth]
    {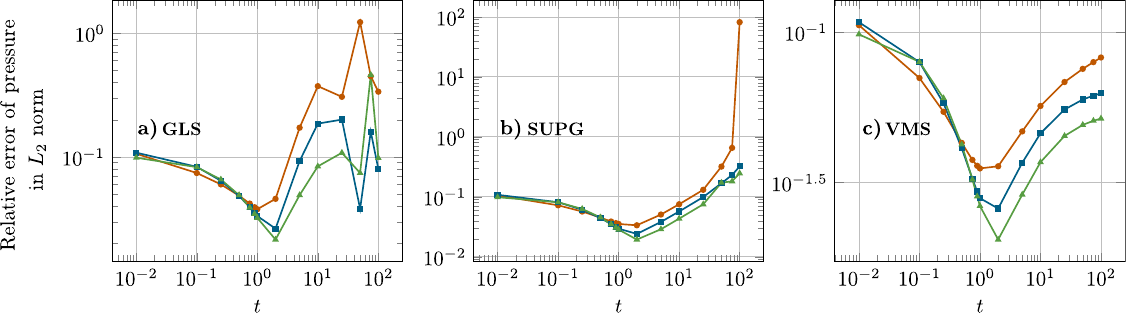}
    \caption{
    Relative pressure \(L^2\)-error for Cook's membrane versus the
    nondimensional stabilization level \(t=\tau_e\Lambda_e\) using \stablepair{residual-based stabilized pair
    \(\mathcal{U}(k,\mathcal{C}^{k-1},h)/\mathcal{P}(k-1,\mathcal{C}^{k-2},h)\)}.
    From left to right, the panels show the GLS \((\chi=+1)\),
    SUPG/PSPG-type \((\chi=0)\), and VMS \((\chi=-1)\) formulations.
    The \ktwo, \kthree, and \kfour{} curves correspond to
    displacement discretizations order.
    The reference solution uses
    \(\mathcal{U}(6,\mathcal{C}^5,h)/\mathcal{P}(5,\mathcal{C}^4,2h)\) on a \(256\times256\) displacement mesh,
    eight times finer in each direction than the \(32\times32\) mesh used
    for the parameter study.
    }
    \label{fig:cook_continuous_tau_relative_error}
\end{figure*}

The influence of the residual-based formulation and stabilization
parameter is examined more directly in
\cref{fig:cook_continuous_tau_relative_error}. The displacement mesh is
fixed at \(32\times32\) elements and the nondimensional stabilization
level \(t=\tau_e\Lambda_e\) is varied over several orders of magnitude.
The relative pressure \(L^2\)-error is evaluated with respect to the
reference solution obtained using
\stablepair{Galerkin pair \(\mathcal{U}(6,\mathcal{C}^5,h)/\allowbreak\mathcal{P}(5,\mathcal{C}^4,2h)\)} on a \(256\times256\) displacement mesh,
which is eight times finer in each direction.

The results clearly distinguish the three residual-based formulations.
For GLS and the SUPG/PSPG-type formulation, increasing \(t\) from the
under-stabilized regime initially reduces the pressure error, but the
error increases rapidly once the useful stabilization range is exceeded.
This behavior is consistent with the conditional stability bounds derived
previously for these formulations. In contrast, VMS maintains a broad
stable range as \(t\) is increased, in agreement with its unconditional
stability with respect to positive \(t\). Nevertheless, the VMS error
also increases for sufficiently large stabilization parameters.
Unconditional stability therefore does not imply parameter-independent
accuracy: excessively large values of \(t\) produce an over-stabilized
solution even though the formulation remains stable.

\subsubsection{Immersed}

\begin{figure*}[!htbp]
    \centering
    \includegraphics[width=\linewidth]{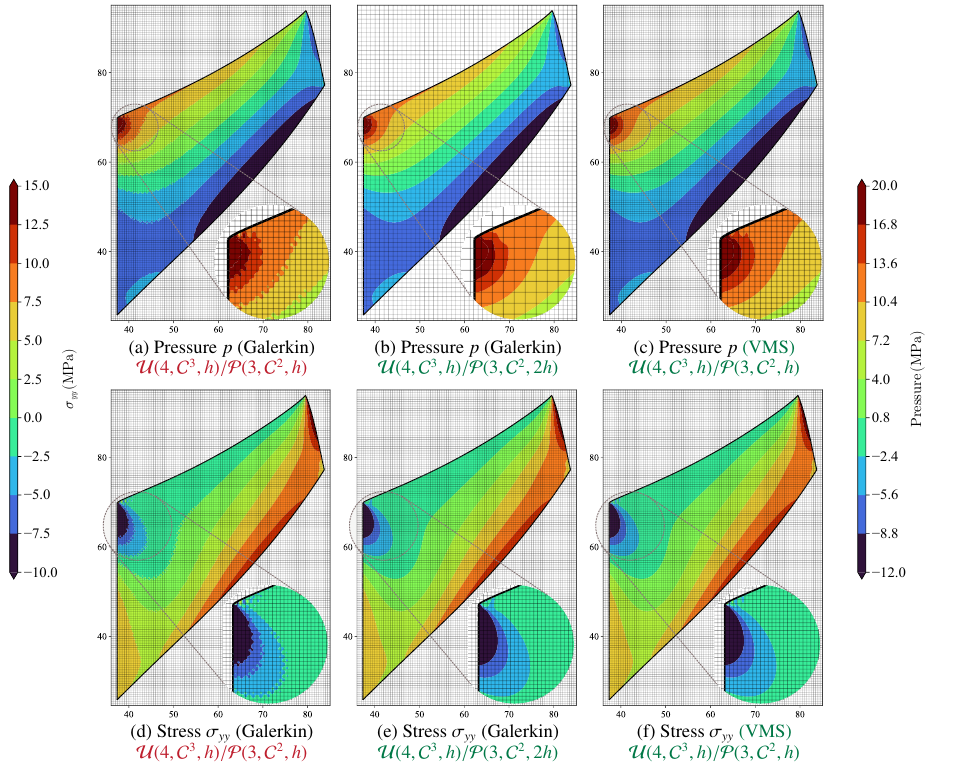}
    \caption{
    Pressure \(p\) and Cauchy stress \(\sigma_{yy}\) for the immersed
    Cook's membrane problem obtained using quartic displacement and
    cubic pressure spaces. Each column presents the pressure and stress
    fields associated with the same approximation pair. The first column
    uses the \unstablepair{Galerkin pair
    \(\mathcal{U}(4,\mathcal{C}^3,h)/\mathcal{P}(3,\mathcal{C}^2,h)\)}, the second uses the coarsened-pressure
   \stablepair{ Galerkin pair
    \(\mathcal{U}(4,\mathcal{C}^3,h)/\mathcal{P}(3,\mathcal{C}^2,2h)\)}, and the third uses the corresponding
    \stablepair{VMS-stabilized pair
    \(\mathcal{U}(4,\mathcal{C}^3,h)/\mathcal{P}(3,\mathcal{C}^2,h)\)}.
    The shared stress colorbar is shown on the left, and the shared
    pressure colorbar is shown on the right.
    }
    \label{fig:cook_immersed_Q4Q3_pressure_stress_comparison}

\end{figure*}

\noindent
We next repeat the continuous-pressure comparison using the immersed
formulation. Representative results are shown in
\cref{fig:cook_immersed_Q4Q3_pressure_stress_comparison} for the same
three discretizations considered above, using a
\(120~\mathrm{mm}\times120~\mathrm{mm}\) Cartesian background domain
discretized into \(240\times240\) displacement-grid cells. Even at this
relatively fine resolution, the immersed results reproduce a similar
behavior observed in the mesh-conforming case. The same-mesh \unstablepair{Galerkin pair
\(\mathcal{U}(4,\mathcal{C}^3,h)/\allowbreak\mathcal{P}(3,\mathcal{C}^2,h)\)} exhibits spurious pressure and stress oscillations,
whereas both the inf-sup-stable coarsened-pressure pair
\stablepair{\(\mathcal{U}(4,\mathcal{C}^3,h)/\allowbreak\mathcal{P}(3,\mathcal{C}^2,2h)\)} and the \stablepair{VMS-stabilized pair \(\mathcal{U}(4,\mathcal{C}^3,h)/\allowbreak\mathcal{P}(3,\mathcal{C}^2,h)\)} produce
smooth fields. The localized behavior near the immersed boundary further
indicates that the cut-cell treatment does not introduce additional
oscillatory pressure modes when the underlying mixed formulation is
properly stabilized.

\begin{figure*}[!htb]
    \centering
    \includegraphics[width=\textwidth]
    {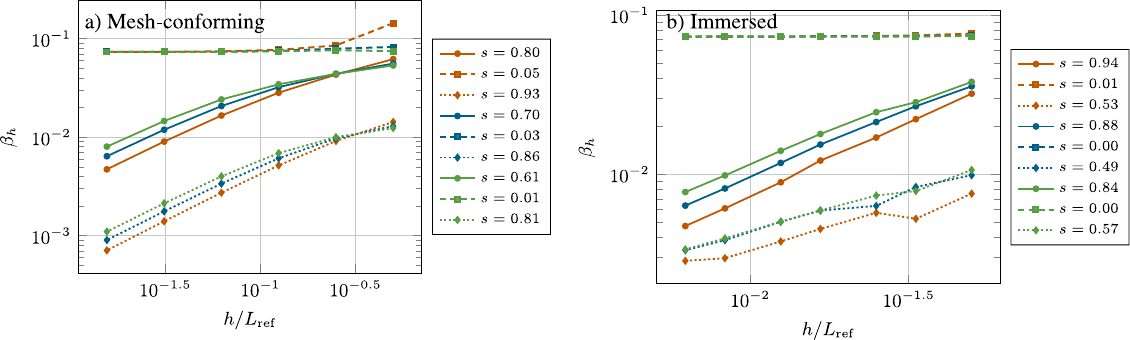}
    \caption{
    Discrete inf-sup constant \(\beta_h\) for the continuous-pressure
    Cook's membrane problem under mesh refinement:
    \textbf{(a)} mesh-conforming and \textbf{(b)} immersed discretizations.
    The mesh size is normalized by
    \(L_{\mathrm{ref}}=60\ \textrm{mm}\), corresponding to the upper height of Cook's
    membrane. The \ktwo, \kthree, and \kfour{} curves correspond to the displacement polynomial order.
    Solid curves with circular markers use
    \unstablepair{\(\mathcal{U}(k,\mathcal{C}^{k-1},h)/\mathcal{P}(k-1,\mathcal{C}^{k-2},h)\)};
    dashed curves with square markers use
    \stablepair{\(\mathcal{U}(k,\mathcal{C}^{k-1},h)/\mathcal{P}(k-1,\mathcal{C}^{k-2},2h)\)};
    and short-dashed curves with diamond markers use
    \unstablepair{\(\mathcal{U}(k,\mathcal{C}^{k-1},h)/\mathcal{P}(k,\mathcal{C}^{k-1},h)\)}.
    The values \(s\) reported in the legends denote the fitted slopes.
    }
    \label{fig:cook_continuous_inf_sup}
\end{figure*}

This conclusion is further supported by the immersed inf-sup results in
\cref{fig:cook_continuous_inf_sup}. The same-mesh pressure degree-reduced \unstablepair{\(\mathcal{U}(k,\mathcal{C}^{k-1},h)/\allowbreak\mathcal{P}(k-1,\mathcal{C}^{k-2},h)\)} and equal-order \unstablepair{\(\mathcal{U}(k,\mathcal{C}^{k-1},h)/\allowbreak\mathcal{P}(k,\mathcal{C}^{k-1},h)\)} Galerkin 
families again exhibit decreasing \(\beta_h\) under refinement, consistent
with the pressure and stress oscillations observed for these unstable
discretizations. In contrast, the \stablepair{\(\mathcal{U}(k,\mathcal{C}^{k-1},h)/\allowbreak\mathcal{P}(k-1,\mathcal{C}^{k-2},2h)\)} family maintains a
mesh-independent inf-sup constant, in agreement with the smooth fields
obtained with this pair. Importantly, the persistence of this inf-sup
stability in the immersed setting demonstrates that the theoretically stable
displacement--pressure pair remains stable even when extended B-splines are
introduced near the immersed boundary. The agreement between the mesh-conforming and immersed stability trends
therefore indicates that the continuous-field extension and cut-cell
integration procedures preserve the mixed stability properties of the
underlying approximation spaces.

\subsection{Constant Pressure}
\label{subsubsec:cook_constant_pressure}

\noindent
Before considering higher-order discontinuous pressure spaces, we isolate
the elementwise constant pressure mode. This step is essential because
the residual-based pressure-gradient stabilization does not act on this
mode: the gradient of an elementwise constant pressure vanishes within
each pressure element. Consequently, the stability of the constant
pressure component must be supplied by the underlying
displacement--pressure pair itself.

We therefore consider
\unstablepair{\(\mathcal{U}(k,\mathcal{C}^{k-1},h)/\allowbreak\mathcal{P}(0,\mathcal{C}^{-1},h)\)} and
\stablepair{\(\mathcal{U}(k,\mathcal{C}^{k-1},h)/\allowbreak\mathcal{P}(0,\mathcal{C}^{-1},2h)\)}, for \(k=2,3,4\).
For the same-mesh family, the constant-mode stability condition gives
\(k-(k-1)=1\), which does not satisfy the strict inf-sup requirement.
Coarsening the pressure grid by a factor of two instead gives
\(2[k-(k-1)]=2>1\), and the constant pressure mode is therefore stable.
The Cook's membrane problem provides a direct numerical test of this
distinction.

\subsubsection{Mesh-Conforming}

\begin{figure*}[!htbp]
    \centering
    \includegraphics[width=\linewidth]{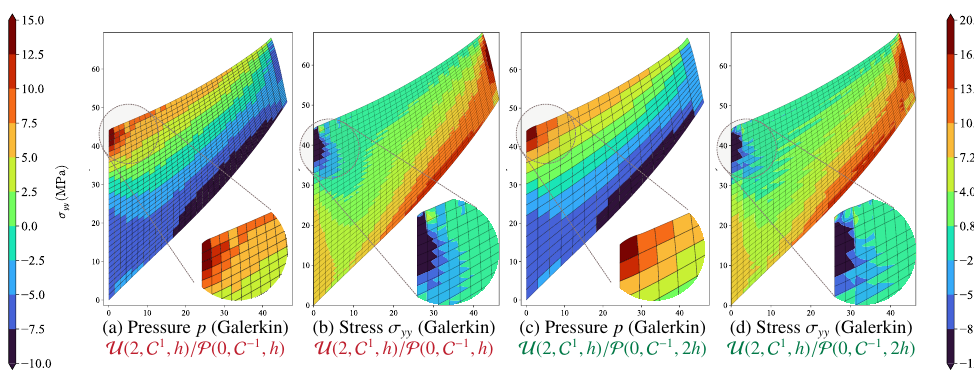}
    \caption{
    Pressure \(p\) and Cauchy stress \(\sigma_{yy}\) for Cook's membrane
    obtained with piecewise-constant discontinuous pressure spaces.
    Panels (a) and (b) use
   \unstablepair{ \(\mathcal{U}(2,\mathcal{C}^1,h)/\mathcal{P}(0,\mathcal{C}^{-1},h)\)}, while panels (c) and (d) use the
    coarsened pressure grid
   \stablepair{ \(\mathcal{U}(2,\mathcal{C}^1,h)/\mathcal{P}(0,\mathcal{C}^{-1},2h)\)}.
    The shared stress colorbar is shown on the left, and the shared
    pressure colorbar is shown on the right.
    }
    \label{fig:cook_constant_pressure_comparison}

\end{figure*}

\noindent
The mesh-conforming fields obtained using the quadratic displacement
space with a $32\times32$ mesh are shown in \cref{fig:cook_constant_pressure_comparison}.
The first two panels use
\unstablepair{\(\mathcal{U}(2,\mathcal{C}^1,h)/\allowbreak\mathcal{P}(0,\mathcal{C}^{-1},h)\)}, whereas the last two use
\stablepair{\(\mathcal{U}(2,\mathcal{C}^1,h)/\allowbreak\mathcal{P}(0,\mathcal{C}^{-1},2h)\)}. The difference between the two pressure
grids is apparent in the pressure field. The same-mesh approximation \unstablepair{\(\mathcal{U}(2,\mathcal{C}^1,h)/\allowbreak\mathcal{P}(0,\mathcal{C}^{-1},h)\)}
exhibits some oscillations associated with the unstable elementwise
constant pressure mode, whereas the coarsened pressure \stablepair{\(\mathcal{U}(2,\mathcal{C}^1,h)/\allowbreak\mathcal{P}(0,\mathcal{C}^{-1},2h)\)} discretization produces a
regular piecewise-constant pressure distribution. The corresponding stress fields retain visible element-scale variations in
both cases because the stress combines the smooth displacement contribution
with the discontinuous piecewise-constant pressure contribution.

\subsubsection{Immersed}

\begin{figure*}[!htbp]
    \centering
    \includegraphics[width=\linewidth]{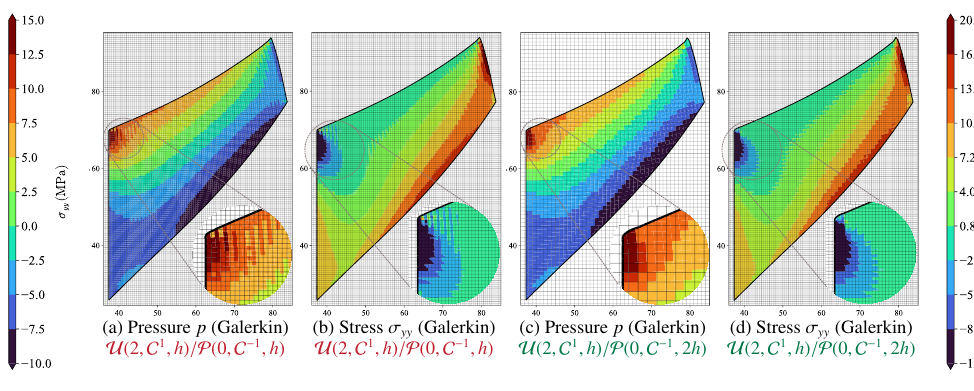}
    \caption{
    Pressure \(p\) and Cauchy stress \(\sigma_{yy}\) for the immersed
    Cook's membrane problem obtained with piecewise-constant
    discontinuous pressure spaces. Panels (a) and (b) use
    \unstablepair{\(\mathcal{U}(2,\mathcal{C}^1,h)/\mathcal{P}(0,\mathcal{C}^{-1},h)\)}, while panels (c) and (d) use the
    coarsened pressure grid
    \stablepair{\(\mathcal{U}(2,\mathcal{C}^1,h)/\mathcal{P}(0,\mathcal{C}^{-1},2h)\)}.
    The shared stress colorbar is shown on the left, and the shared
    pressure colorbar is shown on the right.
    }
    \label{fig:cook_immersed_constant_pressure_comparison}

\end{figure*}

\noindent
The corresponding immersed fields are shown in
\cref{fig:cook_immersed_constant_pressure_comparison} using a
\(120~\mathrm{mm}\times120~\mathrm{mm}\) Cartesian background domain
discretized into \(240\times240\) quadratic displacement-grid cells.
The same stability distinction persists in the presence of cut cells.
The same-mesh pair \unstablepair{\(\mathcal{U}(2,\mathcal{C}^1,h)/\allowbreak\mathcal{P}(0,\mathcal{C}^{-1},h)\)} exhibits visible
oscillations in the piecewise-constant pressure field, whereas the
coarsened-pressure pair \stablepair{\(\mathcal{U}(2,\mathcal{C}^1,h)/\allowbreak\mathcal{P}(0,\mathcal{C}^{-1},2h)\)} produces a
substantially more regular pressure distribution. As discussed previously,
oscillatory variations in the stress field are expected for both
discretizations due to the discontinuous constant-pressure approximation.

\begin{figure*}[!htb]
    \centering
    \includegraphics[width=\textwidth]
    {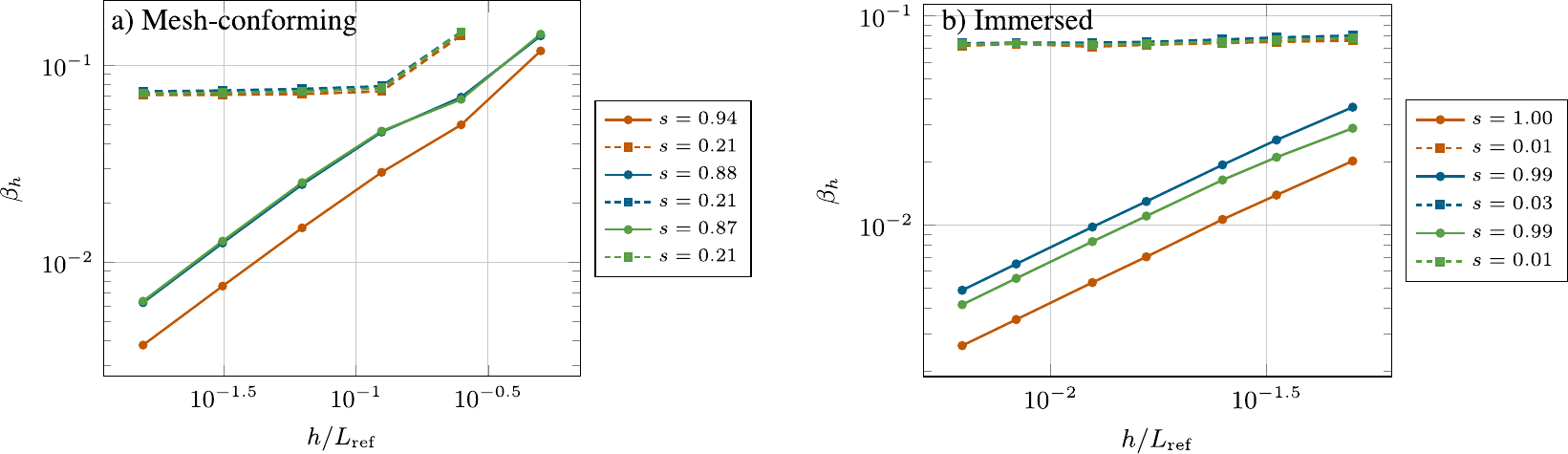}
    \caption{
    Discrete inf-sup constant \(\beta_h\) for the constant-pressure
    Cook's membrane problem under mesh refinement:
    \textbf{(a)} mesh-conforming and \textbf{(b)} immersed discretizations.
    The mesh size is normalized by
    \(L_{\mathrm{ref}}=60\ \textrm{mm}\), corresponding to the upper height of Cook's
    membrane. The \ktwo, \kthree, and \kfour{} curves correspond to the displacement
    polynomial order approximation.
    Solid curves with circular markers use
    \unstablepair{\(\mathcal{U}(k,\mathcal{C}^{k-1},h)/\mathcal{P}(0,\mathcal{C}^{-1},h)\)}, whereas dashed curves with square
    markers use
    \stablepair{\(\mathcal{U}(k,\mathcal{C}^{k-1},h)/\mathcal{P}(0,\mathcal{C}^{-1},2h)\)}.
    The values \(s\) reported in the legends denote the fitted slopes.
    }
    \label{fig:cook_constant_pressure_inf_sup}
\end{figure*}

The discrete inf-sup results in
\cref{fig:cook_constant_pressure_inf_sup} directly support the stability
distinction observed in the pressure contours. For
\unstablepair{\(\mathcal{U}(k,\mathcal{C}^{k-1},h)/\allowbreak\mathcal{P}(0,\mathcal{C}^{-1},h)\)}, \(\beta_h\) decreases under mesh
refinement for all displacement orders considered, consistent with the
spurious oscillations observed in the same-mesh pressure contours. In
contrast, \stablepair{\(\mathcal{U}(k,\mathcal{C}^{k-1},h)/\allowbreak\mathcal{P}(0,\mathcal{C}^{-1},2h)\)} maintains an approximately
mesh-independent inf-sup constant, consistent with the more regular pressure
fields obtained using the coarsened pressure grid. Increasing the displacement
polynomial degree while retaining maximum continuity does not by itself
restore inf-sup stability of the elementwise constant pressure mode.
Coarsening the pressure grid from \(h\) to \(2h\), however, yields a uniformly
stable pair for all displacement orders considered. The immersed results
further show that pressure-cell aggregation and cut-cell treatment preserve
this constant-mode stability.

\subsection{Higher-Order Discontinuous Pressure}

\noindent
We finally consider the VMS-stabilized higher-order discontinuous pressure families
\unstablepair{\(\mathcal{U}(k,\mathcal{C}^{k-1},h)/\allowbreak\mathcal{P}(k-1,\mathcal{C}^{-1},h)\)} and
\stablepair{\(\mathcal{U}(k,\mathcal{C}^{k-1},h)\allowbreak/\allowbreak\mathcal{P}(k-1,\mathcal{C}^{-1},2h)\)}. These discretizations combine the
two stability mechanisms examined separately above. The residual-based
VMS term acts on the nonconstant elementwise pressure modes through the
pressure gradient, whereas the elementwise constant component remains
outside the control of this term and must satisfy the discrete inf-sup
condition.

The constant-pressure study therefore predicts an important distinction
between the two families. For the same-mesh discretization
\unstablepair{\(\mathcal{U}(k,\mathcal{C}^{k-1},h)/\allowbreak\mathcal{P}(k-1,\mathcal{C}^{-1},h)\)}, the constant pressure subspace
remains inf-sup unstable despite stabilization of the higher pressure
modes. In contrast, the coarsened-pressure family
\stablepair{\(\mathcal{U}(k,\mathcal{C}^{k-1},h)/\allowbreak\mathcal{P}(k-1,\mathcal{C}^{-1},2h)\)} satisfies the constant-mode
inf-sup condition while the residual-based term stabilizes the remaining
nonconstant modes. The \(2h\) family therefore combines a higher-order discontinuous pressure approximation with both mechanisms required for stability.

\subsubsection{Mesh-Conforming}

\begin{figure*}[!htbp]
    \centering
    \includegraphics[width=\linewidth]{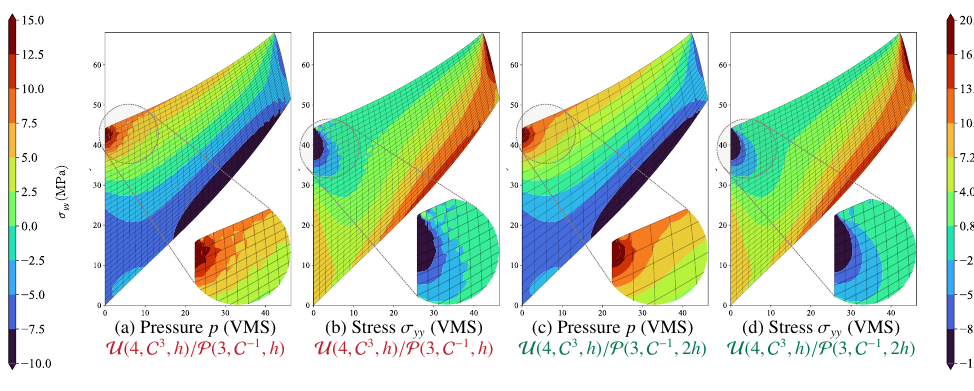}
    \caption{
    Pressure \(p\) and Cauchy stress \(\sigma_{yy}\) for Cook's membrane
    obtained using discontinuous cubic pressure spaces.
    Panels (a) and (b) use
    \unstablepair{\(\mathcal{U}(4,\mathcal{C}^3,h)/\mathcal{P}(3,\mathcal{C}^{-1},h)\)}, while panels (c) and (d) use the
    coarsened pressure grid
    \stablepair{\(\mathcal{U}(4,\mathcal{C}^3,h)/\mathcal{P}(3,\mathcal{C}^{-1},2h)\)}.
    The shared stress colorbar is shown on the left, and the shared
    pressure colorbar is shown on the right.
    }
    \label{fig:cook_dg_q4q3_comparison}

\end{figure*}

\noindent
Representative mesh-conforming VMS solutions on the \(32\times32\) mesh
using quartic displacement and discontinuous cubic pressure spaces are shown
in \cref{fig:cook_dg_q4q3_comparison}. Both the \unstablepair{\(\mathcal{U}(4,\mathcal{C}^{3},h)/\allowbreak\mathcal{P}(3,\mathcal{C}^{-1},h)\)} and \stablepair{\(\mathcal{U}(4,\mathcal{C}^3,h)/\allowbreak\mathcal{P}(3,\mathcal{C}^{-1},2h)\)}
pressure-grid formulations produce substantially improved pressure and stress
fields relative to their unstabilized counterparts, as the VMS term controls
the nonconstant modes of the discontinuous pressure space. An important
distinction nevertheless remains: only the \(2h\) pressure-grid family
provides inf-sup control of the elementwise constant pressure mode. This
difference is also visible in the computed fields, where stress and pressure oscillations
associated with the uncontrolled constant mode remain more pronounced for the
same-mesh discretization, while the \(2h\) formulation produces a noticeably
smoother pressure field.

\begin{figure*}[!htb]
    \centering
    \includegraphics[width=\textwidth]
    {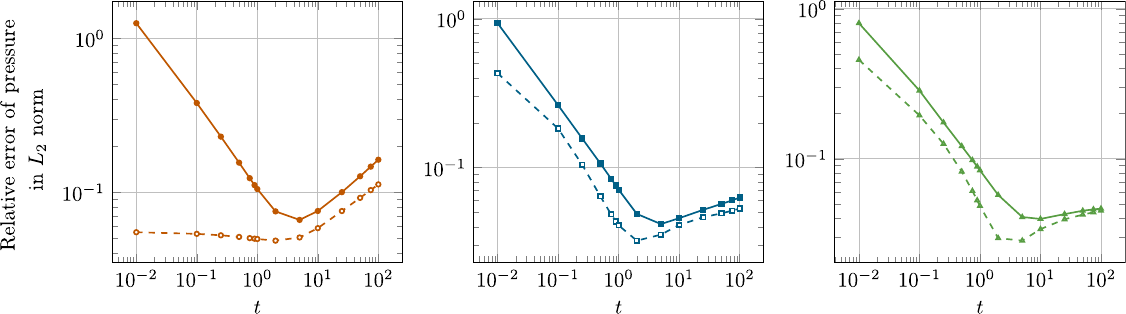}
    \caption{
    Relative pressure \(L^2\)-error for Cook's membrane versus the
    nondimensional VMS stabilization parameter \(t=\tau_e\Lambda_e\) using
    \unstablepair{\(\mathcal{U}(k,\mathcal{C}^{k-1},h)/\mathcal{P}(k-1,\mathcal{C}^{-1},h)\)} (solid curves) and
    \stablepair{\(\mathcal{U}(k,\mathcal{C}^{k-1},h)/\mathcal{P}(k-1,\mathcal{C}^{-1},2h)\)} (dashed curves).
    Panels (a)--(c) correspond to
    \(\textcolor{kTwoColor}{k=2}\),
    \(\textcolor{kThreeColor}{k=3}\), and
    \(\textcolor{kFourColor}{k=4}\), respectively.
    The reference solution uses a \stablepair{Galerkin pair
    \(\mathcal{U}(6,\mathcal{C}^5,h)/\mathcal{P}(5,\mathcal{C}^4,2h)\) on a \(256\times256\)} displacement mesh,
    eight times finer in each direction than the \(32\times32\) mesh used
    for the parameter study.
    }
    \label{fig:cook_dg_vms_tau_relative_error}
\end{figure*}

The influence of the stabilization parameter is shown more clearly in
\cref{fig:cook_dg_vms_tau_relative_error}. For each polynomial order,
the relative pressure \(L^2\)-error is plotted as a function of
\(t=\tau_e\Lambda_e\) for the same-mesh \unstablepair{\(\mathcal{U}(4,\mathcal{C}^3,h)/\allowbreak\mathcal{P}(3,\mathcal{C}^{-1},h)\)} and coarsened-pressure \stablepair{\(\mathcal{U}(4,\mathcal{C}^3,h)/\allowbreak\mathcal{P}(3,\mathcal{C}^{-1},2h)\)} discretizations. The errors are evaluated relative to a reference solution
obtained using the \stablepair{Galerkin pair \(\mathcal{U}(6,\mathcal{C}^5,h)/\allowbreak\mathcal{P}(5,\mathcal{C}^4,2h)\)} on a \(256\times256\)
displacement mesh, which is eight times finer in each direction than the
\(32\times32\) mesh used in the parameter study. At small \(t\), the
nonconstant pressure modes are insufficiently stabilized and the error is
generally large. An exception is the quadratic \(2h\) case, for which
relatively small errors are observed even at weak stabilization. For this
pair, the \revx{full-space} inf-sup criterion \revx{of \cref{eq:discontinuous_stability_condition}} is exactly at the borderline,
\(2(2-1)=1-(-1)=2\), and therefore does not satisfy the required strict
inequality\revx{; the constant-mode criterion, \(2>1\), is satisfied}. Nevertheless, the corresponding unstabilized solution exhibits
only mild pressure oscillations, which explains its comparatively small
error as \(t\) approaches zero. Increasing \(t\) generally reduces the error
until an intermediate range is reached, after which over-stabilization
gradually increases the error. However, even for \(t=100\), the errors
remain relatively small. Moreover, the     \stablepair{\(\mathcal{U}(k,\mathcal{C}^{k-1},h)/\allowbreak\mathcal{P}(k-1,\mathcal{C}^{-1},2h)\)} family consistently attains
a smaller error than the corresponding same-mesh family. This is consistent
with the constant-mode analysis: The VMS stabilization controls the nonconstant elementwise pressure modes in
both cases. In addition, the
\stablepair{\(\mathcal{U}(k,\mathcal{C}^{k-1},h)/\allowbreak\mathcal{P}(k-1,\mathcal{C}^{-1},2h)\)} discretizations satisfy the
strict inf-sup condition for the elementwise constant pressure mode and
therefore provide the additional control required for the constant pressure
component.

\subsubsection{Immersed}

\begin{figure*}[!htbp]
    \centering
    \includegraphics[width=\linewidth]{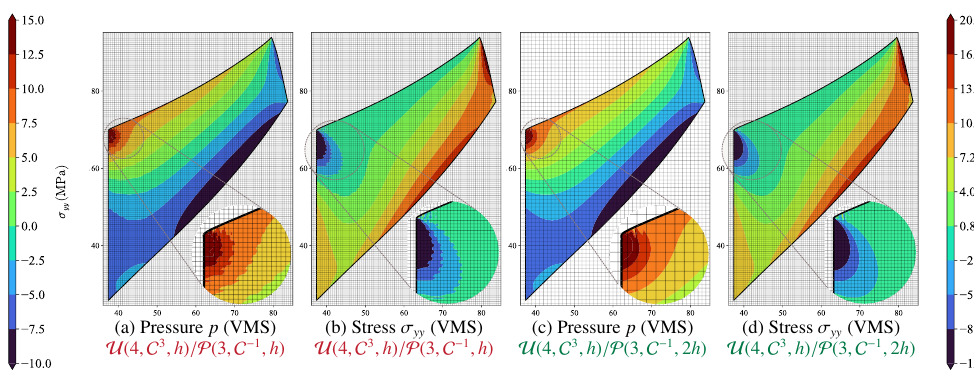}
    \caption{
    Pressure \(p\) and Cauchy stress \(\sigma_{yy}\) for the immersed
    Cook's membrane problem obtained using discontinuous cubic pressure
    spaces. Panels (a) and (b) use
    \unstablepair{\(\mathcal{U}(4,\mathcal{C}^3,h)/\mathcal{P}(3,\mathcal{C}^{-1},h)\)}, while panels (c) and (d) use the
    coarsened pressure grid
   \stablepair{ \(\mathcal{U}(4,\mathcal{C}^3,h)/\mathcal{P}(3,\mathcal{C}^{-1},2h)\)}.
    The shared stress colorbar is shown on the left, and the shared
    pressure colorbar is shown on the right.
    }
    \label{fig:cook_dg_immersed_q4q3_comparison}

\end{figure*}

\noindent
The higher-order discontinuous formulation is finally examined in the
immersed setting. Representative pressure and stress fields are shown in
\cref{fig:cook_dg_immersed_q4q3_comparison} for
\unstablepair{\(\mathcal{U}(4,\mathcal{C}^3,h)/\allowbreak\mathcal{P}(3,\mathcal{C}^{-1},h)\)} and
\stablepair{\(\mathcal{U}(4,\mathcal{C}^3,h)/\allowbreak\mathcal{P}(3,\mathcal{C}^{-1},2h)\)}, using a
\(120~\mathrm{mm}\times120~\mathrm{mm}\) Cartesian background domain
discretized into \(240\times240\) displacement-grid cells. Panels (a) and
(b) show the pressure and stress fields for the same-mesh
\unstablepair{\(\mathcal{U}(4,\mathcal{C}^3,h)/\allowbreak\mathcal{P}(3,\mathcal{C}^{-1},h)\)} discretization, whereas panels (c) and (d)
show the corresponding fields for the coarsened-pressure
\stablepair{\(\mathcal{U}(4,\mathcal{C}^3,h)/\allowbreak\mathcal{P}(3,\mathcal{C}^{-1},2h)\)} discretization.

Even at this relatively fine background-grid resolution, residual
oscillations associated with the lack of inf-sup control of the elementwise
constant pressure mode remain visible in the same-mesh \unstablepair{\(\mathcal{U}(4,\mathcal{C}^3,h)/\allowbreak\mathcal{P}(3,\mathcal{C}^{-1},h)\)} case. In contrast,
the coarsened-pressure \stablepair{ \(\mathcal{U}(4,\mathcal{C}^3,h)/\allowbreak\mathcal{P}(3,\mathcal{C}^{-1},2h)\)} discretization produces smooth pressure and stress
fields, consistent with the constant-mode stability established previously.
Importantly, the stabilized \stablepair{ \(\mathcal{U}(4,\mathcal{C}^3,h)/\allowbreak\mathcal{P}(3,\mathcal{C}^{-1},2h)\)} formulation remains well behaved in the
presence of nonconforming cut cells, including in the localized regions
adjacent to the immersed boundary. These results indicate that the combination of pressure-grid coarsening,
residual-based VMS stabilization, and the immersed cut-cell treatment preserves
the desired behavior of the higher-order discontinuous pressure formulation.


\section{Conclusions}
\label{sec:conclusions}

\noindent This work developed a residual-based stabilized mixed isogeometric
formulation for nearly incompressible elasticity, with particular emphasis
on higher-order discontinuous pressure approximations. The element-local
support of the discontinuous pressure space permits static condensation,
leaving a globally coupled system involving only displacement degrees of freedom. The
central result is that residual-based VMS stabilization controls the nonconstant
elementwise pressure modes, whereas the constant pressure mode must be
controlled separately through the inf--sup compatibility of the underlying
spaces. Consequently, a pressure grid coarsened by a fixed factor of two
relative to the displacement grid is sufficient for all polynomial orders
considered here. This avoids the progressively stronger pressure-grid
coarsening required by the corresponding Galerkin formulation while
preserving maximum displacement continuity and local pressure elimination.
\revx{The unconditional stability of the VMS variant rests on an exact energy identity, \cref{eq:vms_energy_identity}: at the energy pairing, the test operator coincides with the residual operator, so the stabilization contribution is a perfect square. The GLS and SUPG/PSPG variants produce, respectively, a difference of squares and an indefinite cross term, which is the precise origin of their conditional bounds.}
The numerical studies using the Kirsch and Cook's membrane problems support
the theoretical stability analysis for continuous, constant,
and higher-order discontinuous pressure spaces. The discrete inf--sup results distinguish the stable and unstable
approximation pairs even when the instability is not evident from the
convergence curves or from oscillations in the pressure and stress fields,
as observed for the smooth Kirsch problem. In the more complex Cook's membrane problem,
unstable pairs exhibit pronounced pressure and stress oscillations, whereas
the stable \(2h\) families remain well behaved. For higher-order
discontinuous pressure spaces, VMS stabilization effectively controls the
nonconstant pressure modes, while the \(2h\) pressure grid additionally
provides the required inf--sup control of the elementwise constant mode.

\revx{Two points of scope deserve restating. First, the macroelement criteria invoked throughout are sufficient conditions~\cite{8179196,antolin2017isogeometric}; instability of the pairs that fail them is established here by the discrete inf--sup test, not by the criteria themselves. Second, the sufficiency of constant-mode inf--sup stability for the stabilized discontinuous formulation follows from the classical analysis of Franca, Hughes, and Stenberg~\cite{FrancaHughesStenberg1993}. The ingredients of that argument do not require the displacement approximation to have exactly \(\mathcal{C}^0\) inter-element continuity and therefore extend to maximally continuous splines once the corresponding spline inverse estimate and constant-mode inf--sup condition are supplied; the latter is provided here by the macroelement analysis~\cite{antolin2017isogeometric}.}
\revx{On the practical side, the \(\chi\)-family presents a trade-off that should inform the choice of variant. The GLS form preserves the symmetry of the condensed displacement operator and can therefore be handed unchanged to a displacement solver built on conjugate gradients or a Cholesky factorization, at the cost of the conditional bound on the stabilization level. The VMS form is unconditionally stable but yields a nonsymmetric condensed operator. Where solver reuse dominates, GLS with a modest stabilization level is the natural choice; where robustness with respect to the stabilization parameter dominates, VMS is preferable.}

The formulation was further extended to immersed isogeometric analysis using
extended B-splines for continuous fields, pressure-cell aggregation for the discontinuous pressure field, subdivision-based cut-cell
quadrature, and weak enforcement of essential boundary conditions through
Nitsche's method. The immersed discrete inf--sup studies show that these
treatments preserve the intended stability properties of the underlying
displacement--pressure pairs, while the numerical solutions demonstrate
robust behavior in the presence of nonconforming cut cells. The resulting
framework therefore combines high-order approximation, residual
stabilization, local pressure elimination, and immersed geometric
flexibility within a unified formulation for nearly incompressible
elasticity.
Future work will consider extensions to finite-deformation and multiphysics problems in both mesh-conforming and immersed settings.


\backmatter

\bmhead{Acknowledgements}
\noindent The authors gratefully acknowledge funding from the U.S. National Science Foundation (NSF) under Award No.~CMMI-2549543.

\begin{appendices}
\renewcommand*{\theHequation}{appendix.\Alph{section}.\arabic{equation}}
\section{Numerical evaluation of the discrete inf--sup constant}
\label{app:numerical_inf_sup_test}

\noindent
The discrete inf--sup condition in \cref{eq:inf_sup_condition} can be
difficult to verify analytically for a particular family of approximation
spaces. We therefore also evaluate the discrete inf--sup constant numerically
using the generalized eigenvalue procedure of Chapelle and
Bathe~\cite{chapelle1993inf}. The test is performed over a sequence of
uniformly refined meshes to determine whether \(\beta_h\) remains bounded
away from zero as \(h\rightarrow0\).

Let \(\BK_{uu}\), \(\BK_{up}\), and \(\BM_p\) denote the displacement
stiffness matrix, displacement--pressure coupling matrix, and pressure mass
matrix, respectively, introduced in
\cref{eq:galerkin_block_system,eq:galerkin_pressure_mass_block}.
The displacement matrix \(\BK_{uu}\) defines the discrete
displacement-energy norm used in the numerical test. By Korn's inequality,
this norm is equivalent to the \(H^1\)-norm appearing in
\cref{eq:inf_sup_condition}, provided that the homogeneous essential
boundary conditions eliminate the rigid-body modes.

For a pressure coefficient vector \(\Bq\), evaluation of the supremum over
the discrete displacement space leads to the pressure Schur-complement
matrix
\begin{equation}
    \BS_h
    =
    \BK_{up}^{T}
    \BK_{uu}^{-1}
    \BK_{up}.
    \label{eq:numerical_infsup_schur_complement}
\end{equation}
The square of the discrete inf--sup constant is then obtained from the
generalized eigenvalue problem
\begin{equation}
    \BK_{up}^{T}
    \BK_{uu}^{-1}
    \BK_{up}
    \Bq
    =
    \lambda_h
    \BM_p
    \Bq.
    \label{eq:numerical_infsup_eigenproblem}
\end{equation}
Accordingly,
\begin{equation}
    \beta_h
    =
    \sqrt{\lambda_{\min}^{+}},
    \label{eq:numerical_infsup_beta}
\end{equation}
where \(\lambda_{\min}^{+}\) denotes the smallest positive generalized
eigenvalue after restricting the displacement space to eliminate rigid-body
modes.

The matrices entering the numerical inf--sup test correspond to the
underlying Galerkin displacement and pressure spaces; the finite-compressibility
and residual-based stabilization contributions are excluded. Consequently,
the computed value of \(\beta_h\) measures the intrinsic LBB compatibility
of the selected approximation spaces, independently of the additional
pressure control provided by finite compressibility or stabilization.

\revx{For the immersed formulation the essential boundary condition is imposed weakly, so the active displacement space carries no trace condition on \(\Gamma_D\), and the strain-energy matrix alone is singular on the rigid-body modes. The displacement norm used in the immersed test is therefore the one in which the immersed formulation is coercive: the strain energy augmented by the Nitsche boundary contribution of \cref{eq:immersed_nitsche_problem}, assembled on the extended displacement basis,
\begin{equation}
    \BK_{uu}^{\mathrm{imm}}
    =
    \BE_u^T\left(\BK_{uu}+\BK_{uu}^{N}\right)\BE_u,
    \qquad
    \left[\BK_{uu}^{N}\right]_{AB}
    =
    \sum_{K\in\calT_h^{\Gamma_D}}
    \int_{\Gamma_D\cap K}
    \gamma_{N,K}\frac{\mu}{h_{\Gamma,K}}
    N_A^u N_B^u
    \dA,
    \label{eq:immersed_numerical_infsup_Kuu}
\end{equation}
which is positive definite whenever \(\Gamma_D\) has sufficient measure to eliminate the rigid motions. The coupling and pressure mass matrices are assembled on the same extended displacement basis and on the extended or aggregated pressure basis,
\begin{equation}
    \BK_{up}^{\mathrm{imm}}
    =
    \BE_u^T
    \left(
        \BK_{up}+\BK_{up}^{N}
    \right)
    \BE_p,
    \qquad
    \BM_p^{\mathrm{imm}}
    =
    \BE_p^T\BM_p\BE_p,
    \label{eq:immersed_numerical_infsup_matrices}
\end{equation}
where the Nitsche contribution to the displacement--pressure coupling is associated with the boundary term
\begin{equation}
    \int_{\Gamma_D}
    p_h\,\Bn\cdot\Bv_h
    \dA,
\end{equation}
that is,
\begin{equation}
    \left[\BK_{up}^{N}\right]_{(Ai)B}
    =
    \sum_{K\in\calT_h^{\Gamma_D}}
    \int_{\Gamma_D\cap K}
    N_A^u\,n_i\,N_B^p
    \dA .
    \label{eq:immersed_numerical_infsup_KupN}
\end{equation}
Here \(i\) denotes the displacement component, and \(\BE_u\) and \(\BE_p\) are the displacement and pressure transformations introduced in \cref{eq:immersed_extension_matrix}. These three matrices replace \(\BK_{uu}\), \(\BK_{up}\), and \(\BM_p\) in \cref{eq:numerical_infsup_eigenproblem}. In practice the transformations are applied locally to the basis functions during assembly rather than formed explicitly. The resulting inf--sup constants therefore incorporate the effects of the cut-cell quadrature, the displacement extension, the pressure extension or aggregation, and the weak enforcement of the immersed essential boundary conditions.}
A displacement--pressure family is considered uniformly inf--sup stable if
\(\beta_h\) remains bounded away from zero under mesh refinement. In
contrast, a systematic decrease of \(\beta_h\) toward zero indicates a lack
of uniform inf--sup stability. The refinement behavior is quantified by
fitting
\begin{equation}
    \beta_h
    \approx
    C
    \left(
        \frac{h}{L_{\mathrm{ref}}}
    \right)^s,
    \label{eq:numerical_infsup_fitted_slope}
\end{equation}
where \(L_{\mathrm{ref}}\) is the characteristic length used to normalize
the mesh size. A fitted slope \(s\approx0\) indicates an approximately
mesh-independent inf--sup constant, whereas \(s>0\) indicates that
\(\beta_h\) deteriorates under mesh refinement.
\end{appendices}

\bibliography{PaperCitations}

\end{document}